%% file: 00-paper.tex
\documentclass[letter,seceqn,secthm]{elsart}

\usepackage{latexsym}   
\usepackage{amssymb}
\usepackage{amsmath}
\usepackage{stmaryrd}
\usepackage{graphicx}
\usepackage{pst-all} 
\usepackage{times}

\begin{document}

\begin{frontmatter}
\title{Interferometry of non-Abelian Anyons}

\author[StationQ,Caltech]{Parsa Bonderson\corauthref{cor}},
\corauth[cor]{Corresponding author.}
\ead{parsab@microsoft.com}
\author[UCR,Caltech]{Kirill Shtengel},
\ead{kirill.shtengel@ucr.edu}
\author[UCR,Caltech]{J.~K.~Slingerland}
\ead{joost.slingerland@gmail.com}
\address[StationQ]{Microsoft Research, Station Q, CNSI Building, University of California, Santa Barbara, CA 93106, USA}
\address[UCR]{Department of Physics and Astronomy, University of California,
Riverside,  CA 92521, USA}
\address[Caltech]{California Institute of Technology, Pasadena, CA 91125,
  USA}

\begin{abstract}
We develop the general quantum measurement theory of non-Abelian anyons
through interference experiments. The paper starts with a terse introduction
to the theory of anyon models, focusing on the basic formalism necessary to apply
standard quantum measurement theory to such systems. This is then applied to give
a detailed analysis of anyonic charge measurements using a Mach-Zehnder interferometer
for arbitrary anyon models. We find that, as anyonic probes are sent through the legs of the
interferometer, superpositions of the total anyonic charge located in the target region
collapse when they are distinguishable via monodromy with the probe anyons, which also
determines the rate of collapse. We give estimates on the number of probes needed to obtain a desired confidence level for the measurement outcome distinguishing between charges, and explicitly work out
a number of examples for some significant anyon models. We apply the same techniques to describe
interferometry measurements in a double point-contact interferometer realized in fractional quantum
Hall systems. To lowest order in tunneling, these results essentially match those from the Mach-Zehnder interferometer, but we also provide the corrections due to processes involving multiple tunnelings. Finally, we give explicit predictions describing state measurements for experiments in the Abelian hierarchy states,
the non-Abelian Moore-Read state at $\nu=5/2$ and Read-Rezayi state at $\nu = 12/5$.
\end{abstract}

\date{\today}

\end{frontmatter}

\input{01-intro}
\input{02-anyonmodels}
\input{03a-experiment}
\input{03b-oneprobe}
\input{03c-Nprobes}
\input{03d-W}
\input{03e-psandqs}
\input{03f-distinguishability}
\input{03g-targetgeneralizations}
\input{03g-probegeneralizations}
\input{03h-Examples}
\input{04-FQH_2PC}

\section*{Acknowledgements}

We thank L.~Bonderson, A.~Kitaev, I.~Klich, and J.~Preskill for illuminating
discussions, and acknowledge the hospitality of the IQI, Microsoft Station Q,
and the Aspen Center for Physics. This work was supported in part by the NSF
under Grant No.~PHY-0456720 and the ARO under Grant No.~W911NF-05-1-0294.

\bibliography{corr}
\bibliographystyle{elsart-num}
\end{document}

%% file: 01-intro.tex
\section{Introduction}

One of the striking differences between two and three spatial dimensions is
manifested in the allowed exchange statistics of quantum particles. In three
spatial dimensions, the allowed particle types can
be classified according to irreducible representations of the permutation
group. There are only two such one-dimensional representations, the trivial and alternating representations, for which exchange of any two particles introduces a factor of  $1$ and $-1$, respectively, to the wavefunction, corresponding to bosonic and fermionic statistics. Multi-dimensional
representations of the permutation group give rise to what is known as
``parastatistics''~\cite{Green53}, however, it has been shown that
parastatistics can be replaced by bosonic and fermionic statistics, if a
hidden degree of freedom (a non-Abelian isospin group) is introduced~\cite{Druhl70}.

In contrast, for two spatial dimensions the particles are classified
according to representations of the braid group~\cite{Leinaas77}.
The corresponding types of particles have been dubbed
``anyons''~\cite{Wilczek82a,Wilczek82b}, and their exchange statistics are
more precisely referred to as ``braiding statistics.''
Braiding statistics described by multi-dimensional irreducible representations
of the braid group~\cite{Goldin85} give rise to non-Abelian
anyons.\footnote{In this paper, the term ``anyon'' will be used in reference
to both the Abelian and non-Abelian varieties.}

The group representation theory used to characterize particles' braiding statistics,
 however, becomes progressively cumbersome if one attempts to describe a
system with several distinct ``species'' of anyons, especially those
corresponding to multi-dimensional representations.
Furthermore, one would typically like to consider systems in which there are
processes that do not conserve particle number, a notion unsupported by the
group theoretic language. To circumvent these shortcomings for systems with two
spatial dimensions, one may switch over
to the quantum field theoretic-type formalism of anyon models, in which the
topological and algebraic properties of the anyonic system are described by
category theory, rather than group theory. The structures of anyon models
originated from conformal field theory (CFT)~\cite{Moore88,Moore89b} and
Chern-Simons theory~\cite{Witten89}. They were further developed in terms of
algebraic quantum field theory~\cite{Fredenhagen89,Froehlich90}, and made
mathematically rigorous in the language of braided tensor
categories~\cite{Turaev94,Kassel95,Bakalov01}.

Surprisingly, even in our three-dimensional
universe, there are physical systems that are
\emph{effectively} two dimensional and have quasiparticles -- point-like
localized coherent state excitations that behave like particles -- that
appear to possess such exotic braiding statistics. In fact, some of these are
even strongly believed (though, thus far, experimentally unconfirmed) to be non-Abelian
anyons. Anyon models describe the topological behavior of
quasiparticle excitations in two-dimensional, many-body systems with an
energy gap that suppresses (non-topological) long-range interactions, and hence
an anyon model is said to characterize a system's ``topological order.''

The fractional quantum Hall effect is the most prominent example of
anyonic systems, so we will briefly review some relevant facts on the
subject. (For a general introduction into the quantum Hall effect we refer the
reader to Refs.~\cite{Prange87,Karlhede92,DasSarma97,Ezawa00}.)
The quantum Hall effect is an anomalous Hall effect that occurs in
two dimensional electron gases (2DEGs) subjected to strong transverse magnetic
fields ($\sim 10~\text{T}$) at very low
temperatures ($\sim 10~\text{mK}$). Under these conditions, the Hall
resistance
$R_{xy}$ develops plateaus while $R_{xx}$ develops a series of deep minima as
a function of the applied magnetic field.
These plateaus occur at values which are quantized to extreme precision in
integer~\cite{von_Klitzing80} or fractional~\cite{Tsui82} multiples of the
fundamental conductance quantum $e^{2}/h$. These multiples are the
filling fractions, usually denoted $\nu\equiv N_\text{e}/N_{\phi}$
where $N_\text{e}$ is the number of electrons and $N_{\phi}$ is the number of
fundamental flux quanta through the area occupied by the 2DEG at magnetic
field corresponding to the center of a plateau.
At the plateaus, the conductance tensor is off-diagonal, meaning a
dissipationless transverse current flows in response to an applied electric
field. In particular, the electric field generated by threading an additional
localized flux quantum through the system expels a net charge of $\nu e$,
thus creating a quasihole. Consequently, charge and flux are intimately coupled
together in the quantum Hall effect.

In the  fractional quantum Hall (FQH) regime, electrons form an incompressible
fluid state that supports localized excitations (quasiholes and quasiparticles)
which, for the simplest cases, carry one magnetic flux quantum and, hence,
fractional charge $\nu e$. This combination of fractional charge and unit flux
implies that they are anyons, due to their mutual Aharonov--Bohm effect. The
fractional charge of quasiparticles in the $\nu =1/3$ Laughlin state was
first measured in 1995 \cite{Goldman95}. Recently, a series of experiments
asserting verification of the fractional braiding statistics has been reported
\cite{Goldman05a,Camino05a,Camino05b,Camino06a,Camino07a,Camino07b}.

In the bulk of a FQH sample, the long-distance interactions between quasiholes are
purely topological and may be described by an anyon model. Boundary excitations
and currents of the Hall liquid are described by a $1+1$
dimensional conformal field theory~\cite{Wen92b} whose topological order is the same as that
of the bulk, when there is no edge reconstruction~\cite{Tsiper01,Wan02}. These boundary excitations
provide one way of coupling measurement devices to the 2DEG. A further
connection between the physics of the bulk and CFT
can be established following the observation in \cite{Moore91} that the
microscopic trial wavefunction describing the ground state of the
incompressible FQH liquid can be constructed from conformal blocks (CFT
correlators).

For the purpose of this paper, we are particularly interested in
the possibility of non-Abelian statistics existing at several
observed plateaus in the second Landau level
($2\le\nu\le 4$), in particular $\nu=5/2$, $7/2$, and
$12/5$~\cite{Willett87,Pan99,Eisenstein02,Xia04}.
Predictions of non-Abelian statistics in FQH states originated with the
paired state of Moore and Read~\cite{Moore91}, and was generalized by Read and Rezayi
to a series of clustered non-Abelian states~\cite{Read99}. At least for $\nu=5/2$
(the Moore--Read state) and $\nu=12/5$ (the $k=3,M=1$
Read--Rezayi state), these wavefunctions were found to have very good
overlap with the exact ground states obtained by numerical diagonalization of
small systems \cite{Morf98,Rezayi00}.

Detailed investigations of the braiding behavior of quasiholes of the
Moore--Read state were carried out in Ref.~\cite{Nayak96c}, and of the
$\nu=12/5$ state, as well as the other states in the Read--Rezayi
series in Ref.~\cite{Slingerland01}. Owing to the special feature of the
Moore--Read state as a weakly-paired state of a $p+\text{i}p$ superconductor of
composite fermions~\cite{Read00}, alternative explicit calculations of the
non-Abelian exchange statistics of quasiparticles were carried out in the
language of unpaired, zero-energy Majorana modes associated with the vortex
cores~\cite{Ivanov01,Stern04}. (Unfortunately, this language does not readily
adapt to give a similar interpretation for the other states in the Read--Rezayi
series.)

In addition to the proposed fractional quantum Hall states that could
host non-Abelian anyons \cite{Moore91,Read99,Ardonne99}, there are a number of
other more speculative proposals of systems that may be able to exhibit
non-Abelian braiding statistics. These include lattice models \cite%
{Kitaev97,Kitaev06a}, quantum loop gases \cite%
{Freedman03,Freedman04a,Freedman05a,Freedman05b}, string-net gases \cite%
{Turaev92,Levin05a,Fendley05a,Fidkowski07a}, Josephson junction arrays \cite%
{Doucot04}, $p+\text{i}p$ superconductors
\cite{DasSarma06a,Tewari07a,Gurarie05a}, and rapidly rotating bose condensates
\cite{Cooper01a,Cooper04a,Rezayi05}. Since non-Abelian anyons are
representative of an entirely new and exotic phase of matter, their discovery
would be of great importance, in and of itself. However, as additional
motivation, non-Abelian anyons could also turn out to be an invaluable
resource for quantum computing.

The idea to use the non-local, multi-dimensional state space shared by
non-Abelian anyons as a place to encode qubits was put forth by
Kitaev~\cite{Kitaev97}, and further developed in Refs.~\cite%
{Preskill98,Ogburn99,Freedman01,Freedman02a,FKLW,Mochon03,Mochon04}. The
advantage of this scheme, known as ``topological quantum computing,'' is that
the non-local state space is impervious to local perturbations, so the qubit
encoded there is ``topologically'' protected from errors. A model for
topological qubits in the Moore--Read state was proposed in
Ref.~\cite{DasSarma05},
however braiding operations alone in this state are not computationally
universal, severely limiting its usefulness in this regard. Nevertheless, one
may still hope to salvage the situation by supplementing braiding in the
Moore--Read state with topology changing operations~\cite{FNW05a,FNW05b} or
non-topologically protected operations~\cite{Bravyi06} to produce universality.
The greater hope, however, lies in the $k=3$ Read--Rezayi state, for which the
non-Abelian braiding statistics are essentially described by
the computationally universal ``Fibonnaci'' anyon model (see
Section~\ref{sec:Fib}). Consequently, the efforts
in ``topological quantum compiling'' (i.e. designing anyon braids that produce
desired computational gates) for this anyon model
\cite{Bonesteel05,Simon06a,Hormozi07a} may be applied directly.

The primary focus of this paper is to address the measurement theory
of anyonic charge. This provides a key element in detecting non-Abelian
statistics and correctly identifying the topological order of a system.
Furthermore, the ability to perform measurements of anyonic charge is a crucial
component of topological quantum computing, in particular for the purposes of
qubit initialization and readout. Clearly, the most direct way of probing
braiding statistics is through experiments that establish interference between
different braiding operations. In this vein, we will consider interferometry
experiments which probe braiding statistics via Aharonov--Bohm type
interactions~\cite{Aharonov59}, where probe anyons exhibit quantum interference
between homotopically distinct paths traveled around a target, producing
measurement distributions that distinguish different anyonic charges in the target.
This sort of experiment provides a quantum non-demolitional measurement~\cite{Braginsky75}
and is ideally suited for the qubit readout procedure in topological
quantum computing.

The paper is structured as follows:

In Section~\ref{sec:anyon_models}, we provide an introduction to the theory
of anyon models, giving all the essential background needed to understand the
rest of the paper, and establishing the connection with standard concepts of
quantum information theory.

In Section~\ref{sec:MZI}, we analyze a Mach-Zehnder type interferometer for an
arbitrary anyon model. We consider a target anyon allowed to be in a
superposition of anyonic states, and describe its collapse behavior resulting
from interferometry measurements by probe anyons. We find that
probe anyons will collapse any superpositions of states they can
distinguish by monodromy, as well as decohere anyonic charge entanglement that they can
detect between the target and outside anyons. We show how these measurements may
be used to determine the target's anyonic charge and/or help identify the
topological order of a system. We conclude this section by applying the results
to a few particularly relevant examples.

In Section~\ref{sec:FQH_2PC}, we consider a double point-contact interferometer
designed for fractional quantum Hall systems. We give the evolution operator
to all orders in tunneling, and apply the methods and results of
Section~\ref{sec:MZI} to describe how superpositions in the target
anyon state collapse as a result of interferometry measurements, and how to
determine the anyonic charge of the target. We give detailed predictions for the
Abelian hierarchy states, the Moore--Read state ($\nu=5/2,7/2$), and the $k=3,M=1$
Read--Rezayi state ($\nu=12/5$). 

%% file: 02-anyonmodels.tex
\section{Anyon Models}
  \label{sec:anyon_models}

In this section, we briefly review aspects of the theory of anyon models which are relevant to the rest of the paper. We follow the relatively concrete approach found in Refs.~\cite{Kitaev06a,Preskill-lectures},
and develop some concepts in this formalism that are essential for the treatment of the measurement
problem, such as the density matrix description of states and the partial trace and partial quantum trace.

\subsection{Fusion and Quantum Dimensions}

An anyon model has a finite set $\mathcal{C}$ of superselection sector
labels called topological or anyonic charges. These conserved charges obey a
commutative, associative fusion algebra%
\begin{equation}
a\times b=\sum\limits_{c\in \mathcal{C}}N_{ab}^{c}c
\end{equation}%
where the fusion multiplicities $N_{ab}^{c}$ are non-negative integers which
indicate the number of different ways the charges $a$ and $b$ can be
combined to produce the charge $c$. There is a unique trivial ``vacuum''
charge $1 \in \mathcal{C}$ for which $N_{a1}^{c}=\delta _{ac}$, and each charge
$a$ has a unique conjugate charge, or ``antiparticle,'' $\bar{a}\in \mathcal{C}$
such that $N_{ab}^{1}=\delta _{b\bar{a}}$. ($1=\bar{1}$ and $\bar{\bar{a}}=a$.)

In order to have a non-Abelian representation of the braid group (details on braiding follow), there must be at least one pair of charges $a$ and $b$ in the theory which have multiple fusion channels, i.e.%
\begin{equation}
\sum\limits_{c}N_{ab}^{c}>1,
\end{equation}
The domain of a sum will henceforth be left implicit when it runs over all
possible labels. Charges $a$ which have $\sum\nolimits_{c}N_{ab}^{c}=1$ for every $b$ must correspond to Abelian anyons (possibly bosons or fermions). Abelian and non-Abelian charges may also be distinguished by their \emph{quantum dimensions}. The quantum dimension $d_a$ of a charge $a$ is a measure for the amount of entropy contributed to the system by the presence of a particle of type $a$. It may be found from the fusion multiplicities by considering the asymptotic scaling of the number of possible fusion channels for $n$ anyons of charge $a$. For large $n$, this scales as $d_{a}^{n}$. Abelian charges have quantum dimension equal to $1$, while non-Abelian charges quantum dimensions strictly larger than $1$.
The \emph{total quantum dimension} of an anyon model is defined as
\begin{equation}
\mathcal{D}=\sqrt{\sum\limits_{a}d_{a}^{2}}.
\end{equation}

\subsection{States, Operators and Inner Product}

To each fusion product, there is assigned a fusion vector
space $V_{ab}^{c}$ with $\dim V_{ab}^{c}=N_{ab}^{c}$, and a corresponding
splitting space $V_{c}^{ab}$, which is the dual space. We pick some orthonormal
set of basis vectors $\left| a,b;c,\mu \right\rangle \in V_{c}^{ab}$\ ($%
\left\langle a,b;c,\mu \right| \in V_{ab}^{c}$) for these spaces, where $\mu
=1,\ldots ,N_{ab}^{c}$. If $N_{ab}^{c}=0$, then $V_{c}^{ab}$ is zero-dimensional and
it has no basis elements. We will sometimes use the notation
$c \in \left\{ a \times b \right\}$ to mean $c$ such that $N_{ab}^{c} \neq 0$.
Allowed splitting and fusion spaces involving the vacuum
charge have dimension one, and so we will leave their basis vector labels $%
\mu =1$ implicit.

It is extremely useful to employ a diagrammatic formalism for anyon models.
Each anyonic charge label is associated with an oriented line. It is useful
in some contexts to think of these lines as the anyons' worldlines (we will
consider time as increasing in the upward direction), however, such an
interpretation is not necessary nor even always appropriate. Reversing the
orientation of a line is equivalent to conjugating the charge labeling it, i.e.
\begin{equation}
\pspicture[shift=-0.7](0.4,-0.1)(1.05,1.5)
  \small
  \psset{linewidth=0.9pt,linecolor=black,arrowscale=1.5,arrowinset=0.15}
  \psline(0.6,0.1)(0.6,1.3)
  \psline{->}(0.6,0.1)(0.6,0.9)
  \rput[bl]{0}(0.75,0.41){$a$}
  \endpspicture
=
\pspicture[shift=-0.7](0.4,-0.1)(1.05,1.5)
  \small
  \psset{linewidth=0.9pt,linecolor=black,arrowscale=1.5,arrowinset=0.15}
  \psline(0.6,0.1)(0.6,1.3)
  \psline{-<}(0.6,0.1)(0.6,0.8)
  \rput[bl]{0}(0.75,0.41){$\bar{a}$}
  \endpspicture
.
\end{equation}%
The fusion and splitting states are assigned to trivalent vertices with the
appropriately corresponding anyonic charges:
\begin{equation}
\left( d_{c} / d_{a}d_{b} \right) ^{1/4}
\pspicture[shift=-0.6](-0.1,-0.2)(1.5,-1.2)
  \small
  \psset{linewidth=0.9pt,linecolor=black,arrowscale=1.5,arrowinset=0.15}
  \psline{-<}(0.7,0)(0.7,-0.35)
  \psline(0.7,0)(0.7,-0.55)
  \psline(0.7,-0.55) (0.25,-1)
  \psline{-<}(0.7,-0.55)(0.35,-0.9)
  \psline(0.7,-0.55) (1.15,-1)	
  \psline{-<}(0.7,-0.55)(1.05,-0.9)
  \rput[tl]{0}(0.4,0){$c$}
  \rput[br]{0}(1.4,-0.95){$b$}
  \rput[bl]{0}(0,-0.95){$a$}
 \scriptsize
  \rput[bl]{0}(0.85,-0.5){$\mu$}
  \endpspicture
=\left\langle a,b;c,\mu \right| \in
V_{ab}^{c} ,
\label{eq:bra}
\end{equation}
\begin{equation}
\left( d_{c} / d_{a}d_{b}\right) ^{1/4}
\pspicture[shift=-0.65](-0.1,-0.2)(1.5,1.2)
  \small
  \psset{linewidth=0.9pt,linecolor=black,arrowscale=1.5,arrowinset=0.15}
  \psline{->}(0.7,0)(0.7,0.45)
  \psline(0.7,0)(0.7,0.55)
  \psline(0.7,0.55) (0.25,1)
  \psline{->}(0.7,0.55)(0.3,0.95)
  \psline(0.7,0.55) (1.15,1)	
  \psline{->}(0.7,0.55)(1.1,0.95)
  \rput[bl]{0}(0.4,0){$c$}
  \rput[br]{0}(1.4,0.8){$b$}
  \rput[bl]{0}(0,0.8){$a$}
 \scriptsize
  \rput[bl]{0}(0.85,0.35){$\mu$}
  \endpspicture
=\left| a,b;c,\mu \right\rangle \in
V_{c}^{ab},
\label{eq:ket}
\end{equation}
where the normalization factors $\left( d_{c}/d_{a}d_{b}\right) ^{1/4}$ are
included so that diagrams are in the isotopy invariant convention throughout
this paper. Isotopy invariance means that the value of a (labeled)
diagram is not changed by continuous deformations, so long as open endpoints are
held fixed and lines are not passed through each other or around open endpoints.
Open endpoints should be thought of as ending on some boundary (e.g. a
timeslice or an edge of the system) through which
isotopy is not permitted. Building in isotopy invariance is a bit more
complicated than just making this normalization change, but for the purposes of
this paper, we can ignore the full details (which can be found
in~\cite{Kitaev06a,Bonderson07b}).

Inner products are formed diagrammatically by stacking
vertices so the fusing/splitting lines connect%
\begin{equation}
\label{eq:inner_product}
  \pspicture[shift=-0.95](-0.2,-0.35)(1.2,1.75)
  \small
  \psarc[linewidth=0.9pt,linecolor=black,border=0pt] (0.8,0.7){0.4}{120}{240}
  \psarc[linewidth=0.9pt,linecolor=black,arrows=<-,arrowscale=1.4,
    arrowinset=0.15] (0.8,0.7){0.4}{165}{240}
  \psarc[linewidth=0.9pt,linecolor=black,border=0pt] (0.4,0.7){0.4}{-60}{60}
  \psarc[linewidth=0.9pt,linecolor=black,arrows=->,arrowscale=1.4,
    arrowinset=0.15] (0.4,0.7){0.4}{-60}{15}
  \psset{linewidth=0.9pt,linecolor=black,arrowscale=1.5,arrowinset=0.15}
  \psline(0.6,1.05)(0.6,1.55)
  \psline{->}(0.6,1.05)(0.6,1.45)
  \psline(0.6,-0.15)(0.6,0.35)
  \psline{->}(0.6,-0.15)(0.6,0.25)
  \rput[bl]{0}(0.07,0.55){$a$}
  \rput[bl]{0}(0.94,0.55){$b$}
  \rput[bl]{0}(0.26,1.25){$c$}
  \rput[bl]{0}(0.24,-0.05){$c'$}
 \scriptsize
  \rput[bl]{0}(0.7,1.05){$\mu$}
  \rput[bl]{0}(0.7,0.15){$\mu'$}
  \endpspicture
=\delta _{c c ^{\prime }}\delta _{\mu \mu ^{\prime }} \sqrt{\frac{d_{a}d_{b}}{d_{c}}}
  \pspicture[shift=-0.95](0.15,-0.35)(0.8,1.75)
  \small
  \psset{linewidth=0.9pt,linecolor=black,arrowscale=1.5,arrowinset=0.15}
  \psline(0.6,-0.15)(0.6,1.55)
  \psline{->}(0.6,-0.15)(0.6,0.85)
  \rput[bl]{0}(0.75,1.25){$c$}
  \endpspicture
\end{equation}%
and this generalizes to more complicated diagrams. An important feature of this
relation is that it diagrammatically encodes charge conservation, and, in
particular, forbids tadpole diagrams. An important special case is $c=1$, which shows that
an unknotted loop carrying charge $a$ evaluates to its quantum dimension
\begin{equation}
\label{eq:loop=d}
  \pspicture[shift=-0.35](-0.08,0.25)(1.35,1.25)
  \small
  \psarc[linewidth=0.9pt,linecolor=black,arrows=<-,arrowscale=1.5,
    arrowinset=0.15] (0.8,0.7){0.5}{165}{363}
  \psarc[linewidth=0.9pt,linecolor=black,border=0pt]
(0.8,0.7){0.5}{0}{170}
  \rput[bl]{0}(-0.03,0.55){$a$}
 \endpspicture
=d_{a}=d_{\bar{a}}.
\end{equation}
The completeness relation for the identity operator on a pair of anyons with charges $a$ and $b$
respectively is written diagrammatically as
\begin{equation}
\label{eq:Id}
\pspicture[shift=-0.65](0,-0.5)(1.1,1.1)
  \small
  \psset{linewidth=0.9pt,linecolor=black,arrowscale=1.5,arrowinset=0.15}
  \psline(0.3,-0.45)(0.3,1)
  \psline{->}(0.3,-0.45)(0.3,0.50)
  \psline(0.8,-0.45)(0.8,1)
  \psline{->}(0.8,-0.45)(0.8,0.50)
  \rput[br]{0}(1.05,0.8){$b$}
  \rput[bl]{0}(0,0.8){$a$}
  \endpspicture
 = \sum\limits_{c,\mu }
\sqrt{\frac{d_{c}}{d_{a}d_{b}}} \;
 \pspicture[shift=-0.6](-0.1,-0.45)(1.4,1)
  \small
  \psset{linewidth=0.9pt,linecolor=black,arrowscale=1.5,arrowinset=0.15}
  \psline{->}(0.7,0)(0.7,0.45)
  \psline(0.7,0)(0.7,0.55)
  \psline(0.7,0.55) (0.25,1)
  \psline{->}(0.7,0.55)(0.3,0.95)
  \psline(0.7,0.55) (1.15,1)
  \psline{->}(0.7,0.55)(1.1,0.95)
  \rput[bl]{0}(0.38,0.2){$c$}
  \rput[br]{0}(1.4,0.8){$b$}
  \rput[bl]{0}(0,0.8){$a$}
  \psline(0.7,0) (0.25,-0.45)
  \psline{-<}(0.7,0)(0.35,-0.35)
  \psline(0.7,0) (1.15,-0.45)
  \psline{-<}(0.7,0)(1.05,-0.35)
  \rput[br]{0}(1.4,-0.4){$b$}
  \rput[bl]{0}(0,-0.4){$a$}
\scriptsize
  \rput[bl]{0}(0.85,0.4){$\mu$}
  \rput[bl]{0}(0.85,-0.03){$\mu$}
  \endpspicture
\; ,
\end{equation}
Any diagrammatic equation, such as this, is also valid as a local relation
within larger, more complicated diagrams.
Using Eq.~(\ref{eq:Id}), Eq.~(\ref{eq:inner_product}) and isotopy, we get the
following important relation, which expresses the compatibility of the quantum
dimensions with fusion.
\begin{equation}
d_{a}d_{b}=
  \pspicture[shift=-0.35](-0.08,0.25)(1.35,1.25)
  \small
  \psarc[linewidth=0.9pt,linecolor=black,arrows=<-,arrowscale=1.5,
    arrowinset=0.15] (0.8,0.7){0.5}{165}{363}
  \psarc[linewidth=0.9pt,linecolor=black,border=0pt] (0.8,0.7){0.5}{0}{170}
  \psarc[linewidth=0.9pt,linecolor=black,arrows=<-,arrowscale=1.3,
    arrowinset=0.15] (0.8,0.7){0.3}{162}{363}
  \psarc[linewidth=0.9pt,linecolor=black,border=0pt] (0.8,0.7){0.3}{0}{180}
  \rput[bl]{0}(-0.03,0.55){$a$}
  \rput[bl]{0}(0.64,0.55){$b$}
  \endpspicture
= \sum\limits_{ c,\mu } \sqrt{\frac{d_{c}}{d_{a}d_{b}}}
  \pspicture[shift=-1.05](-0.1,-0.45)(1.45,1.85)
  \small
  \psarc[linewidth=0.9pt,linecolor=black,border=0pt] (0.8,0.7){0.4}{120}{240}
  \psarc[linewidth=0.9pt,linecolor=black,arrows=<-,arrowscale=1.4,
    arrowinset=0.15] (0.8,0.7){0.4}{165}{240}
  \psarc[linewidth=0.9pt,linecolor=black,border=0pt] (0.4,0.7){0.4}{-60}{60}
  \psarc[linewidth=0.9pt,linecolor=black,arrows=->,arrowscale=1.4,
    arrowinset=0.15] (0.4,0.7){0.4}{-60}{15}
  \psset{linewidth=0.9pt,linecolor=black,arrowscale=1.5,arrowinset=0.15}
  \psline{->}(0.6,1.05)(0.6,1.45)
  \psline(0.6,0)(0.6,0.35)
  \psline(1.4,0)(1.4,1.4)
  \psarc[linewidth=0.9pt,linecolor=black,border=0pt] (1,1.4){0.4}{0}{180}
  \psarc[linewidth=0.9pt,linecolor=black,border=0pt] (1,0){0.4}{180}{360}
  \rput[bl]{0}(0.07,0.55){$a$}
  \rput[bl]{0}(0.94,0.55){$b$}
  \rput[bl]{0}(0.26,1.25){$c$}
\scriptsize
  \rput[bl]{0}(0.7,1.06){$\mu$}
  \rput[bl]{0}(0.7,0.17){$\mu$}
  \endpspicture
=\sum\limits_{c}N_{ab}^{c}d_{c}
\, .
\label{eq:two-circles}
\end{equation}
For general operators, we introduce the notation
\begin{equation}
 \pspicture[shift=-1.4](-1,-1.5)(0.9,1.5)
  \small
  \psframe[linewidth=0.9pt,linecolor=black,border=0](-0.8,-0.5)(0.8,0.5)
  \rput[bl]{0}(-0.15,-0.1){$X$}
  \rput[bl]{0}(-0.22,0.7){$\mathbf{\ldots}$}
  \rput[bl]{0}(-0.22,-0.75){$\mathbf{\ldots}$}
  \psset{linewidth=0.9pt,linecolor=black,arrowscale=1.5,arrowinset=0.15}
  \psline(0.6,0.5)(0.6,1)
  \psline(-0.6,0.5)(-0.6,1)
  \psline(0.6,-0.5)(0.6,-1)
  \psline(-0.6,-0.5)(-0.6,-1)
  \psline{->}(0.6,0.5)(0.6,0.9)
  \psline{->}(-0.6,0.5)(-0.6,0.9)
  \psline{-<}(0.6,-0.5)(0.6,-0.9)
  \psline{-<}(-0.6,-0.5)(-0.6,-0.9)
  \rput[bl](-0.8,1.05){$A_1$}
  \rput[bl](0.4,1.05){$A_m$}
  \rput[tl](-0.8,-1.05){$A'_1$}
  \rput[tl](0.4,-1.05){$A'_n$}
 \endpspicture
 = X \in V_{A_{1}^{\prime },\ldots ,A_{n}^{\prime }}^{A_{1},\ldots
,A_{m}}=\sum\limits_{\substack{ a_{1},\ldots ,a_{m} \\ a_{1}^{\prime
},\ldots ,a_{n}^{\prime }}}V_{a_{1}^{\prime },\ldots ,a_{n}^{\prime
}}^{a_{1},\ldots ,a_{m}}
\end{equation}
where a capitalized anyonic charge label means a (direct) sum over all
possible charges, so that the operator $X$ is defined for acting on any
$n$ anyon input and $m$ anyon output. The box stands for a linear combination of diagrams describing the action of the operator. The indices on
operators will be left implicit when they are contextually clear (and
unnecessary). Conjugation of a diagram or operator is carried out by simultaneously reflecting the
diagram across the horizontal plane and reversing the orientation of arrows.

Tensoring together two operators (on separate sets of anyons) is simply executed
by juxtaposition of their diagrams:
\begin{equation}
 \pspicture[shift=-1.4](-1.4,-1.5)(1.3,1.5)
  \small
  \psframe[linewidth=0.9pt,linecolor=black,border=0](-1.2,-0.5)(1.2,0.5)
  \rput[bl]{0}(-0.55,-0.15){$X \otimes Y$}
  \rput[bl]{0}(0.4,0.7){$\mathbf{\ldots}$}
  \rput[bl]{0}(0.4,-0.75){$\mathbf{\ldots}$}
  \rput[bl]{0}(-0.8,0.7){$\mathbf{\ldots}$}
  \rput[bl]{0}(-0.8,-0.75){$\mathbf{\ldots}$}
  \psset{linewidth=0.9pt,linecolor=black,arrowscale=1.5,arrowinset=0.15}
  \psline(1.0,0.5)(1.0,1)
  \psline(0.2,0.5)(0.2,1)
  \psline(1.0,-0.5)(1.0,-1)
  \psline(0.2,-0.5)(0.2,-1)
  \psline(-1.0,0.5)(-1.0,1)
  \psline(-0.2,0.5)(-0.2,1)
  \psline(-1.0,-0.5)(-1.0,-1)
  \psline(-0.2,-0.5)(-0.2,-1)
  \psline{->}(1.0,0.5)(1.0,0.9)
  \psline{->}(0.2,0.5)(0.2,0.9)
  \psline{-<}(1.0,-0.5)(1.0,-0.9)
  \psline{-<}(0.2,-0.5)(0.2,-0.9)
  \psline{->}(-1.0,0.5)(-1.0,0.9)
  \psline{->}(-0.2,0.5)(-0.2,0.9)
  \psline{-<}(-1.0,-0.5)(-1.0,-0.9)
  \psline{-<}(-0.2,-0.5)(-0.2,-0.9)
 \endpspicture
 =
  \pspicture[shift=-1.4](-1,-1.5)(0.9,1.5)
  \small
  \psframe[linewidth=0.9pt,linecolor=black,border=0](-0.8,-0.5)(0.8,0.5)
  \rput[bl]{0}(-0.15,-0.12){$X$}
  \rput[bl]{0}(-0.22,0.7){$\mathbf{\ldots}$}
  \rput[bl]{0}(-0.22,-0.75){$\mathbf{\ldots}$}
  \psset{linewidth=0.9pt,linecolor=black,arrowscale=1.5,arrowinset=0.15}
  \psline(0.6,0.5)(0.6,1)
  \psline(-0.6,0.5)(-0.6,1)
  \psline(0.6,-0.5)(0.6,-1)
  \psline(-0.6,-0.5)(-0.6,-1)
  \psline{->}(0.6,0.5)(0.6,0.9)
  \psline{->}(-0.6,0.5)(-0.6,0.9)
  \psline{-<}(0.6,-0.5)(0.6,-0.9)
  \psline{-<}(-0.6,-0.5)(-0.6,-0.9)
\endpspicture
 \pspicture[shift=-1.4](-1,-1.5)(0.9,1.5)
  \small
  \psframe[linewidth=0.9pt,linecolor=black,border=0](-0.8,-0.5)(0.8,0.5)
  \rput[bl]{0}(-0.15,-0.12){$Y$}
  \rput[bl]{0}(-0.22,0.7){$\mathbf{\ldots}$}
  \rput[bl]{0}(-0.22,-0.75){$\mathbf{\ldots}$}
  \psset{linewidth=0.9pt,linecolor=black,arrowscale=1.5,arrowinset=0.15}
  \psline(0.6,0.5)(0.6,1)
  \psline(-0.6,0.5)(-0.6,1)
  \psline(0.6,-0.5)(0.6,-1)
  \psline(-0.6,-0.5)(-0.6,-1)
  \psline{->}(0.6,0.5)(0.6,0.9)
  \psline{->}(-0.6,0.5)(-0.6,0.9)
  \psline{-<}(0.6,-0.5)(0.6,-0.9)
  \psline{-<}(-0.6,-0.5)(-0.6,-0.9)
 \endpspicture
\end{equation}

\subsection{Associativity}

The splitting of three anyons with charges
$a,b,c$ from the charge $d$ corresponds to a space $V_{d}^{abc}$ which can be
decomposed into tensor products of two anyon splitting spaces by matching
the intermediate charge. This can be done in two isomorphic ways
\begin{equation}
V_{d}^{abc}\cong \bigoplus\limits_{e}V_{e}^{ab}\otimes V_{d}^{ec}\cong
\bigoplus\limits_{f}V_{d}^{af}\otimes V_{f}^{bc}.
\end{equation}
To incorporate the notion of associativity at the level of splitting spaces,
we need to specify a set of unitary isomorphisms between different
decompositions that are to be considered simply
a change of basis. These isomorphisms (called $F$-moves) are written
diagrammatically as
\begin{equation}
  \pspicture[shift=-1.0](0,-0.45)(1.8,1.8)
  \small
  \psset{linewidth=0.9pt,linecolor=black,arrowscale=1.5,arrowinset=0.15}
  \psline(0.2,1.5)(1.4,0)
  \psline(1.8,1.5) (1,0.5)
  \psline(0.6,1) (1,1.5)
   \psline{->}(0.6,1)(0.3,1.375)
   \psline{->}(0.6,1)(0.9,1.375)
   \psline{->}(1,0.5)(1.7,1.375)
   \psline{->}(1,0.5)(0.7,0.875)
   \psline{->}(1.4,0)(1.1,0.375)
   \rput[bl]{0}(0.05,1.6){$a$}
   \rput[bl]{0}(0.95,1.6){$b$}
   \rput[bl]{0}(1.75,1.6){${c}$}
   \rput[bl]{0}(0.5,0.5){$e$}
   \rput[bl]{0}(1.35,-0.3){$d$}
 \scriptsize
   \rput[bl]{0}(0.3,0.8){$\alpha$}
   \rput[bl]{0}(0.78,0.2){$\beta$}
  \endpspicture
= \sum_{f,\mu,\nu} \left[F_d^{abc}\right]_{(e,\alpha,\beta)(f,\mu,\nu)}
 \pspicture[shift=-1.0](0,-0.45)(1.8,1.8)
  \small
  \psset{linewidth=0.9pt,linecolor=black,arrowscale=1.5,arrowinset=0.15}
  \psline(0.2,1.5)(1.4,0)
  \psline(1.8,1.5) (1,0.5)
  \psline(1.4,1) (1,1.5)
   \psline{->}(0.6,1)(0.3,1.375)
   \psline{->}(1.4,1)(1.1,1.375)
   \psline{->}(1,0.5)(1.7,1.375)
   \psline{->}(1,0.5)(1.3,0.875)
   \psline{->}(1.4,0)(1.1,0.375)
   \rput[bl]{0}(0.05,1.6){$a$}
   \rput[bl]{0}(0.95,1.6){$b$}
   \rput[bl]{0}(1.75,1.6){${c}$}
   \rput[bl]{0}(1.25,0.45){$f$}
   \rput[bl]{0}(1.35,-0.3){$d$}
 \scriptsize
   \rput[bl]{0}(1.5,0.8){$\mu$}
   \rput[bl]{0}(0.78,0.2){$\nu$}
  \endpspicture
  .
\end{equation}
The same notion
of associativity is, of course, true for fusion of three anyons. The
associativity for fusion is given by
$F^{\dagger }$, and together with unitarity, we have
\begin{equation}
\left[ \left( F_{d}^{abc}\right) ^{\dagger }\right]
_{\left( f,\mu ,\nu \right) \left( e,\alpha ,\beta \right) } =
\left[ F_{d}^{abc}\right] _{\left( e,\alpha ,\beta \right) \left( f,\mu
,\nu \right) }^{\ast } =
\left[ \left( F_{d}^{abc}\right) ^{-1}\right] _{\left( f,\mu
,\nu \right) \left( e,\alpha ,\beta \right) }
.
\end{equation}
For fusion and splitting of more anyons, one does the obvious iteration of such
decompositions. For this to be consistent, the $F$-symbols must satisfy a
constraint called the Pentagon equation. One also imposes the physical
requirement that fusion and splitting with the vacuum charge does not change the
state. This means in particular that in diagrams, we may move, add, and delete
vacuum lines at will (in fact, we already did this in Eq.~\ref{eq:loop=d}).
Despite the constraints, the $F$-symbols have a certain amount of ``gauge
freedom,'' which comes from the fact that we are free to choose
bases for the vertex spaces.

We will also need the $F$-move with one of its legs bent down
\begin{equation}
\pspicture[shift=-0.6](0,-0.4)(1.1,1.2)
\small
  \psset{linewidth=0.9pt,linecolor=black,arrowscale=1.5,arrowinset=0.15}
  \psline(0.3,-0.45)(0.3,1)
  \psline{->}(0.3,-0.45)(0.3,-0.05)
  \psline{->}(0.3,0.5)(0.3,0.85)
  \psline(0.8,-0.45)(0.8,1)
  \psline{->}(0.8,-0.45)(0.8,-0.05)
  \psline{->}(0.8,0)(0.8,0.85)
  \psline(0.8,0.05)(0.3,0.45)
  \psline{->}(0.8,0.05)(0.45,0.33)
  \rput[bl]{0}(0.48,0.38){$e$}
  \rput[bl]{0}(0.93,0.8){$b$}
  \rput[bl]{0}(0,0.8){$a$}
  \rput[bl]{0}(0.95,-0.4){$d$}
  \rput[bl]{0}(-0.05,-0.4){$c$}
  \scriptsize
  \rput[bl]{0}(0.02,0.35){$\alpha$}
  \rput[bl]{0}(0.87,-0.1){$\beta$}
  \endpspicture
=\sum\limits_{f,\mu
,\nu }\left[ F_{cd}^{ab}\right] _{\left( e,\alpha ,\beta \right) \left(
f,\mu ,\nu \right)}
 \pspicture[shift=-0.6](-0.1,-0.45)(1.4,1)
  \small
  \psset{linewidth=0.9pt,linecolor=black,arrowscale=1.5,arrowinset=0.15}
  \psline{->}(0.7,0)(0.7,0.45)
  \psline(0.7,0)(0.7,0.55)
  \psline(0.7,0.55) (0.25,1)
  \psline{->}(0.7,0.55)(0.3,0.95)
  \psline(0.7,0.55) (1.15,1)
  \psline{->}(0.7,0.55)(1.1,0.95)
  \rput[bl]{0}(0.38,0.1){$f$}
  \rput[br]{0}(1.4,0.8){$b$}
  \rput[bl]{0}(0,0.8){$a$}
  \psline(0.7,0) (0.25,-0.45)
  \psline{-<}(0.7,0)(0.35,-0.35)
  \psline(0.7,0) (1.15,-0.45)
  \psline{-<}(0.7,0)(1.05,-0.35)
  \rput[br]{0}(1.4,-0.4){$d$}
  \rput[bl]{0}(0,-0.4){$c$}
\scriptsize
  \rput[bl]{0}(0.85,0.4){$\mu$}
  \rput[bl]{0}(0.85,-0.03){$\nu$}
  \endpspicture
\end{equation}
which is also a unitary transformation. {}From Eq.~(\ref{eq:Id}), we immediately
find that
\begin{equation}
\left[ F_{ab}^{ab}\right] _{1 \left( c,\mu ,\nu \right) }=
\left[ \left( F_{ab}^{ab}\right) ^{-1}\right] _{\left( c,\mu ,\nu \right) 1}
=\sqrt{\frac{d_{c}}{%
d_{a}d_{b}}}\;\delta _{\mu \nu }
,
\label{eq:idF}
\end{equation}
and more generally, applying Eqs.~(\ref{eq:Id}) and (\ref{eq:inner_product})
gives a relation between the two types of $F$-symbols
\begin{equation}
\left[ F_{cd}^{ab}\right] _{\left( e,\alpha ,\beta \right) \left( f,\mu
,\nu \right) } =\sqrt{\frac{d_{e}d_{f}}{d_{a}d_{d}}}\left[ F_{f}^{ceb}%
\right] _{\left( a,\alpha ,\mu \right) \left( d,\beta ,\nu \right) }^{\ast }
.
\end{equation}

\subsection{Trace and Partial Trace}

The trace over operators formed from bras and kets is defined in the usual way.
To translate the trace into the diagrammatic formalism, one defines the
\emph{quantum trace}, denoted $\widetilde{\text{Tr}}$, by closing the diagram
with loops that match the outgoing lines with the
respective incoming lines at the same position%
\begin{equation}
\widetilde{\text{Tr}}X =
\widetilde{\text{Tr}}
\left[
 \pspicture[shift=-1.4](-1,-1.5)(1,1.5)
  \small
  \psframe[linewidth=0.9pt,linecolor=black,border=0](-0.8,-0.5)(0.8,0.5)
  \rput[bl]{0}(-0.15,-0.1){$X$}
  \rput[bl]{0}(-0.22,0.7){$\mathbf{\ldots}$}
  \rput[bl]{0}(-0.22,-0.75){$\mathbf{\ldots}$}
  \psset{linewidth=0.9pt,linecolor=black,arrowscale=1.5,arrowinset=0.15}
  \psline(0.6,0.5)(0.6,1)
  \psline(-0.6,0.5)(-0.6,1)
  \psline(0.6,-0.5)(0.6,-1)
  \psline(-0.6,-0.5)(-0.6,-1)
  \psline{->}(0.6,0.5)(0.6,0.9)
  \psline{->}(-0.6,0.5)(-0.6,0.9)
  \psline{-<}(0.6,-0.5)(0.6,-0.9)
  \psline{-<}(-0.6,-0.5)(-0.6,-0.9)
  \rput[bl](-0.8,1.05){$A_1$}
  \rput[bl](0.4,1.05){$A_n$}
  \rput[tl](-0.8,-1.05){$A'_1$}
  \rput[tl](0.4,-1.05){$A'_n$}
 \endpspicture
\right]
=
 \pspicture[shift=-1.1](-1.1,-1.2)(2.2,1.2)
  \small
  \psframe[linewidth=0.9pt,linecolor=black,border=0](-0.8,-0.5)(0.8,0.5)
  \rput[bl]{0}(-0.15,-0.1){$X$}
  \rput[bl]{0}(-0.4,0.7){$\mathbf{\ldots}$}
  \rput[bl]{0}(-0.22,-0.75){$\mathbf{\ldots}$}
  \rput[bl]{0}(1.52,0){$\mathbf{\ldots}$}
  \psset{linewidth=0.9pt,linecolor=black,arrowscale=1.5,arrowinset=0.15}
  \psarc(1.0,0.5){0.4}{0}{180}
  \psarc(1.0,-0.5){0.4}{180}{360}
  \psarc(0,0.5){0.6}{90}{180}
  \psarc(0,-0.5){0.6}{180}{270}
  \psarc(1.5,0.5){0.6}{0}{90}
  \psarc(1.5,-0.5){0.6}{270}{360}
  \psline(1.4,-0.5)(1.4,0.5)
  \psline(0,1.1)(1.5,1.1)
  \psline(0,-1.1)(1.5,-1.1)
  \psline(2.1,-0.5)(2.1,0.5)
  \psline{->}(1.4,0.2)(1.4,-0.1)
  \psline{->}(2.1,0.2)(2.1,-0.1)
  \rput[bl](-1.07,0.6){$A_1$}
  \rput[bl](0.1,0.6){$A_n$}
 \endpspicture
.
\end{equation}%
Connecting the endpoints of two lines labeled by different anyonic charges
violates charge conservation, so such diagrams evaluate to zero. The operator
$X\in V_{A_{1}^{\prime}\ldots A_{n}^{\prime }}^{A_{1}\ldots A_{n}}$ may be
written as%
\begin{equation}
X =\sum\limits_{c}X_{c}, \quad \quad
X_{c} \in V_{c}^{A_{1}\ldots A_{n}}\otimes
V_{A_{1}^{\prime }\ldots A_{n}^{\prime }}^{c}
\end{equation}%
(note that this decomposition is basis independent), which may be used to
relate the quantum trace and the standard trace of bras and kets via%
\begin{equation}
\text{Tr}X =\sum\limits_{c}\frac{1}{d_{c}}\widetilde{\text{Tr}}X_{c},
\quad \quad
\widetilde{\text{Tr}}X =\sum\limits_{c}d_{c}\text{Tr}X_{c}
.
\end{equation}
Note that these are the same when the overall charge of the system is the vacuum charge
(or any Abelian charge for that matter).

We also need to define the \emph{partial} traces for anyons. Since we have not
yet introduced braiding, in order to take the partial trace
over a single anyon $B$, the planar structure requires that it must be one of
the two outer anyons (i.e. the first or last in the lineup). Physically,
this corresponds to the fact that one cannot treat the subsystem excluding
$B$ as independent of $B$ if this anyon is still located in the midst of the
remaining anyons. The partial quantum trace over $B$ of an operator $X\in
V_{A_{1}^{\prime },\ldots ,A_{n}^{\prime },B^{\prime }}^{A_{1},\ldots
,A_{n},B}$ is defined by looping only the line for anyon $B$ back on itself%
\begin{equation}
\widetilde{\text{Tr}}_{B}X =
 \pspicture[shift=-1.35](-0.9,-1.45)(1.9,1.4)
  \small
  \psframe[linewidth=0.9pt,linecolor=black,border=0](-0.8,-0.5)(0.8,0.5)
  \rput[bl]{0}(-0.15,-0.1){$X$}
  \rput[bl]{0}(-0.32,0.7){$\mathbf{\ldots}$}
  \rput[bl]{0}(-0.32,-0.75){$\mathbf{\ldots}$}
  \psset{linewidth=0.9pt,linecolor=black,arrowscale=1.5,arrowinset=0.15}
  \psline(0.4,0.5)(0.4,1)
  \psline(-0.6,0.5)(-0.6,1)
  \psline(0.4,-0.5)(0.4,-1)
  \psline(-0.6,-0.5)(-0.6,-1)
 \psline(1.4,-0.5)(1.4,0.5)
  \psline{->}(0.4,0.5)(0.4,0.9)
  \psline{->}(-0.6,0.5)(-0.6,0.9)
  \psline{-<}(0.4,-0.5)(0.4,-0.9)
  \psline{-<}(-0.6,-0.5)(-0.6,-0.9)
  \psline{->}(1.4,0.2)(1.4,-0.15)
  \psarc(1.0,0.5){0.4}{0}{180}
  \psarc(1.0,-0.5){0.4}{180}{360}
  \rput[tl](-0.76,1.35){$A_1$}
  \rput[tl](0.25,1.35){$A_n$}
  \rput[br](1.82,-0.1){$B$}
  \rput[tl](-0.76,-1.05){$A'_1$}
  \rput[tl](0.25,-1.05){$A'_n$}
 \endpspicture
\end{equation}
and for $X\in V_{B^{\prime },A_{1}^{\prime },\ldots ,A_{n}^{\prime
}}^{B,A_{1},\ldots ,A_{n}}$ as
\begin{equation}
\widetilde{\text{Tr}}_{B}X =
 \pspicture[shift=-1.35](-1.9,-1.45)(0.9,1.4)
  \small
  \psframe[linewidth=0.9pt,linecolor=black,border=0](-0.8,-0.5)(0.8,0.5)
  \rput[bl]{0}(-0.15,-0.1){$X$}
  \rput[bl]{0}(-0.12,0.7){$\mathbf{\ldots}$}
  \rput[bl]{0}(-0.12,-0.75){$\mathbf{\ldots}$}
  \psset{linewidth=0.9pt,linecolor=black,arrowscale=1.5,arrowinset=0.15}
  \psline(-0.4,0.5)(-0.4,1)
  \psline(0.6,0.5)(0.6,1)
  \psline(-0.4,-0.5)(-0.4,-1)
  \psline(0.6,-0.5)(0.6,-1)
 \psline(-1.4,-0.5)(-1.4,0.5)
  \psline{->}(-0.4,0.5)(-0.4,0.9)
  \psline{->}(0.6,0.5)(0.6,0.9)
  \psline{-<}(-0.4,-0.5)(-0.4,-0.9)
  \psline{-<}(0.6,-0.5)(0.6,-0.9)
  \psline{->}(-1.4,0.2)(-1.4,-0.15)
  \psarc(-1.0,0.5){0.4}{0}{180}
  \psarc(-1.0,-0.5){0.4}{180}{360}
  \rput[tl](0.46,1.35){$A_n$}
  \rput[tl](-0.55,1.35){$A_1$}
  \rput[bl](-1.82,-0.1){$B$}
  \rput[tl](0.46,-1.05){$A'_n$}
  \rput[tl](-0.55,-1.05){$A'_1$}
 \endpspicture
.
\end{equation}
To relate the partial quantum trace to the partial trace, we implement
factors for the quantum dimensions of the overall charges of the operator
before \emph{and} after the partial trace
\begin{equation}
\text{Tr}_{B}X =\sum\limits_{c,f}\frac{d_{f}}{d_{c}}\left[
\widetilde{\text{Tr}}_{B}X_{c}%
\right] _{f}, \quad \quad
\widetilde{\text{Tr}}_{B}X =\sum\limits_{c,f}\frac{d_{c}}{d_{f}}\left[
\text{Tr}_{B}X_{c} \right] _{f}
,
\end{equation}
where
\begin{equation}
\widetilde{\text{Tr}}_{B}X_{c} =\sum\limits_{f}\left[
\widetilde{\text{Tr}}_{B}X_{c}\right] _{f}, \quad \quad
\left[ \widetilde{\text{Tr}}_{B}X_{c}\right] _{f} \in V_{f}^{A_{1},\ldots
,A_{n}}\otimes V_{A_{1}^{\prime },\ldots ,A_{n}^{\prime }}^{f}
.
\end{equation}%
The partial trace and partial quantum trace over the subsystem of anyons
\newline $B=\left\{ B_{1},\ldots ,B_{n}\right\} $ that are sequential outer
lines (on either, possibly alternating, sides) of an operator is defined by
iterating the partial quantum trace on the $B$ anyons
\begin{equation}
\text{Tr}_{B} =\text{Tr}_{B_{1}}\ldots \text{Tr}_{B_{n}}, \quad \quad
\widetilde{\text{Tr}}_{B} =\widetilde{\text{Tr}}_{B_{1}}\ldots
\widetilde{\text{Tr}}_{B_{n}}
\end{equation}
Iterating these over all the anyons of a system returns the trace and quantum
trace, respectively, as they should.

Using Eq.~(\ref{eq:idF}) and the fact that tadpole diagrams
evaluate to zero, we may calculate the partial quantum trace for basis elements
of two-particle operators
\begin{eqnarray}
\widetilde{\text{Tr}}_{B} \left[
 \pspicture[shift=-1.2](-0.1,-1)(1.45,1.3)
 \small
  \psset{linewidth=0.9pt,linecolor=black,arrowscale=1.5,arrowinset=0.15}
  \psline{->}(0.7,0)(0.7,0.45)
  \psline(0.7,0)(0.7,0.55)
  \psline(0.7,0.55) (0.2,1.05)
  \psline{->}(0.7,0.55)(0.3,0.95)
  \psline(0.7,0.55) (1.2,1.05)
  \psline{->}(0.7,0.55)(1.1,0.95)
  \rput[bl]{0}(0.33,0.15){$c$}
  \rput[bl]{0}(1.12,1.13){$b$}
  \rput[bl]{0}(0.1,1.13){$a$}
  \psline(0.7,0) (0.2,-0.5)
  \psline{-<}(0.7,0)(0.35,-0.35)
  \psline(0.7,0) (1.2,-0.5)
  \psline{-<}(0.7,0)(1.05,-0.35)
  \rput[bl]{0}(1.12,-0.91){$b'$}
  \rput[bl]{0}(0.1,-0.91){$a'$}
  \scriptsize
  \rput[bl]{0}(0.82,0.38){$\mu$}
  \rput[bl]{0}(0.82,-0.02){$\mu'$}
  \endpspicture
\right] &=&
 \pspicture[shift=-1.2](-0.1,-1)(1.95,1.3)
  \small
  \psset{linewidth=0.9pt,linecolor=black,arrowscale=1.5,arrowinset=0.15}
  \psline{->}(0.7,0)(0.7,0.45)
  \psline(0.7,0)(0.7,0.55)
  \psline(0.7,0.55) (0.2,1.05)
  \psline{->}(0.7,0.55)(0.3,0.95)
  \psline(0.7,0.55) (1.2,1.05)
  \psline{->}(0.7,0.55)(1.1,0.95)
  \rput[bl]{0}(0.33,0.15){$c$}
  \rput[bl]{0}(1.02,1.13){$b$}
  \rput[bl]{0}(0.1,1.13){$a$}
  \psline(0.7,0) (0.2,-0.5)
  \psline{-<}(0.7,0)(0.35,-0.35)
  \psline(0.7,0) (1.2,-0.5)
  \psline{-<}(0.7,0)(1.05,-0.35)
 \psarc[linewidth=0.9pt,linecolor=black,border=0pt] (1.45,0.8){0.35}{0}{135}
 \psarc[linewidth=0.9pt,linecolor=black,border=0pt]
   (1.45,-0.25){0.35}{225}{360}
 \psline(1.8,-0.25)(1.8,0.8)
\psline{->}(1.8,0.8)(1.8,0.1)
  \rput[bl]{0}(1.02,-0.91){$b'$}
  \rput[bl]{0}(0.1,-0.91){$a'$}
  \scriptsize
  \rput[bl]{0}(0.82,0.38){$\mu$}
  \rput[bl]{0}(0.82,-0.02){$\mu'$}
  \endpspicture
= \sum\limits_{e,\alpha ,\beta}
\left[ \left( F_{a^{\prime} b^{\prime}}^{ab}\right) ^{-1}\right] _{\left( c,\mu ,\mu^{\prime} \right) \left( e,\alpha ,\beta \right)}
\pspicture[shift=-1.16](-0.8,-1.25)(0.8,1.15)
  \small
  \psellipse[linewidth=0.9pt,linecolor=black,border=0](0.2,0.0)(0.18,0.4)
  \psset{linewidth=0.9pt,linecolor=black,arrowscale=1.4,arrowinset=0.15}
  \psline{-<}(0.358,-0.1)(0.365,0.1)
  \psset{linewidth=0.9pt,linecolor=black,arrowscale=1.5,arrowinset=0.15}
  \psline(-0.45,-0.775)(-0.45,0.775)
  \psline{>-}(-0.45,0.3)(-0.45,0.6)
  \psline{<-}(-0.45,-0.3)(-0.45,-0.6)
  \psline(-0.45,0.15)(0.03,-0.05)
  \psline{-<}(-0.45,0.15)(-0.09,0.0)
  \rput[br]{0}(-0.35,0.875){$a$}
  \rput[br]{0}(-0.32,-1.2){$a'$}
  \rput[br]{0}(-0.08,0.17){$e$}
  \rput[bl]{0}(0.5,-0.1){$b$}
  \scriptsize
   \rput[br]{0}(-0.5,0.08){$\alpha$}
   \rput[bl]{0}(-0.18,-0.38){$\beta$}
  \endpspicture
\nonumber \\
&=&
\left[ \left( F_{ab}^{ab}\right) ^{-1}\right] _{\left( c,\mu ,\mu^{\prime} \right) 1}
\pspicture[shift=-1.06](-0.8,-1.15)(0.8,1.15)
  \small
  \psellipse[linewidth=0.9pt,linecolor=black,border=0](0.2,0.0)(0.18,0.4)
  \psset{linewidth=0.9pt,linecolor=black,arrowscale=1.4,arrowinset=0.15}
  \psline{-<}(0.358,-0.1)(0.365,0.1)
  \psset{linewidth=0.9pt,linecolor=black,arrowscale=1.5,arrowinset=0.15}
  \psline(-0.35,-0.775)(-0.35,0.775)
  \psline{->}(-0.35,-0.3)(-0.35,0.2)
  \rput[br]{0}(-0.5,-0.1){$a$}
  \rput[bl]{0}(0.5,-0.1){$b$}
  \endpspicture
= \sqrt{\frac{d_{b}d_{c}}{d_{a}}} \delta_{\mu \mu^{\prime}}
\pspicture[shift=-1.06](0.1,-1.15)(0.8,1.15)
  \small
  \psset{linewidth=0.9pt,linecolor=black,arrowscale=1.5,arrowinset=0.15}
  \psline(0.3,-0.775)(0.3,0.775)
  \psline{->}(0.3,-0.3)(0.3,0.2)
  \rput[bl]{0}(0.5,-0.1){$a$}
  \endpspicture
\label{eq:tracebub}
.
\end{eqnarray}%
From this, we see that the partial trace acting on ket-bra elements seems to
behave as the usual partial trace, and one might think it should be treated as
such. However, things are a bit more subtle than this, since these bras and kets
do not have the usual tensor product structure. When considering
tensor products of operators, it is the partial quantum trace that behaves in
the appropriate manner for a partial traces (i.e. as in the usual basis
independent definition of partial trace), indicating that this should be
treated as the usual notion of partial trace. Specifically, tracing over the set
of anyons $B$ on which the operator $Y$ acts, we have%
\begin{equation}
\label{eq:tensor_qtrace}
\widetilde{\text{Tr}}_{B} \left[ X \otimes Y \right]
= X \widetilde{\text{Tr}}Y
\end{equation}
\begin{equation}
\text{Tr}_{B} \left[ X \otimes Y \right]
= \sum\limits_{a,b,c} N_{ab}^{c} X_{a} \text{Tr}Y_{b}.
\end{equation}%

\subsection{Braiding}

The unitary braiding operations of pairs of anyons, also called
$R$-moves, are written as
\begin{equation}
R_{ab}=
\pspicture[shift=-0.55](-0.1,-0.2)(1.3,1.05)
\small
  \psset{linewidth=0.9pt,linecolor=black,arrowscale=1.5,arrowinset=0.15}
  \psline(0.96,0.05)(0.2,1)
  \psline{->}(0.96,0.05)(0.28,0.9)
  \psline(0.24,0.05)(1,1)
  \psline[border=2pt]{->}(0.24,0.05)(0.92,0.9)
  \rput[bl]{0}(-0.02,0){$a$}
  \rput[br]{0}(1.2,0){$b$}
  \endpspicture
,\qquad
R_{ab}^{\dag}= R_{ab}^{-1}=
\pspicture[shift=-0.55](-0.1,-0.2)(1.3,1.05)
\small
  \psset{linewidth=0.9pt,linecolor=black,arrowscale=1.5,arrowinset=0.15}
  \psline{->}(0.24,0.05)(0.92,0.9)
  \psline(0.24,0.05)(1,1)
  \psline(0.96,0.05)(0.2,1)
  \psline[border=2pt]{->}(0.96,0.05)(0.28,0.9)
  \rput[bl]{0}(-0.01,0){$b$}
  \rput[bl]{0}(1.06,0){$a$}
  \endpspicture
,
\end{equation}
which are defined through their application to basis vectors:
\begin{eqnarray}
\qquad \qquad
R_{ab}\left| a,b;c,\mu \right\rangle = \sum\limits_{\nu }\left[ R_{c}^{ab}%
\right] _{\mu \nu }\left| b,a;c,\nu \right\rangle
\\
\pspicture[shift=-0.65](-0.1,-0.2)(1.5,1.2)
  \small
  \psset{linewidth=0.9pt,linecolor=black,arrowscale=1.5,arrowinset=0.15}
  \psline{->}(0.7,0)(0.7,0.43)
  \psline(0.7,0)(0.7,0.5)
  \psarc(0.8,0.6732051){0.2}{120}{240}
  \psarc(0.6,0.6732051){0.2}{-60}{35}
  \psline (0.6134,0.896410)(0.267,1.09641)
  \psline{->}(0.6134,0.896410)(0.35359,1.04641)
  \psline(0.7,0.846410) (1.1330,1.096410)	
  \psline{->}(0.7,0.846410)(1.04641,1.04641)
  \rput[bl]{0}(0.4,0){$c$}
  \rput[br]{0}(1.35,0.85){$a$}
  \rput[bl]{0}(0.05,0.85){$b$}
 \scriptsize
  \rput[bl]{0}(0.82,0.35){$\mu$}
  \endpspicture
= \sum\limits_{\nu }\left[ R_{c}^{ab}\right] _{\mu \nu}
\pspicture[shift=-0.65](-0.1,-0.2)(1.5,1.2)
  \small
  \psset{linewidth=0.9pt,linecolor=black,arrowscale=1.5,arrowinset=0.15}
  \psline{->}(0.7,0)(0.7,0.45)
  \psline(0.7,0)(0.7,0.55)
  \psline(0.7,0.55) (0.25,1)
  \psline{->}(0.7,0.55)(0.3,0.95)
  \psline(0.7,0.55) (1.15,1)	
  \psline{->}(0.7,0.55)(1.1,0.95)
  \rput[bl]{0}(0.4,0){$c$}
  \rput[br]{0}(1.4,0.8){$a$}
  \rput[bl]{0}(0,0.8){$b$}
 \scriptsize
  \rput[bl]{0}(0.82,0.37){$\nu$}
  \endpspicture
\end{eqnarray}
and similarly for $R^{-1}$, which, by unitarity, satisfy
$\left[ \left( R_{c}^{ab}\right)^{-1} \right] _{\mu \nu} =
\left[ R_{c}^{ba}\right] _{\nu \mu}^{\ast}$.

For braiding to be consistent with fusion, it must satisfy constraints (the Hexagon
equations), which essentially impose the property that lines may be passed over or under
vertices respectively (i.e. braiding commutes with fusion), and which imply the
usual Yang-Baxter relation for braids.

With the ability to braid, one also gains the ability to trace out any anyon in a system, not just those situated at one of the two outer positions of a planar diagram. To do so, one simply uses a series of braiding operations to move the anyon to one of the outside positions. In general, if one applies a
different series of braids to move the anyon to the outside position before tracing, it will give a different outcome. Consequently, the braiding path applied to an anyon before closing its charge line should be
included as part of the definition of the partial (quantum) trace. Physically, this corresponds to specifying the path (with respect to the other anyons) by which the traced anyon is removed from the system in consideration. In this paper, we will diagrammatically indicate the removal path of the anyon being traced out whenever the issue arises.

The braiding matrices satisfy the ribbon property
\begin{equation}
\sum\limits_{\lambda }\left[ R_{c}^{ab}\right] _{\mu \lambda }\left[
R_{c}^{ba}\right] _{\lambda \nu }=\frac{\theta _{c}}{\theta _{a}\theta _{b}%
}\delta _{\mu \nu }
\end{equation}%
where $\theta _{a}$ is a root of unity called the topological spin of $a$,
defined by%
\begin{equation}
\theta _{a}=\theta _{\bar{a}}=d_{a}^{-1}\widetilde{\text{Tr}}R_{aa}
=\sum\limits_{c,\mu } \frac{d_{c}}{d_{a}}\left[ R_{c}^{aa}\right] _{\mu \mu }
= \frac{1}{d_{a}}
\pspicture[shift=-0.5](-1.3,-0.6)(1.3,0.6)
\small
  \psset{linewidth=0.9pt,linecolor=black,arrowscale=1.5,arrowinset=0.15}
  \psarc[linewidth=0.9pt,linecolor=black] (0.7071,0.0){0.5}{-135}{135}
  \psarc[linewidth=0.9pt,linecolor=black] (-0.7071,0.0){0.5}{45}{315}
  \psline(-0.3536,0.3536)(0.3536,-0.3536)
  \psline[border=2.3pt](-0.3536,-0.3536)(0.3536,0.3536)
  \psline[border=2.3pt]{->}(-0.3536,-0.3536)(0.0,0.0)
  \rput[bl]{0}(-0.2,-0.5){$a$}
  \endpspicture.
\end{equation}
When applicable, this is related to $s_{a}$, the (ordinary angular momentum)
spin or CFT conformal scaling dimension of $a$, by
\begin{equation}
\theta _{a}=e^{i2\pi s_{a}}.
\end{equation}
The topological $S$-matrix is defined by
\begin{equation}
S_{ab}=\mathcal{D}^{-1}\widetilde{\text{Tr}}\left[ R_{ba}R_{ab}\right] =\mathcal{D}^{-1}\sum%
\limits_{c}N_{ab}^{c}\frac{\theta _{c}}{\theta _{a}\theta _{b}}d_{c}
=\frac{1}{\mathcal{D}}
\pspicture[shift=-0.4](0.0,0.2)(2.4,1.3)
\small
  \psarc[linewidth=0.9pt,linecolor=black,arrows=<-,arrowscale=1.5,
arrowinset=0.15] (1.6,0.7){0.5}{165}{363}
  \psarc[linewidth=0.9pt,linecolor=black] (0.9,0.7){0.5}{0}{180}
  \psarc[linewidth=0.9pt,linecolor=black,border=3pt,arrows=->,arrowscale=1.5,
arrowinset=0.15] (0.9,0.7){0.5}{180}{375}
  \psarc[linewidth=0.9pt,linecolor=black,border=3pt] (1.6,0.7){0.5}{0}{160}
  \psarc[linewidth=0.9pt,linecolor=black] (1.6,0.7){0.5}{155}{170}
  \rput[bl]{0}(0.15,0.3){$a$}
  \rput[bl]{0}(2.15,0.3){$b$}
  \endpspicture
  .
\end{equation}
One can see from this that $S_{ab}=S_{ba}=S_{\bar{a}b}^{\ast}$ and
$d_{a}=S_{1a}/S_{11}$. A useful property for removing loops from
lines is
\begin{equation}
\pspicture[shift=-0.55](-0.25,-0.1)(0.9,1.3)
\small
  \psset{linewidth=0.9pt,linecolor=black,arrowscale=1.5,arrowinset=0.15}
  \psline(0.4,0)(0.4,0.22)
  \psline(0.4,0.45)(0.4,1.2)
  \psellipse[linewidth=0.9pt,linecolor=black,border=0](0.4,0.5)(0.4,0.18)
  \psset{linewidth=0.9pt,linecolor=black,arrowscale=1.4,arrowinset=0.15}
  \psline{->}(0.2,0.37)(0.3,0.34)
\psline[linewidth=0.9pt,linecolor=black,border=2.5pt,arrows=->,arrowscale=1.5,
arrowinset=0.15](0.4,0.5)(0.4,1.1)
  \rput[bl]{0}(-0.15,0.15){$a$}
  \rput[tl]{0}(0.55,1.2){$b$}
  \endpspicture
=\frac{S_{ab}}{S_{1b}}
\pspicture[shift=-0.55](0.05,-0.1)(1,1.3)
\small
  \psset{linewidth=0.9pt,linecolor=black,arrowscale=1.5,arrowinset=0.15}
  \psline(0.4,0)(0.4,1.2)
\psline[linewidth=0.9pt,linecolor=black,arrows=->,arrowscale=1.5,
arrowinset=0.15](0.4,0.5)(0.4,0.9)
  \rput[tl]{0}(0.52,1.0){$b$}
  \endpspicture
\label{eq:loopaway}
\end{equation}%
An anyon model is ``modular'' and corresponds to a TQFT (topological quantum field
theory), if its monodromy is non-degenerate, i.e. for each $a \neq 1$, there is
some $b$ such that $R_{ba}R_{ab}\neq \mathbb{I}_{ab}$, which is the case iff the
topological $S$-matrix is unitary. For such theories, the $S$-matrix, together
with $T_{ab}=\theta_{a} \delta_{ab}$ represent the generators of the modular
group $\text{PSL}\left(2,\mathbb{Z} \right)$.

The monodromy scalar component
\begin{equation}
M_{ab}=\frac{\widetilde{\text{Tr}}\left[
R_{ba}R_{ab}\right]}{\widetilde{\text{Tr}}\mathbb{I}_{ab}}
=\frac{1}{d_{a}d_{b}}
\pspicture[shift=-0.4](0.0,0.2)(2.4,1.3)
\small
  \psarc[linewidth=0.9pt,linecolor=black,arrows=<-,arrowscale=1.5,
arrowinset=0.15] (1.6,0.7){0.5}{165}{363}
  \psarc[linewidth=0.9pt,linecolor=black] (0.9,0.7){0.5}{0}{180}
  \psarc[linewidth=0.9pt,linecolor=black,border=3pt,arrows=->,arrowscale=1.5,
arrowinset=0.15] (0.9,0.7){0.5}{180}{375}
  \psarc[linewidth=0.9pt,linecolor=black,border=3pt] (1.6,0.7){0.5}{0}{160}
  \psarc[linewidth=0.9pt,linecolor=black] (1.6,0.7){0.5}{155}{170}
  \rput[bl]{0}(0.15,0.3){$a$}
  \rput[bl]{0}(2.15,0.3){$b$}
  \endpspicture
=\frac{S_{ab}S_{11}}{S_{1a}S_{1b}}
\label{eq:monodromy}
\end{equation}
is an important quantity, typically arising in interference terms, such as those
occurring in experiments that probe anyonic charge. It may also be written the $1,1$ component of the operator $B^{2}=F^{-1}R^{2}F$. Since $B^2$ is a unitary operator, we must have
$\left| M_{ab}\right| \leq 1$. Indeed unitarity implies that when $\left| M_{ab}\right| =1$, only the $1,1$
element of $B^{2}$ is non-zero, hence
\begin{equation}
  \pspicture[shift=-0.6](0.0,-0.05)(1.1,1.45)
  \small
  \psarc[linewidth=0.9pt,linecolor=black,border=0pt] (0.8,0.7){0.4}{120}{225}
  \psarc[linewidth=0.9pt,linecolor=black,arrows=<-,arrowscale=1.4,
    arrowinset=0.15] (0.8,0.7){0.4}{165}{225}
  \psarc[linewidth=0.9pt,linecolor=black,border=0pt] (0.4,0.7){0.4}{-60}{45}
  \psarc[linewidth=0.9pt,linecolor=black,arrows=->,arrowscale=1.4,
    arrowinset=0.15] (0.4,0.7){0.4}{-60}{15}
  \psarc[linewidth=0.9pt,linecolor=black,border=0pt]
(0.8,1.39282){0.4}{180}{225}
  \psarc[linewidth=0.9pt,linecolor=black,border=0pt]
(0.4,1.39282){0.4}{-60}{0}
  \psarc[linewidth=0.9pt,linecolor=black,border=0pt]
(0.8,0.00718){0.4}{120}{180}
  \psarc[linewidth=0.9pt,linecolor=black,border=0pt]
(0.4,0.00718){0.4}{0}{45}
  \rput[bl]{0}(0.1,1.2){$a$}
  \rput[br]{0}(1.06,1.2){$b$}
  \endpspicture
= M_{ab}
\pspicture[shift=-0.6](0,-0.45)(1.0,1.1)
  \small
  \psset{linewidth=0.9pt,linecolor=black,arrowscale=1.5,arrowinset=0.15}
  \psline(0.3,-0.4)(0.3,1)
  \psline{->}(0.3,-0.4)(0.3,0.50)
  \psline(0.7,-0.4)(0.7,1)
  \psline{->}(0.7,-0.4)(0.7,0.50)
  \rput[br]{0}(0.96,0.8){$b$}
  \rput[bl]{0}(0,0.8){$a$}
  \endpspicture
\end{equation}
so that the braiding of $a$ and $b$ is Abelian. The monodromy of $a$ and $b$
is trivial if $M_{ab}=1$. If $N_{ab}^{c}\neq 0$ and $\left| M_{be}\right| =1$
for some $e$, then the relation%
\begin{equation}
M_{ce}=M_{ae}M_{be}
\label{eq:fused_monodromy}
\end{equation}%
follows from the diagrammatic equation%
\begin{equation}
 \pspicture[shift=-1.3](-0.05,-0.45)(1.45,2.45)
  \small
  \psarc[linewidth=0.9pt,linecolor=black,border=0pt] (0.8,0.7){0.4}{120}{240}
  \psarc[linewidth=0.9pt,linecolor=black,arrows=<-,arrowscale=1.4,
    arrowinset=0.15] (0.8,0.7){0.4}{165}{240}
  \psarc[linewidth=0.9pt,linecolor=black,border=0pt] (0.4,0.7){0.4}{-60}{60}
  \psarc[linewidth=0.9pt,linecolor=black,arrows=->,arrowscale=1.4,
    arrowinset=0.15] (0.4,0.7){0.4}{-60}{15}
  \psellipse[linewidth=0.9pt,linecolor=black,border=0](0.6,1.59)(0.4,0.18)
  \psset{linewidth=0.9pt,linecolor=black,arrowscale=1.4,arrowinset=0.15}
  \psline{->}(0.4,1.46)(0.5,1.43)
  \psset{linewidth=0.9pt,linecolor=black,arrowscale=1.5,arrowinset=0.15}
  \psline(0.6,1.05)(0.6,1.35)
  \psline[border=2pt]{->}(0.6,1.5)(0.6,2.05)
  \psline(0.6,0)(0.6,0.35)
  \psline(1.4,0)(1.4,2.0)
  \psarc[linewidth=0.9pt,linecolor=black,border=0pt] (1,2.0){0.4}{0}{180}
  \psarc[linewidth=0.9pt,linecolor=black,border=0pt] (1,0){0.4}{180}{360}
  \rput[bl]{0}(0.05,0.55){$a$}
  \rput[bl]{0}(0.9,0.55){$b$}
  \rput[bl]{0}(0,1.4){$e$}
  \rput[bl]{0}(0.32,1.95){$c$}
\scriptsize
  \rput[bl]{0}(0.7,1.1){$\mu$}
  \rput[bl]{0}(0.7,0.15){$\mu$}
  \endpspicture
=M_{be}
 \pspicture[shift=-1.3](-0.5,-0.45)(1.45,2.45)
  \small
  \psellipse[linewidth=0.9pt,linecolor=black,border=0](0.25,1.0)(0.4,0.18)
  \psset{linewidth=0.9pt,linecolor=black,arrowscale=1.4,arrowinset=0.15}
  \psline{->}(0.05,0.87)(0.15,0.84)
  \psarc[linewidth=0.9pt,linecolor=black,border=0pt](0.95,1.0){0.7}{199}{240}
  \psarc[linewidth=0.9pt,linecolor=black,border=2pt,arrows=<-,arrowscale=1.4,
    arrowinset=0.15] (0.95,1.0){0.7}{135}{187}
  \psarc[linewidth=0.9pt,linecolor=black,border=0pt](0.95,1.0){0.7}{120}{138}
  \psarc[linewidth=0.9pt,linecolor=black,border=0pt] (0.25,1.0){0.7}{-60}{60}
  \psarc[linewidth=0.9pt,linecolor=black,arrows=->,arrowscale=1.4,
    arrowinset=0.15] (0.25,1.0){0.7}{-60}{10}
  \psset{linewidth=0.9pt,linecolor=black,arrowscale=1.5,arrowinset=0.15}
  \psline{->}(0.6,1.62)(0.6,2.05)
  \psline(0.6,0)(0.6,0.38)
  \psline(1.4,0)(1.4,2.0)
  \psarc[linewidth=0.9pt,linecolor=black,border=0pt] (1,2.0){0.4}{0}{180}
  \psarc[linewidth=0.9pt,linecolor=black,border=0pt] (1,0){0.4}{180}{360}
  \rput[bl]{0}(0.05,1.4){$a$}
  \rput[bl]{0}(0.9,1.35){$b$}
  \rput[bl]{0}(-0.35,0.81){$e$}
  \rput[bl]{0}(0.32,1.95){$c$}
\scriptsize
  \rput[bl]{0}(0.7,1.65){$\mu$}
  \rput[bl]{0}(0.7,0.2){$\mu$}
  \endpspicture
\end{equation}

\subsection{States and Density Matrices}

To describe the state of anyons in a system using a state vector, one must specify all the splitting channels starting from vacuum. For example, in order to have anyons with charges $a$ and $b$ with overall charge $c$, one must also have a $\bar{c}$ charge somewhere, and one would write a general state of this form as
\begin{eqnarray}
\qquad \qquad \quad
\left| \Psi \right\rangle &=& \sum\limits_{a,b,c,\mu}
\psi_{a,b,c,\mu}\left| a,b;c,\mu \right\rangle \left| c,\bar{c};1 \right\rangle \notag \\
&=& \sum\limits_{a,b,c,\mu} \frac{\psi_{a,b,c,\mu}}{\left(d_{a}d_{b}d_{c} \right)^{1/4}}
\pspicture[shift=-0.55](0,0.4)(1.8,1.65)
  \small
  \psset{linewidth=0.9pt,linecolor=black,arrowscale=1.5,arrowinset=0.15}
  \psline(0.2,1.5)(1,0.5)
  \psline(1.8,1.5) (1,0.5)
  \psline(0.6,1) (1,1.5)
   \psline{->}(0.6,1)(0.3,1.375)
   \psline{->}(0.6,1)(0.9,1.375)
   \psline{->}(1,0.5)(1.7,1.375)
   \psline{->}(1,0.5)(0.7,0.875)
   \rput[bl]{0}(0.05,1.6){$a$}
   \rput[bl]{0}(0.95,1.6){$b$}
   \rput[bl]{0}(1.75,1.6){$\bar{c}$}
   \rput[bl]{0}(0.5,0.5){$c$}
 \scriptsize
   \rput[bl]{0}(0.3,0.8){$\mu$}
  \endpspicture
.
\end{eqnarray}
Unfortunately, describing states in this manner can become cumbersome, and, as usual, does not
naturally accommodate the restriction to subsystems, so it is better for us to use the density
matrix formalism.

The density matrix for an arbitrary two anyon system is%
\begin{eqnarray}
\qquad \qquad
\rho  &=&\sum\limits_{\substack{ a,a^{\prime },b,b^{\prime } \\ c,\mu ,\mu
^{\prime }}}\rho _{\left( a,b,c,\mu \right) \left( a^{\prime },b^{\prime
},c,\mu ^{\prime }\right) } \frac{1}{d_c}  \left| a,b;c,\mu \right\rangle
\left\langle a^{\prime },b^{\prime };c,\mu ^{\prime }\right|  \nonumber \\
&=&\sum\limits_{\substack{ a,a^{\prime },b,b^{\prime } \\ c,\mu ,\mu
^{\prime }}} \frac{ \rho _{\left( a,b,c,\mu \right) \left( a^{\prime },b^{\prime
},c,\mu ^{\prime }\right) } }{ \left( d_{a}d_{b}d_{a^{\prime}}d_{b^{\prime }}d_{c}^{2} \right) ^{1/4}}\;
 \pspicture[shift=-0.6](0,-0.45)(1.5,1)
 \small
  \psset{linewidth=0.9pt,linecolor=black,arrowscale=1.5,arrowinset=0.15}
  \psline{->}(0.7,0)(0.7,0.45)
  \psline(0.7,0)(0.7,0.55)
  \psline(0.7,0.55) (0.25,1)
  \psline{->}(0.7,0.55)(0.3,0.95)
  \psline(0.7,0.55) (1.15,1)
  \psline{->}(0.7,0.55)(1.1,0.95)
  \rput[bl]{0}(0.38,0.2){$c$}
  \rput[br]{0}(1.4,0.8){$b$}
  \rput[bl]{0}(0,0.8){$a$}
  \psline(0.7,0) (0.25,-0.45)
  \psline{-<}(0.7,0)(0.35,-0.35)
  \psline(0.7,0) (1.15,-0.45)
  \psline{-<}(0.7,0)(1.05,-0.35)
  \rput[br]{0}(1.45,-0.4){$b'$}
  \rput[bl]{0}(-0.05,-0.4){$a'$}
\scriptsize
  \rput[bl]{0}(0.85,0.38){$\mu$}
  \rput[bl]{0}(0.85,-0.05){$\mu'$}
  \endpspicture
  .
\end{eqnarray}%
The overall charge $c$ must match up between the bra and the ket because of
charge conservation. The normalization is chosen such that the trace condition
for density matrices takes the form%
\begin{equation}
\label{eq:tracecondition}
\widetilde{\text{Tr}} \left[ \rho \right] = \sum\limits_{a,b,c,\mu} \rho
_{\left( a,b,c,\mu \right) \left( a,b,c,\mu \right) } = 1
\end{equation}
The factor $1/d_{c}$ could, of course, be absorbed into
$\rho _{\left( a,b,c,\mu \right) \left( a^{\prime },b^{\prime
},c,\mu ^{\prime }\right) }$ (as a matter of convention), but then the $d_{c}$
would appear in the summand of Eq.~(\ref{eq:tracecondition}). Furthermore, the
density matrix is written this way so that one can naturally think of it as
$\rho=\widetilde{\text{\text{Tr}}}_{\overline{C}}\left[ \rho ^{\prime }\right] $,
the partial quantum trace over $\overline{C}$ of a density matrix that describes
the actual entire system
\begin{eqnarray}
\rho ^{\prime } &=&\sum\limits_{\substack{ a,b,c,\mu  \\ a^{\prime
},b^{\prime },c^{\prime },\mu ^{\prime }}}\rho _{\left( a,b,c,\mu \right)
\left( a^{\prime },b^{\prime },c^{\prime },\mu ^{\prime }\right) }\left|
a,b;c,\mu \right\rangle \left| c,\bar{c};1\right\rangle \langle
c^{\prime},\overline{c^{\prime}};1| \left\langle
a^{\prime},b^{\prime};c^{\prime},\mu ^{\prime}\right|
\nonumber  \\
&=&\sum\limits_{\substack{ a,b,c,\mu  \\ a^{\prime },b^{\prime },c^{\prime
},\mu ^{\prime }}} \frac{\rho _{\left( a,b,c,\mu \right) \left( a^{\prime
},b^{\prime },c^{\prime },\mu ^{\prime }\right) }}{\left(
d_{a}d_{b}d_{c}d_{a^{\prime }}d_{b^{\prime }}d_{c^{\prime }}\right) ^{1/4}}\;
 \pspicture[shift=-1.55](0,-1.45)(1.8,1.65)
  \small
  \psset{linewidth=0.9pt,linecolor=black,arrowscale=1.5,arrowinset=0.15}
  \psline(0.2,1.5)(1,0.5)
  \psline(1.8,1.5) (1,0.5)
  \psline(0.6,1) (1,1.5)
   \psline{->}(0.6,1)(0.3,1.375)
   \psline{->}(0.6,1)(0.9,1.375)
   \psline{->}(1,0.5)(1.7,1.375)
   \psline{->}(1,0.5)(0.7,0.875)
   \rput[bl]{0}(0.05,1.6){$a$}
   \rput[bl]{0}(0.95,1.6){$b$}
   \rput[bl]{0}(1.75,1.6){$\bar{c}$}
   \rput[bl]{0}(0.5,0.5){$c$}
  \psset{linewidth=0.9pt,linecolor=black,arrowscale=1.5,arrowinset=0.15}
  \psline(0.2,-1)(1,0)
  \psline(1.8,-1) (1,0)
  \psline(0.6,-0.5) (1,-1)
   \psline{-<}(0.6,-0.5)(0.3,-0.875)
   \psline{-<}(0.6,-0.5)(0.9,-0.875)
   \psline{-<}(1,0)(1.7,-0.875)
   \psline{-<}(1,0)(0.7,-0.375)
   \rput[bl]{0}(0.05,-1.4){$a'$}
   \rput[bl]{0}(0.95,-1.4){$b'$}
   \rput[bl]{0}(1.75,-1.4){$\overline{c'}$}
   \rput[bl]{0}(0.55,-0.15){$c'$}
 \scriptsize
   \rput[bl]{0}(0.3,0.8){$\mu$}
   \rput[bl]{0}(0.25,-0.6){$\mu'$}
  \endpspicture
\end{eqnarray}
which only has vacuum overall charge. In other words, the entire system
really has trivial total anyonic charge, but by restricting our attention to
some subset of anyons, we have a reduced subsystem with overall charge $c$.
Tracing over the $\overline{C}$ anyon (which imposes $c=c^{\prime }$)
physically represents the fact that it is no longer included in the system of
interest, and cannot be brought back to interact with the $A$ and $B$ anyons.
Because of this, we are restricted to a subsystem which may only have incoherent
superpositions of different overall charges $c$ (i.e. one must keep track of
the $\overline{C}$ anyon to allow access to coherent superpositions). The
manifestation of this property in $\rho $ is exhibited by the charge $c$
matching in the bra and the ket (or diagrammatically as the charge $c$ line
connecting $\mu $ and $\mu ^{\prime }$). The generalization to density
matrices of arbitrary numbers of anyons should be clear.

When considering the combination of two sets of anyons $A=\left\{
A_{1},\ldots ,A_{m}\right\} $ and $B=\left\{ B_{1},\ldots ,B_{n}\right\} $,
the anyons of system $A$ are unentangled with those of
system $B$ if the density matrix of the combined system is the tensor
product (in some basis) of
density matrices of the two systems $\rho ^{AB}=\rho ^{A}\otimes \rho ^{B}$. 
This essentially means the creation histories of the two different systems do not
involve each other. There is a specific aspect of entanglement in anyonic systems 
that we will call \textit{anyonic charge entanglement}, which is encoded 
in the anyonic charge lines connecting anyons. The systems of anyons $A$ and $B$ 
are said to have no anyonic charge entanglement between them if
$\rho ^{AB} \in V_{A_{1}^{\prime },\ldots ,A_{m}^{\prime }}^{A_{1},\ldots
,A_{m}} \otimes V_{B_{1}^{\prime },\ldots ,B_{n}^{\prime }}^{B_{1},\ldots
,B_{n}}$. This is represented diagrammatically as being able to write the combined state 
such that there are no non-trivial charge lines connecting the anyons of $A$ with those of $B$.

%% file: 03a-experiment.tex
\section{Mach-Zehnder Interferometer}
  \label{sec:MZI}

\begin{figure}[t!]
\begin{center}
  \includegraphics[scale=0.7]{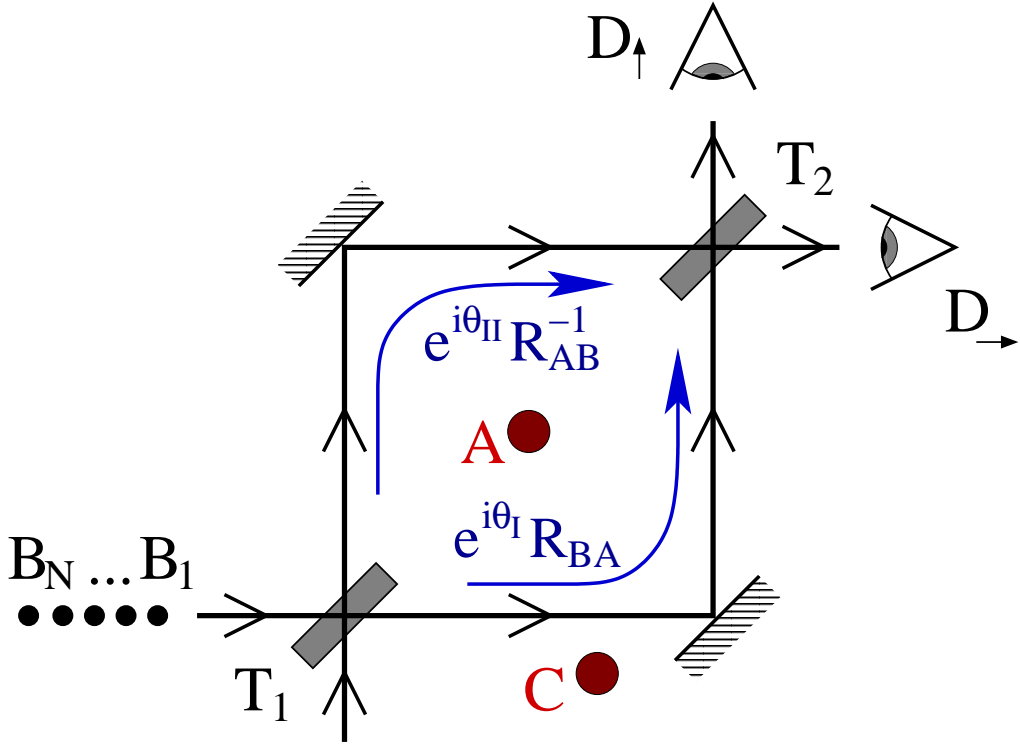}
  \caption{A Mach-Zehnder interferometer for an anyonic system. The target anyon(s) $A$ in the central region shares entanglement only with the anyon(s) $C$ outside this region. A beam of probe anyons $B_{1},\ldots ,B_{N}$ is sent through the interferometer, where $T_{j}$ are beam splitters, and detected at one of the two possible outputs by $D_{s}$.}
  \label{fig:interferometer}
\end{center}
\end{figure}

In this section, we consider, in detail, a Mach-Zehnder type
interferometer~\cite{Zehnder1891,Mach1892} (see Fig.~\ref{fig:interferometer})
for quasiparticles with non-Abelian anyonic braiding statistics,
extending the analysis begun in Ref.~\cite{Bonderson07a}. This
will serve as a prototypic model of interferometry experiments with anyons,
and the methods used in its analysis readily apply to other classes of
interferometers (e.g. the FQH double point-contact interferometer considered in
Section~\ref{sec:FQH_2PC}). This interferometer was first considered for
non-Abelian anyons in Ref.~\cite{Overbosch01}, but only for anyon models
described by a discrete gauge theory-type formalism in
which individual particles are assumed to have internal Hilbert spaces, and
which use probe anyons that are all identical and have trivial self-braiding.
Unfortunately, this excludes perhaps the most important class of anyon
models -- those describing the fractional quantum Hall states -- so we must
dispense with such restrictions. We will abstract to an idealized system that
supports an arbitrary anyon model and also allows for a number of desired
manipulations to be effected. Specifically, without concern for ways to
physically actualize such manipulations, we posit the experimental abilities to: (1) produce,
isolate, and position desired anyons, (2) provide anyons with some manner of
propulsion to produce a beam of probe anyons, (3) construct lossless
beam-splitters and mirrors, and (4) detect the presence of a probe anyon at the
output legs of the interferometer.

The target anyon $A$ is the composite of all anyons $A_{1},A_{2},\ldots $
that are located inside the central interferometry region, and so may be in a
superposition of states with different total anyonic charges. Since these
anyons are treated collectively by the experiment, we ignore their
individuality and consider them as a single anyon $A$ capable of charge
superposition. We will similarly allow the probe anyons, $B_{1},\ldots ,B_{N}$ to
be treated as capable of charge superposition (though this would certainly
be more difficult to physically realize). The probe anyons are sent as a beam
into the interferometer through two possible input channels. They pass through a
beam splitter $T_{1}$, are reflected by mirrors around the central target
region, pass through a second beam splitter $T_{2}$, and then are detected at
one of the two possible output channels by the detectors $D_{s}$. When a probe
anyon $B$ passes through the bottom path of the interferometer, the state
acquires the phase $e^{i\theta_{\text{I}}}$, which results from background
Aharonov-Bohm interactions~\cite{Aharonov59}, path length differences, phase
shifters, etc., and is also acted upon by the braiding operator $R_{BA}$, which
is strictly due to the braiding statistics between the probe and target anyons.
Similarly, when the probe passes through the top path of the interferometer, the
state acquires the phase $e^{i\theta _{\text{II}}}$ and is acted on by
$R_{AB}^{-1}$.
\begin{figure}[t!]
\begin{center}
  \includegraphics[scale=0.7]{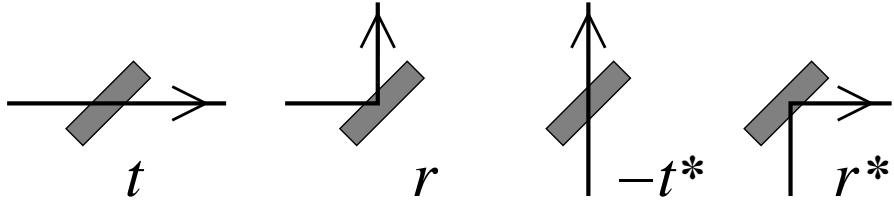}
  \caption{The transmission and reflection coefficients for a beam splitter.}
  \label{fig:splitters}
\end{center}
\end{figure}

Using the two-component vector notation%
\begin{equation}
\left(
\begin{array}{c}
1 \\
0%
\end{array}%
\right) =\left| \shortrightarrow \right\rangle ,\quad \left(
\begin{array}{c}
0 \\
1%
\end{array}%
\right) =\left| \shortuparrow \right\rangle
\end{equation}%
to indicate the direction (horizontal or vertical) a probe anyon is
traveling through the interferometer at any point, the lossless beam
splitters~\cite{Zeilinger81} (see Fig.~\ref{fig:splitters}) are represented by%
\begin{equation}
T_{j}=\left[
\begin{array}{cc}
t_{j} & r_{j}^{\ast } \\
r_{j} & -t_{j}^{\ast }%
\end{array}%
\right]
\end{equation}%
(for $j=1,2$), where $\left| t_{j}\right| ^{2}+\left| r_{j}\right| ^{2}=1$.
We note that these matrices could be multiplied by
overall phases without affecting any of the results, since such phases are
not distinguished by the two paths.

When considering operations involving non-Abelian anyons, it is important to
keep track of all other anyons with which there is non-trivial entanglement.
Indeed, if these additional particles are not tracked or are
physically inaccessible, one should trace them out of the system, forgoing
the ability to use them to form coherent superpositions of anyonic charge.
We assume that the target anyon has no initial entanglement with the probe
anyons, so their systems will be combined as tensor products, with no
non-trivial charge lines connecting them before they interact in the
interferometer.

The target system involves the target anyon $A$ and the anyon $C$ which is
the only one entangled with $A$ that is kept physically accessible. Recall
that these anyons may really represent multiple quasiparticles that are
being treated collectively, but as long as we are not interested in
operations involving the individual quasiparticles, they can be treated as a
single anyon. The density matrix of the target system is%
\begin{eqnarray}
\qquad
\rho ^{A} &=&\sum\limits_{a,a^{\prime },c,c^{\prime },f,\mu ,\mu ^{\prime
}}\rho _{\left( a,c;f,\mu \right) \left( a^{\prime },c^{\prime };f,\mu
^{\prime }\right) }^{A} \frac{1}{d_f} \left| a,c;f,\mu \right\rangle
\left\langle a^{\prime},c^{\prime };f,\mu ^{\prime }\right| \nonumber \\
&=&\sum\limits_{a,a^{\prime },c,c^{\prime },f,\mu ,\mu ^{\prime }}
\frac{ \rho_{\left( a,c;f,\mu \right) \left( a^{\prime },c^{\prime
};f,\mu^{\prime}\right) }^{A} }{ \left( d_{a}d_{a^{\prime
}}d_{c}d_{c^{\prime}}d_{f}^{2} \right) ^{1/4} }
 \pspicture[shift=-0.6](-0.15,-0.45)(1.5,1)
 \small
  \psset{linewidth=0.9pt,linecolor=black,arrowscale=1.5,arrowinset=0.15}
  \psline{->}(0.7,0)(0.7,0.45)
  \psline(0.7,0)(0.7,0.55)
  \psline(0.7,0.55) (0.25,1)
  \psline{->}(0.7,0.55)(0.3,0.95)
  \psline(0.7,0.55) (1.15,1)
  \psline{->}(0.7,0.55)(1.1,0.95)
  \rput[bl]{0}(0.38,0.1){$f$}
  \rput[bl]{0}(1.22,0.8){$c$}
  \rput[bl]{0}(-0.05,0.8){$a$}
  \psline(0.7,0) (0.25,-0.45)
  \psline{-<}(0.7,0)(0.35,-0.35)
  \psline(0.7,0) (1.15,-0.45)
  \psline{-<}(0.7,0)(1.05,-0.35)
  \rput[bl]{0}(1.22,-0.4){$c'$}
  \rput[bl]{0}(-0.05,-0.4){$a'$}
  \scriptsize
  \rput[bl]{0}(0.82,0.38){$\mu$}
  \rput[bl]{0}(0.82,-0.02){$\mu'$}
  \endpspicture
.
\end{eqnarray}

We will assume that the probe anyons are also not entangled with each other,
and that they are all identical (or, more accurately, belong to an ensemble of
particles all described by the density matrix $\rho^{B}$). We will consider
generalizations of the probe anyons in Section~\ref{sec:probegen}. Such
generalizations complicate the bookkeeping of the calculation, but will have
qualitatively similar results. A probe system involves the probe anyon
$B$, which is sent through the interferometer entering the horizontal leg
$s=\shortrightarrow$, and the anyon $D$ which is entangled with $B$ and will be
sent off to the (left) side. We will write the directional index $s$ of the
probe particle as a subscript on its anyonic charge label, i.e. $b_s$. The
density matrix of a probe system is%
\begin{eqnarray}
\quad
\rho ^{B} &=&\sum\limits_{b,b^{\prime },d,d^{\prime },h,\lambda ,\lambda
^{\prime }}\rho _{\left( d,b_{\shortrightarrow};h,\lambda \right) \left(
d^{\prime },b^{\prime}_{\shortrightarrow};h,\lambda ^{\prime }\right)
}^{B} \frac{1}{d_h} \left| d,b_{\shortrightarrow};h,\lambda \right\rangle
\left\langle
d^{\prime },b^{\prime }_{\shortrightarrow};h,\lambda ^{\prime
} \right| \nonumber \\
&=&\sum\limits_{b,b^{\prime },d,d^{\prime },h,\lambda ,\lambda ^{\prime }}%
\frac{\rho _{\left( d,b_{\shortrightarrow};h,\lambda \right) \left( d^{\prime
},b^{\prime
}_{\shortrightarrow};h,\lambda ^{\prime }\right) }^{B}}
{ \left(d_{d}d_{d^{\prime
}}d_{b}d_{b^{\prime }}d_{h}^{2} \right) ^{1/4} }
 \pspicture[shift=-0.6](-0.15,-0.45)(1.7,1)
 \small
  \psset{linewidth=0.9pt,linecolor=black,arrowscale=1.5,arrowinset=0.15}
  \psline{->}(0.7,0)(0.7,0.45)
  \psline(0.7,0)(0.7,0.55)
  \psline(0.7,0.55) (0.25,1)
  \psline{->}(0.7,0.55)(0.3,0.95)
  \psline(0.7,0.55) (1.15,1)
  \psline{->}(0.7,0.55)(1.1,0.95)
  \rput[bl]{0}(0.32,0.2){$h$}
  \rput[br]{0}(1.65,0.72){$b_{\shortrightarrow}$}
  \rput[bl]{0}(-0.05,0.8){$d$}
  \psline(0.7,0) (0.25,-0.45)
  \psline{-<}(0.7,0)(0.35,-0.35)
  \psline(0.7,0) (1.15,-0.45)
  \psline{-<}(0.7,0)(1.05,-0.35)
  \rput[br]{0}(1.65,-0.48){$b^{\prime }_{\shortrightarrow}$}
  \rput[bl]{0}(-0.05,-0.4){$d^{\prime }$}
  \scriptsize
  \rput[bl]{0}(0.82,0.38){$\lambda$}
  \rput[bl]{0}(0.82,-0.02){$\lambda'$}
  \endpspicture
.
\end{eqnarray}

The unitary operator representing a probe anyon passing through the
interferometer is given by%
\begin{equation}
U = T_{2}\Sigma T_{1}
\end{equation}
\begin{equation}
\Sigma = \left[
\begin{array}{cc}
0 & e^{i\theta _{\text{II}}}R_{AB}^{-1} \\
e^{i\theta _{\text{I}}}R_{BA} & 0%
\end{array}%
\right] .
\end{equation}%
This can be written diagrammatically as%
\begin{equation}
\label{eq:Udiag}
\pspicture[shift=-0.4](-0.2,0)(1,1)
\rput[tl](0,0){$B_{s'}$}
\rput[tl](0,1.2){$A$}
\rput[tl](0.85,1.2){$B_{s}$}
\rput[tl](0.85,0){$A$}
 \psline[linewidth=0.9pt](0.4,0)(0.4,0.24)
 \psline[linewidth=0.9pt](0.78,0)(0.78,0.24)
 \psline[linewidth=0.9pt](0.4,0.78)(0.4,1)
 \psline[linewidth=0.9pt](0.78,0.78)(0.78,1)
\rput[bl](0.29,0.24){\psframebox{$U$}}
  \endpspicture
=e^{i\theta _{\text{I}}}\left[
\begin{array}{cc}
t_{1}r_{2}^{\ast } & r_{1}^{\ast }r_{2}^{\ast } \\
-t_{1}t_{2}^{\ast } & -r_{1}^{\ast }t_{2}^{\ast }
\end{array}
\right]_{s,s'}
\pspicture[shift=-0.4](-0.2,0)(1.25,1)
  \psset{linewidth=0.9pt,linecolor=black,arrowscale=1.5,arrowinset=0.15}
  \psline(0.92,0.1)(0.2,1)
  \psline{->}(0.92,0.1)(0.28,0.9)
  \psline(0.28,0.1)(1,1)
  \psline[border=2pt]{->}(0.28,0.1)(0.92,0.9)
  \rput[tl]{0}(-0.2,0.2){$B$}
  \rput[tr]{0}(1.25,0.2){$A$}
  \endpspicture
+e^{i\theta _{\text{II}}}\left[
\begin{array}{cc}
r_{1}t_{2} & -t_{1}^{\ast }t_{2} \\
r_{1}r_{2} & -t_{1}^{\ast }r_{2}
\end{array}
\right]_{s,s'}
\pspicture[shift=-0.4](-0.2,0)(1.25,1)
  \psset{linewidth=0.9pt,linecolor=black,arrowscale=1.5,arrowinset=0.15}
  \psline{->}(0.28,0.1)(0.92,0.9)
  \psline(0.28,0.1)(1,1)
  \psline(0.92,0.1)(0.2,1)
  \psline[border=2pt]{->}(0.92,0.1)(0.28,0.9)
  \rput[tl]{0}(-0.2,0.2){$B$}
  \rput[tr]{0}(1.25,0.2){$A$}
  \endpspicture
  .
\end{equation}%
The position of the anyon $C$ with respect to the other anyons must be
specified, and we will take it to be located below the central
interferometry region and slightly to the right of $A$. (The specification
``slightly to the right'' merely indicates how the diagrams are to be drawn, and
has no physical consequence.) For this choice of positioning, the operator%
\begin{equation}
V=\left[
\begin{array}{cc}
R_{CB}^{-1} & 0 \\
0 & R_{CB}^{-1}%
\end{array}%
\right]
=
\pspicture[shift=-0.4](-0.2,0)(1.25,1)
  \psset{linewidth=0.9pt,linecolor=black,arrowscale=1.5,arrowinset=0.15}
  \psline{->}(0.28,0.1)(0.92,0.9)
  \psline(0.28,0.1)(1,1)
  \psline(0.92,0.1)(0.2,1)
  \psline[border=2pt]{->}(0.92,0.1)(0.28,0.9)
  \rput[tl]{0}(-0.2,0.2){$B$}
  \rput[tr]{0}(1.25,0.2){$C$}
  \endpspicture
\end{equation}%
represents the braiding of $C$ with the probe. In Section~\ref{sec:targetgen}, we 
will give the results for situating the anyon(s) $C$ in different locations outside 
the central interferometry region, and find that the qualitative behavior is essentially 
the same.

After a probe anyon $B$ is measured at one of the detectors, it no longer
interests us, and we remove it along with its entangled partner $D$ from the
vicinity of the target anyon system. Mathematically, this means we take the
tensor product of the probe and target systems, evolve them with $VU$ (which
sends the probe through the interferometer) to get%
\begin{equation}
\rho =VU\left( \rho ^{B}\otimes \rho ^{A}\right) U^{\dagger }V^{\dagger},
\label{eq:rho_VU}
\end{equation}
apply the usual orthogonal measurement collapse projection
\begin{equation}
\Pr \left( s\right)  = \widetilde{\text{Tr}}\left[ \rho \Pi _{s}\right]
\end{equation}
\begin{equation}
\rho  \mapsto \frac{1}{\Pr \left( s\right) }\Pi _{s}\rho \Pi _{s}
\label{eq:measurment_projection}
\end{equation}
with $\Pi _{s}=\left| s\right\rangle \left\langle s\right| $ for the outcome
$s$, and then finally trace out the anyons $B$ and $D$. Since the probe
anyons are all initially unentangled, we may obtain their effect on the target
system by considering that of each probe individually. 

%% file: 03b-oneprobe.tex
\subsection{One Probe}

We begin by considering the effect of a single probe with definite
anyonic charge $b$, i.e. $\rho ^{b}=\left| \bar{b},b_{\shortrightarrow};1
\right\rangle \left\langle \bar{b},b_{\shortrightarrow};1 \right| $, and
return to general $\rho ^{B}$ immediately afterwards. For a particular
component of the target anyons' density matrix, the relevant diagram that
must be evaluated for a single probe measurement is%
\begin{equation}
 \pspicture[shift=-3](-1.1,-3.1)(3.6,3.1)
  \small
  \psframe[linewidth=0.9pt,linecolor=black,border=0](0.6,0.9)(1.7,1.4)
  \psframe[linewidth=0.9pt,linecolor=black,border=0](0.6,-0.9)(1.7,-1.4)
  \rput[bl]{0}(1.0,1.01){$U$}
  \rput[tl]{0}(0.95,-0.97){$U^{\dag}$}
  \psset{linewidth=0.9pt,linecolor=black,arrowscale=1.5,arrowinset=0.15}
  \psline(0.8,-1.4)(0.8,-2.8)
  \psline(0.8,1.4)(0.8,2.8)
  \psarc[linewidth=0.9pt,linecolor=black,border=0pt] (1.5,0.9){0.7}{-60}{0}
  \psarc[linewidth=0.9pt,linecolor=black,arrows=->,arrowscale=1.4,
    arrowinset=0.15] (1.5,0.9){0.7}{-60}{-10}
  \psarc[linewidth=0.9pt,linecolor=black,border=0pt]
(2.2,0.9){0.7}{180}{240}
  \psarc[linewidth=0.9pt,linecolor=black,arrows=<-,arrowscale=1.4,
    arrowinset=0.15] (2.2,0.9){0.7}{190}{240}
  \psarc[linewidth=0.9pt,linecolor=black,border=0pt](1.5,-0.9){0.7}{0}{60}
  \psarc[linewidth=0.9pt,linecolor=black,arrows=->,arrowscale=1.4,
    arrowinset=0.15] (1.5,-0.9){0.7}{0}{35}
  \psarc[linewidth=0.9pt,linecolor=black,border=0pt]
(2.2,-0.9){0.7}{120}{180}
  \psarc[linewidth=0.9pt,linecolor=black,arrows=<-,arrowscale=1.4,
    arrowinset=0.15] (2.2,-0.9){0.7}{145}{180}
  \psarc[linewidth=0.9pt,linecolor=black,border=0pt] (0.1,0.9){0.7}{-60}{0}
  \psarc[linewidth=0.9pt,linecolor=black,arrows=->,arrowscale=1.4,
    arrowinset=0.15] (0.1,0.9){0.7}{-60}{-10}
  \psarc[linewidth=0.9pt,linecolor=black,border=0pt](0.8,0.9){0.7}{180}{240}
  \psarc[linewidth=0.9pt,linecolor=black,arrows=<-,arrowscale=1.4,
    arrowinset=0.15] (0.8,0.9){0.7}{190}{240}
  \psarc[linewidth=0.9pt,linecolor=black,border=0pt](0.1,-0.9){0.7}{0}{60}
  \psarc[linewidth=0.9pt,linecolor=black,arrows=->,arrowscale=1.4,
    arrowinset=0.15] (0.1,-0.9){0.7}{0}{35}
  \psarc[linewidth=0.9pt,linecolor=black,border=0pt](0.8,-0.9){0.7}{120}{180}
  \psarc[linewidth=0.9pt,linecolor=black,arrows=<-,arrowscale=1.4,
    arrowinset=0.15] (0.8,-0.9){0.7}{145}{180}
  \psarc[linewidth=0.9pt,linecolor=black,border=0pt](-0.3,2.6){0.4}{0}{180}
  \psarc[linewidth=0.9pt,linecolor=black,border=0pt](-0.3,-2.6){0.4}{-180}{0}
  \psline(0.1,0.9)(0.1,2.6)
  \psline(0.1,-0.9)(0.1,-2.6)
  \psline(-0.7,2.6)(-0.7,-2.6)
  \psarc[linewidth=0.9pt,linecolor=black,border=0pt] (1.9,1.3){0.4}{90}{170}
  \psarc[linewidth=0.9pt,linecolor=black,border=0pt] (1.9,-1.3){0.4}{190}{270}
  \psline(1.9,1.7)(2.5,1.7)
  \psline(1.9,-1.7)(2.5,-1.7)
  \psarc[linewidth=0.9pt,linecolor=black,border=0pt](2.5,2.1){0.4}{-90}{0}
  \psarc[linewidth=0.9pt,linecolor=black,border=0pt](2.5,-2.1){0.4}{0}{90}
  \psframe[linewidth=0.9pt,linecolor=black,border=0](2.6,2.1)(3.2,2.6)
  \psframe[linewidth=0.9pt,linecolor=black,border=0](2.6,-2.1)(3.2,-2.6)
  \rput[bl]{0}(2.7,2.18){$\Pi_s$}
  \rput[bl]{0}(2.7,-2.52){$\Pi_s$}
  \psarc[linewidth=0.9pt,linecolor=black,border=0pt](3.3,2.6){0.4}{0}{180}
  \psarc[linewidth=0.9pt,linecolor=black,border=0pt](3.3,-2.6){0.4}{-180}{0}
  \psline(3.7,2.6)(3.7,-2.6)
  \psline(1.85,-0.3)(1.85,0.3)
  \psline[border=2pt](2.2,-0.9)(2.2,-2.8)
  \psline[border=2pt](2.2,0.9)(2.2,2.8)
  \psline(0.45,0.28)(0.45,0.306)
  \psline(0.45,-0.28)(0.45,-0.306)
  \psline{->}(0.8,-2.8)(0.8,-2.1)
  \psline{->}(0.8,1.4)(0.8,2.4)
  \psline{->}(2.2,-2.8)(2.2,-2.1)
  \psline{->}(2.2,1.4)(2.2,2.4)
  \psline{->}(-0.7,0.4)(-0.7,-0.1)
  \psline{->}(3.7,0.4)(3.7,-0.1)
  \psline{->}(1.9,1.69)(2.02,1.71)
  \psline{<-}(1.8,-1.69)(2.02,-1.7)
  \psline{->}(1.85,-0.3)(1.85,0.12)
  \rput[bl]{0}(0.7,2.9){$a$}
  \rput[tl]{0}(0.7,-2.8){$a'$}
  \rput[bl]{0}(2.1,2.9){$c$}
  \rput[tl]{0}(2.1,-2.8){$c'$}
  \rput[bl]{0}(0.72,0.08){$b_{\shortrightarrow}$}
  \rput[bl]{0}(-0.03,0.15){$\bar{b}$}
  \rput[bl]{0}(0.72,-0.52){$b_{\shortrightarrow}$}
  \rput[bl]{0}(-0.03,-0.45){$\bar{b}$}
  \rput[bl]{0}(-1.05,-0.1){$\bar{b}$}
  \rput[bl]{0}(3.85,-0.1){$b_{s}$}
  \rput[bl]{0}(1.55,-0.2){$f$}
  \rput[bl]{0}(1.27,0.4){$a$}
  \rput[bl]{0}(1.23,-0.6){$a'$}
  \scriptsize
  \rput[bl]{0}(1.76,0.5){$\mu$}
  \rput[bl]{0}(1.71,-0.7){$\mu'$}
 \endpspicture
\end{equation}%
For the outcome $s=\shortrightarrow $, this is%
\begin{eqnarray}
&&
 \pspicture[shift=-1.95](0.3,-2.1)(2.85,1.9)
  \small
  \psframe[linewidth=0.9pt,linecolor=black,border=0](0.8,0.7)(1.7,1.2)
  \psframe[linewidth=0.9pt,linecolor=black,border=0](0.8,-0.7)(1.7,-1.2)
  \rput[bl]{0}(1.1,0.81){$U$}
  \rput[tl]{0}(1.05,-0.77){$U^{\dag}$}
  \psset{linewidth=0.9pt,linecolor=black,arrowscale=1.5,arrowinset=0.15}
  \psline(1,-0.7)(1,0.7)
  \psline(1,-1.2)(1,-1.7)
  \psline(1,1.2)(1,1.7)
  \psarc[linewidth=0.9pt,linecolor=black,border=0pt] (1.8,-0.6){0.3}{0}{180}
  \psarc[linewidth=0.9pt,linecolor=black,border=0pt] (1.8,0.6){0.3}{180}{360}
  \psarc[linewidth=0.9pt,linecolor=black,border=0pt] (1.9,1.1){0.4}{0}{170}
  \psarc[linewidth=0.9pt,linecolor=black,border=0pt] (1.9,-1.1){0.4}{190}{360}
  \psline(1.8,-0.3)(1.8,0.3)
  \psline(1.5,0.6)(1.5,0.7)
  \psline(1.5,-0.6)(1.5,-0.7)
  \psline[border=2pt](2.1,-0.6)(2.1,-1.7)
  \psline[border=2pt](2.1,0.6)(2.1,1.7)
  \psline(2.3,-1.1)(2.3,1.1)
  \psline{->}(1,-0.7)(1,0.12)
  \psline{->}(1,-1.7)(1,-1.3)
  \psline{->}(1,1.2)(1,1.6)
  \psline{->}(2.3,0.4)(2.3,-0.1)
  \psline{->}(2.1,0.6)(2.1,1.0)
  \psline{->}(2.1,-1.1)(2.1,-0.8)
  \psline{->}(1.8,-0.2)(1.8,0.12)
  \psarc[linewidth=0.9pt,linecolor=black,arrows=->,arrowscale=1.4,
    arrowinset=0.15]{<-}(1.8,-0.6){0.3}{120}{180}
  \psarc[linewidth=0.9pt,linecolor=black,arrows=->,arrowscale=1.4,
    arrowinset=0.15]{<-}(1.8,0.6){0.3}{190}{270}
  \rput[bl]{0}(0.9,1.8){$a$}
  \rput[tl]{0}(0.9,-1.7){$a'$}
  \rput[bl]{0}(2.0,1.8){$c$}
  \rput[tl]{0}(2.0,-1.7){$c'$}
  \rput[bl]{0}(0.4,-0.1){$b_{\shortrightarrow}$}
  \rput[bl]{0}(2.45,-0.1){$b_{\shortrightarrow}$}
  \rput[bl]{0}(1.5,-0.2){$f$}
  \rput[bl]{0}(1.25,0.35){$a$}
  \rput[bl]{0}(1.2,-0.55){$a'$}
  \scriptsize
  \rput[bl]{0}(1.75,0.45){$\mu$}
  \rput[bl]{0}(1.71,-0.65){$\mu'$}
 \endpspicture
=\sum\limits_{ e,\alpha ,\beta }\left[ \left( F_{a^{\prime
}c^{\prime }}^{ac}\right) ^{-1}\right] _{\left( f,\mu ,\mu ^{\prime
}\right) \left( e,\alpha ,\beta \right) }
\pspicture[shift=-1.95](0.3,-2.1)(3.00,1.9)
 \small
  \psframe[linewidth=0.9pt,linecolor=black,border=0](0.8,0.7)(1.7,1.2)
  \psframe[linewidth=0.9pt,linecolor=black,border=0](0.8,-0.7)(1.7,-1.2)
  \rput[bl]{0}(1.1,0.81){$U$}
  \rput[tl]{0}(1.05,-0.77){$U^{\dag}$}
  \psset{linewidth=0.9pt,linecolor=black,arrowscale=1.5,arrowinset=0.15}
  \psline(1,-0.7)(1,0.7)
  \psline(1,-1.7)(1,-1.2)
  \psline(1,1.2)(1,1.7)
  \psline(1.5,-0.7)(1.5,0.7)
  \psarc[linewidth=0.9pt,linecolor=black,border=0pt] (2.1,1.1){0.4}{0}{90}
  \psline(1.9,1.5)(2.1,1.5)
  \psarc[linewidth=0.9pt,linecolor=black,border=0pt] (1.9,1.1){0.4}{90}{170}
  \psarc[linewidth=0.9pt,linecolor=black,border=0pt] (1.9,-1.1){0.4}{190}{270}
  \psline(1.9,-1.5)(2.1,-1.5)
  \psarc[linewidth=0.9pt,linecolor=black,border=0pt] (2.1,-1.1){0.4}{270}{360}
  \psline(2.1,-1.7)(2.1,1.7)
  \psline[border=2pt](2.1,0.7)(2.1,1.7)
  \psline[border=2pt](2.1,-1.7)(2.1,-0.7)
  \psline(2.5,-1.1)(2.5,1.1)
  \psline(1.5,0.1)(2.1,-0.1)
  \psline{->}(1,-0.7)(1,0.12)
  \psline{->}(1,-1.7)(1,-1.3)
  \psline{->}(1,1.2)(1,1.6)
  \psline{->}(2.5,0.4)(2.5,-0.1)
  \psline{->}(2.1,0.7)(2.1,1.1)
  \psline{->}(2.1,-1.2)(2.1,-0.9)
  \psline{->}(2.1,-0.1)(1.633,0.05)
  \psline{->}(1.5,-0.7)(1.5,-0.2)
  \psline{->}(1.5,0.1)(1.5,0.52)
  \rput[bl]{0}(0.9,1.8){$a$}
  \rput[tl]{0}(0.9,-1.7){$a'$}
  \rput[bl]{0}(2.0,1.8){$c$}
  \rput[tl]{0}(2.0,-1.7){$c'$}
  \rput[bl]{0}(0.4,-0.1){$b_{\shortrightarrow}$}
  \rput[bl]{0}(2.65,-0.1){$b_{\shortrightarrow}$}
  \rput[bl]{0}(1.7,-0.35){$e$}
  \rput[bl]{0}(1.2,0.3){$a$}
  \rput[bl]{0}(1.15,-0.45){$a'$}
  \scriptsize
  \rput[bl]{0}(1.25,0.04){$\alpha$}
  \rput[bl]{0}(2.15,-0.27){$\beta$}
\endpspicture
\nonumber \\
&=&\sum\limits_{ e,\alpha ,\beta }\left[ \left( F_{a^{\prime
}c^{\prime }}^{ac}\right) ^{-1}\right] _{\left( f,\mu ,\mu ^{\prime
}\right) \left( e,\alpha ,\beta \right) }
\nonumber \\
&& \times \left\{ \left|
t_{1}\right| ^{2}\left| r_{2}\right| ^{2}
\pspicture[shift=-1.1](0,-1.2)(1.9,1.15)
  \small
  \psellipse[linewidth=0.9pt,linecolor=black,border=0](1.19,0)(0.18,0.4)
  \psset{linewidth=0.9pt,linecolor=black,arrowscale=1.4,arrowinset=0.15}
  \psline{>-}(1.08,-0.30)(1.03,-0.10)
  \psset{linewidth=0.9pt,linecolor=black,arrowscale=1.5,arrowinset=0.15}
  \psline(0.35,-0.775)(0.35,0.775)
  \psline{>-}(0.35,0.3)(0.35,0.6)
  \psline{<-}(0.35,-0.3)(0.35,-0.6)
  \psline(1.55,-0.775)(1.55,0.775)
  \psline{>-}(1.55,0.3)(1.55,0.6)
  \psline{<-}(1.55,-0.3)(1.55,-0.6)
  \psline(1.13,-0.03)(1.55,-0.1)
  \psline[border=2pt](1.13,-0.03)(1.49,-0.09)
   \psline{-<}(0.35,0.1)(0.83,0.02)
   \psline(0.35,0.1)(0.92,0.005)
  \rput[bl]{0}(0.25,0.875){$a$}
  \rput[bl]{0}(1.45,0.875){${c}$}
  \rput[bl]{0}(0.25,-1.075){$a'$}
  \rput[bl]{0}(1.45,-1.075){${c'}$}
  \rput[br]{0}(0.9,0.2){$e$}
  \rput[br]{0}(0.97,-0.52){$b$}
  \scriptsize
  \rput[br]{0}(0.28,0.05){$\alpha$}
  \rput[br]{0}(1.82,-0.25){$\beta$}
  \endpspicture
\right.
+t_{1}r_{1}^{\ast }r_{2}^{\ast }t_{2}^{\ast }e^{i\left( \theta _{\text{I}}-\theta
_{\text{II}}\right) }
 \pspicture[shift=-1.1](-0.25,-1)(1.8,1.35)
  \small
  \psellipse[linewidth=0.9pt,linecolor=black,border=0](0.35,0.6)(0.4,0.18)
  \psset{linewidth=0.9pt,linecolor=black,arrowscale=1.4,arrowinset=0.15}
  \psline{->}(0.1,0.725)(0.3,0.77)
  \psset{linewidth=0.9pt,linecolor=black,arrowscale=1.5,arrowinset=0.15}
  \psline(0.35,-0.6)(0.35,0.69)
  \psline(0.35,0.84)(0.35,0.95)
  \psline[border=3pt]{>-}(0.35,0.3)(0.35,0.6)
  \psline{<-}(0.35,-0.3)(0.35,-0.6)
  \psline(1.35,-0.6)(1.35,0.95)
  \psline{>-}(1.35,0.3)(1.35,0.6)
  \psline{<-}(1.35,-0.3)(1.35,-0.6)
  \psline(0.35,0.1)(1.35,-0.1)
  \psline{->}(1.35,-0.1)(0.7,0.03)
  \rput[bl]{0}(0.25,1.05){$a$}
  \rput[bl]{0}(1.25,1.05){${c}$}
  \rput[bl]{0}(0.25,-0.9){$a'$}
  \rput[bl]{0}(1.25,-0.9){${c'}$}
  \rput[br]{0}(0.95,0.13){$e$}
  \rput[br]{0}(-0.05,0.7){$b$}
  \scriptsize
  \rput[br]{0}(0.28,0.05){$\alpha$}
  \rput[br]{0}(1.62,-0.25){$\beta$}
 \endpspicture
\nonumber \\
&&+t_{1}^{\ast }r_{1}t_{2}r_{2}e^{-i\left( \theta _{\text{I}}-\theta _{\text{II}}\right)
}
 \pspicture[shift=-1.13](-0.25,-1.4)(1.8,1)
  \small
  \psellipse[linewidth=0.9pt,linecolor=black,border=0](0.35,-0.6)(0.4,0.18)
  \psset{linewidth=0.9pt,linecolor=black,arrowscale=1.4,arrowinset=0.15}
  \psline{-<}(0.07,-0.715)(0.27,-0.76)
  \psset{linewidth=0.9pt,linecolor=black,arrowscale=1.5,arrowinset=0.15}
  \psline(0.35,-0.69)(0.35,0.6)
  \psline(0.35,-0.84)(0.35,-0.95)
  \psline{>-}(0.35,0.3)(0.35,0.6)
  \psline[border=3pt]{>-}(0.35,-0.55)(0.35,-0.3)
  \psline(1.35,0.6)(1.35,-0.95)
  \psline{>-}(1.35,0.3)(1.35,0.6)
  \psline{>-}(1.35,-0.55)(1.35,-0.3)
  \psline(0.35,0.1)(1.35,-0.1)
  \psline{->}(1.35,-0.1)(0.7,0.03)
  \rput[bl]{0}(0.25,0.7){$a$}
  \rput[bl]{0}(1.25,0.7){${c}$}
  \rput[bl]{0}(0.25,-1.25){$a'$}
  \rput[bl]{0}(1.25,-1.25){${c'}$}
  \rput[br]{0}(0.95,0.13){$e$}
  \rput[br]{0}(-0.05,-0.9){$b$}
  \scriptsize
  \rput[br]{0}(0.28,0.05){$\alpha$}
  \rput[br]{0}(1.62,-0.25){$\beta$}
 \endpspicture
\left. +\left| r_{1}\right| ^{2}\left| t_{2}\right|
^{2}
\pspicture[shift=-1.1](-0.7,-1.2)(1.75,1.15)
  \small
  \psellipse[linewidth=0.9pt,linecolor=black,border=0](-0.2,0.0)(0.18,0.4)
  \psset{linewidth=0.9pt,linecolor=black,arrowscale=1.4,arrowinset=0.15}
  \psline{->}(-0.358,-0.1)(-0.365,0.1)
  \psset{linewidth=0.9pt,linecolor=black,arrowscale=1.5,arrowinset=0.15}
  \psline(0.35,-0.775)(0.35,0.775)
  \psline{>-}(0.35,0.3)(0.35,0.6)
  \psline{<-}(0.35,-0.3)(0.35,-0.6)
  \psline(1.35,-0.775)(1.35,0.775)
  \psline{>-}(1.35,0.3)(1.35,0.6)
  \psline{<-}(1.35,-0.3)(1.35,-0.6)
  \psline(0.35,0.1)(1.35,-0.1)
  \psline{->}(1.35,-0.1)(0.7,0.03)
  \rput[bl]{0}(0.25,0.875){$a$}
  \rput[bl]{0}(1.25,0.875){${c}$}
  \rput[bl]{0}(0.25,-1.075){$a'$}
  \rput[bl]{0}(1.25,-1.075){${c'}$}
  \rput[br]{0}(0.95,0.13){$e$}
  \rput[br]{0}(-0.5,-0.3){$b$}
  \scriptsize
  \rput[br]{0}(0.28,0.05){$\alpha$}
  \rput[br]{0}(1.62,-0.25){$\beta$}
  \endpspicture
\right\}
\nonumber \\
&=& d_b \sum\limits_{ e,\alpha ,\beta }\left[ \left(
F_{a^{\prime
}c^{\prime }}^{ac}\right) ^{-1}\right] _{\left( f,\mu ,\mu ^{\prime
}\right) \left( e,\alpha ,\beta \right) }p_{aa^{\prime }e,b}^{\shortrightarrow
}
\pspicture[shift=-1.1](-0.1,-1.2)(1.8,1.15)
  \small
  \psset{linewidth=0.9pt,linecolor=black,arrowscale=1.5,arrowinset=0.15}
  \psline(0.35,-0.775)(0.35,0.775)
  \psline{>-}(0.35,0.3)(0.35,0.6)
  \psline{<-}(0.35,-0.3)(0.35,-0.6)
  \psline(1.35,-0.775)(1.35,0.775)
  \psline{>-}(1.35,0.3)(1.35,0.6)
  \psline{<-}(1.35,-0.3)(1.35,-0.6)
  \psline(0.35,0.1)(1.35,-0.1)
   \psline{->}(1.35,-0.1)(0.7,0.03)
  \rput[bl]{0}(0.25,0.875){$a$}
  \rput[bl]{0}(1.25,0.875){${c}$}
  \rput[bl]{0}(0.25,-1.075){$a'$}
  \rput[bl]{0}(1.25,-1.075){${c'}$}
  \rput[br]{0}(0.95,0.13){$e$}
  \scriptsize
  \rput[br]{0}(0.28,0.05){$\alpha$}
  \rput[br]{0}(1.62,-0.25){$\beta$}
  \endpspicture
\nonumber \\
&=& d_b \sum\limits_{\substack{ e,\alpha ,\beta  \\
f^{\prime },\nu ,\nu ^{\prime } }}\left[ \left( F_{a^{\prime
}c^{\prime }}^{ac}\right) ^{-1}\right] _{\left( f,\mu ,\mu ^{\prime
}\right) \left( e,\alpha ,\beta \right) }\left[ F_{a^{\prime }c^{\prime
}}^{ac}\right] _{\left( e,\alpha ,\beta \right) \left( f^{\prime },\nu ,\nu
^{\prime }\right) }p_{aa^{\prime }e,b}^{\shortrightarrow}
 \pspicture[shift=-1.1](0,-0.85)(1.3,1.3)
 \small
  \psset{linewidth=0.9pt,linecolor=black,arrowscale=1.5,arrowinset=0.15}
  \psline{->}(0.7,0)(0.7,0.45)
  \psline(0.7,0)(0.7,0.55)
  \psline(0.7,0.55) (0.2,1.05)
  \psline{->}(0.7,0.55)(0.3,0.95)
  \psline(0.7,0.55) (1.2,1.05)
  \psline{->}(0.7,0.55)(1.1,0.95)
  \rput[bl]{0}(0.28,0.1){$f'$}
  \rput[bl]{0}(1.1,1.15){$c$}
  \rput[bl]{0}(0.1,1.15){$a$}
  \psline(0.7,0) (0.2,-0.5)
  \psline{-<}(0.7,0)(0.35,-0.35)
  \psline(0.7,0) (1.2,-0.5)
  \psline{-<}(0.7,0)(1.05,-0.35)
  \rput[bl]{0}(1.1,-0.8){$c'$}
  \rput[bl]{0}(0.1,-0.8){$a'$}
  \scriptsize
  \rput[bl]{0}(0.82,0.45){$\nu$}
  \rput[bl]{0}(0.82,-0.0){$\nu'$}
  \endpspicture
\end{eqnarray}%
where we have defined%
\begin{eqnarray}
\label{eq:p_right}
\qquad \qquad
p_{aa^{\prime }e,b}^{\shortrightarrow } &=&\left| t_{1}\right| ^{2}\left|
r_{2}\right| ^{2}M_{eb}+t_{1}r_{1}^{\ast }r_{2}^{\ast }t_{2}^{\ast
}e^{i\left( \theta _{\text{I}}-\theta _{\text{II}}\right) }M_{ab}  \notag \\
&&+t_{1}^{\ast }r_{1}t_{2}r_{2}e^{-i\left( \theta _{\text{I}}-\theta _{\text{II}}\right)
}M_{a^{\prime }b}^{\ast }+\left| r_{1}\right| ^{2}\left| t_{2}\right| ^{2}
\end{eqnarray}%
and have used Eqs.~(\ref{eq:loopaway},\ref{eq:monodromy}) to remove the $b$
loops. A similar calculation for
the $s=\shortuparrow $ outcome gives%
\begin{eqnarray}
\label{eq:p_up}
\qquad \qquad
p_{aa^{\prime }e,b}^{\shortuparrow } &=&\left| t_{1}\right| ^{2}\left|
t_{2}\right| ^{2}M_{eb}-t_{1}r_{1}^{\ast }r_{2}^{\ast }t_{2}^{\ast
}e^{i\left( \theta _{\text{I}}-\theta _{\text{II}}\right) }M_{ab}  \notag \\
&&-t_{1}^{\ast }r_{1}t_{2}r_{2}e^{-i\left( \theta _{\text{I}}-\theta _{\text{II}}\right)
}M_{a^{\prime }b}^{\ast }+\left| r_{1}\right| ^{2}\left| r_{2}\right| ^{2}.
\end{eqnarray}
{}From this, inserting the appropriate coefficients and normalization factors,
we find the reduced density matrix of the target anyons after a
single probe measurement with outcome $s$:%
\begin{eqnarray}
\rho ^{A}\left( s \right) &=& \frac{1}{\Pr \left( s\right) }
\widetilde{\text{Tr}}_{\overline{B},B}\left[\Pi _{s}\rho \Pi _{s}\right]
\nonumber \\
&=&\sum\limits_{\substack{ a,a^{\prime
},c,c^{\prime },f,\mu ,\mu ^{\prime } \\ e,\alpha ,\beta,
f^{\prime },\nu ,\nu ^{\prime } }} \frac{ \rho _{\left(
a,c;f,\mu \right) \left( a^{\prime },c^{\prime };f,\mu ^{\prime }\right)
}^{A} }{ \left( d_{a}d_{a^{\prime }}d_{c}d_{c^{\prime}}d_{f}^{2} \right)^{1/4}
}
\frac{p_{aa^{\prime }e,b}^{s}}{\Pr \left( s\right) }
\nonumber \\
& &\times \left[ \left( F_{a^{\prime }c^{\prime }}^{ac}\right) ^{-1}\right]
_{\left( f,\mu ,\mu ^{\prime }\right) \left( e,\alpha ,\beta \right) }\left[
F_{a^{\prime }c^{\prime }}^{ac}\right] _{\left( e,\alpha ,\beta \right)
\left( f^{\prime },\nu ,\nu ^{\prime }\right) }
 \pspicture[shift=-0.75](-0.15,-0.55)(1.5,1.1)
 \small
  \psset{linewidth=0.9pt,linecolor=black,arrowscale=1.5,arrowinset=0.15}
  \psline{->}(0.7,0)(0.7,0.45)
  \psline(0.7,0)(0.7,0.55)
  \psline(0.7,0.55) (0.25,1)
  \psline{->}(0.7,0.55)(0.3,0.95)
  \psline(0.7,0.55) (1.15,1)
  \psline{->}(0.7,0.55)(1.1,0.95)
  \rput[bl]{0}(0.28,0.1){$f'$}
  \rput[bl]{0}(1.22,0.8){$c$}
  \rput[bl]{0}(-0.05,0.8){$a$}
  \psline(0.7,0) (0.25,-0.45)
  \psline{-<}(0.7,0)(0.35,-0.35)
  \psline(0.7,0) (1.15,-0.45)
  \psline{-<}(0.7,0)(1.05,-0.35)
  \rput[bl]{0}(1.22,-0.4){$c'$}
  \rput[bl]{0}(-0.05,-0.4){$a'$}
  \scriptsize
  \rput[bl]{0}(0.82,0.45){$\nu$}
  \rput[bl]{0}(0.82,-0.0){$\nu'$}
  \endpspicture
\nonumber \\
&=&\sum\limits_{\substack{ a,a^{\prime
},c,c^{\prime },f,\mu ,\mu ^{\prime } \\ e,\alpha ,\beta,
f^{\prime },\nu ,\nu ^{\prime } }} \frac{\rho _{\left(
a,c;f,\mu \right) \left( a^{\prime },c^{\prime };f,\mu ^{\prime }\right)
}^{A}}{\left( d_{f} d_{f^{\prime }} \right)^{1/2}}
\frac{p_{aa^{\prime }e,b}^{s}}{\Pr \left( s\right) }
\left[ \left( F_{a^{\prime }c^{\prime }}^{ac}\right)
^{-1}\right]
_{\left( f,\mu ,\mu ^{\prime }\right) \left( e,\alpha ,\beta \right) }
\nonumber \\
& &\times \left[
F_{a^{\prime }c^{\prime }}^{ac}\right] _{\left( e,\alpha ,\beta \right)
\left( f^{\prime },\nu ,\nu ^{\prime }\right) }\left| a,c;f^{\prime },\nu
\right\rangle \left\langle a^{\prime
},c^{\prime };f^{\prime },\nu ^{\prime }\right|
\end{eqnarray}%
where the probability of measurement outcome $s$ is found by additionally taking
the quantum trace of the target system, which projects onto the $e=1$
components,
giving%
\begin{equation}
\Pr \left( s\right)=\widetilde{\text{Tr}}\left[ \rho \Pi_{s} \right]
=\sum\limits_{a,c,f,\mu }\rho _{\left( a,c;f,\mu \right)
\left( a,c;f,\mu \right) }^{A}p_{aa1,b}^{s}.
\label{eq:single_probability}
\end{equation}%
We note that%
\begin{eqnarray}
\qquad
p_{aa1,b}^{\shortrightarrow } &=&\left| t_{1}\right| ^{2}\left| r_{2}\right|
^{2}+\left| r_{1}\right| ^{2}\left| t_{2}\right| ^{2}+2\text{Re}\left\{
t_{1}r_{1}^{\ast }r_{2}^{\ast }t_{2}^{\ast }e^{i\left( \theta _{\text{I}}-\theta
_{\text{II}}\right) }M_{ab}\right\}  \\
p_{aa1,b}^{_{\shortuparrow }} &=&\left| t_{1}\right| ^{2}\left| t_{2}\right|
^{2}+\left| r_{1}\right| ^{2}\left| r_{2}\right| ^{2}-2\text{Re}\left\{
t_{1}r_{1}^{\ast }r_{2}^{\ast }t_{2}^{\ast }e^{i\left( \theta _{\text{I}}-\theta
_{\text{II}}\right) }M_{ab}\right\}
\end{eqnarray}%
give a well-defined probability distribution in $s$ (i.e. $0\leq p_{aa1,b}^{s}\leq 1
$ and $p_{aa1,b}^{\shortrightarrow }+p_{aa1,b}^{_{\shortuparrow }}=1$).

The quantity%
\begin{equation}
t_{1}r_{1}^{\ast }t_{2}^{\ast }r_{2}^{\ast }e^{i\left( \theta _{\text{I}}-\theta
_{\text{II}}\right) }\equiv Te^{i\theta }
\label{eq:visibility}
\end{equation}
determines the visibility of quantum interference in this experiment, where
varying $\theta $ allows one to observe the interference term modulation.
The amplitude $T=\left| t_{1}r_{1}t_{2}r_{2}\right| $ is maximized by $%
\left| t_{j}\right| =\left| r_{j}\right| =1/\sqrt{2}$. In realistic
experiments, the experimental parameters $t_{j}$, $r_{j}$, $\theta _{\text{I}}$,
and $\theta _{\text{II}}$ will have some variance, even for a single probe, that
gives rise to some degree of phase incoherence.
Averaging over some distribution in $\theta$, one finds that $e^{i\theta }$
in the interference terms should effectively be replaced by
$\left\langle e^{i\theta }\right\rangle = Qe^{i\theta_{\ast }}$. In this
expression, $e^{i\theta _{\ast }}$ is the resulting effective phase, and
$Q\in \left[ 0,1 \right] $ is a suppression factor that reflects the
interferometer's lack of coherence, and reduces the visibility of quantum
interference. For the rest of the paper, we will ignore this issue and assume
$Q=1$, but it should always be kept in mind that success of any interferometry
experiment is crucially dependent on $Q$ being made as large as possible.

We can now obtain the result for general $\rho ^{B}$ by simply
replacing $p_{aa^{\prime}e,b}^{s}$ everywhere with%
\begin{equation}
\label{eq:p_B}
p_{aa^{\prime }e,B}^{s} = \sum\limits_{b}\Pr\nolimits_{B}\left( b\right)
p_{aa^{\prime }e,b}^{s}
\end{equation}
\begin{equation}
\Pr\nolimits_{B}\left( b\right) = \sum\limits_{d,h,\lambda }\rho _{\left(
d,b_{\shortrightarrow};h,\lambda \right) \left(
d,b_{\shortrightarrow};h,\lambda \right) }^{B}
.
\end{equation}%
We will also use the notation $M_{aB}=\sum\nolimits_{b}\Pr\nolimits_{B}%
\left( b\right) M_{ab}$. That this replacement gives the appropriate results
follows from the fact that we trace out the $D$ anyon, and may be seen from%
\begin{eqnarray}
\qquad \quad
\widetilde{\text{Tr}}_{D}\left[ \rho ^{B}\right]  &=& \sum\limits_{b,b^{\prime
},d,h,\lambda,\lambda ^{\prime }}
\frac{ \rho _{\left( d,b_{\shortrightarrow };h,\lambda\right)
\left( d,b_{\shortrightarrow }^{\prime };h,\lambda ^{\prime }\right)}^{B}}
{ \left(d_{d}^{2}d_{b}d_{b^{\prime }}d_{h}^{2} \right) ^{1/4} }
\pspicture[shift=-1.2](-0.6,-1)(1.45,1.3)
 \small
  \psset{linewidth=0.9pt,linecolor=black,arrowscale=1.5,arrowinset=0.15}
  \psline{->}(0.7,0)(0.7,0.45)
  \psline(0.7,0)(0.7,0.55)
  \psline(0.7,0.55) (0.2,1.05)
  \psline{->}(0.7,0.55)(0.3,0.95)
  \psline(0.7,0.55) (1.2,1.05)
  \psline{->}(0.7,0.55)(1.1,0.95)
  \rput[bl]{0}(0.28,0.15){$h$}
  \rput[bl]{0}(1.1,1.07){$b_\shortrightarrow$}
  \rput[bl]{0}(0.2,1.13){$d$}
  \psline(0.7,0) (0.2,-0.5)
  \psline{-<}(0.7,0)(0.35,-0.35)
  \psline(0.7,0) (1.2,-0.5)
  \psline{-<}(0.7,0)(1.05,-0.35)
 \psarc[linewidth=0.9pt,linecolor=black,border=0pt] (-0.05,0.8){0.35}{45}{180}
 \psarc[linewidth=0.9pt,linecolor=black,border=0pt]
   (-0.05,-0.25){0.35}{180}{315}
 \psline(-0.4,-0.25)(-0.4,0.8)
\psline{->}(-0.4,0.8)(-0.4,0.1)
  \rput[bl]{0}(1.1,-0.91){$b'_\shortrightarrow$}
  \rput[bl]{0}(0.2,-0.81){$d$}
  \scriptsize
  \rput[bl]{0}(0.82,0.38){$\lambda$}
  \rput[bl]{0}(0.82,-0.02){$\lambda'$}
  \endpspicture
\nonumber
\\
&=& \sum\limits_{b,d,h,\lambda }\rho _{\left( d,b_{\shortrightarrow
};h,\lambda
\right) \left( d,b_{\shortrightarrow };h,\lambda \right) }^{B}
\frac{1}{d_{b}}
\pspicture[shift=-1.02](0.15,-1.15)(0.9,1.15)
  \small
  \psset{linewidth=0.9pt,linecolor=black,arrowscale=1.5,arrowinset=0.15}
  \psline(0.35,-0.775)(0.35,0.775)
  \psline{->}(0.35,-0.3)(0.35,0.2)
  \rput[bl]{0}(0.5,-0.1){$b_\shortrightarrow$}
  \endpspicture
\nonumber \\
&=& \sum\limits_{b,d,h,\lambda }\rho _{\left( d,b_{\shortrightarrow
};h,\lambda
\right) \left( d,b_{\shortrightarrow };h,\lambda \right) }^{B}
\frac{1}{d_{b}}
\pspicture[shift=-1.25](-0.6,-1)(1.45,1.4)
 \small
  \psset{linewidth=0.9pt,linecolor=black,arrowscale=1.5,arrowinset=0.15}
  \psline(0.7,0.55) (0.2,1.05)
  \psline{->}(0.7,0.55)(0.3,0.95)
  \psline(0.7,0.55) (1.2,1.05)
  \psline{->}(0.7,0.55)(1.1,0.95)
  \rput[bl]{0}(1.1,1.07){$b_\shortrightarrow$}
  \rput[bl]{0}(0.25,1.13){$\bar{b}$}
  \psline(0.7,0) (0.2,-0.5)
  \psline{-<}(0.7,0)(0.35,-0.35)
  \psline(0.7,0) (1.2,-0.5)
  \psline{-<}(0.7,0)(1.05,-0.35)
 \psarc[linewidth=0.9pt,linecolor=black,border=0pt] (-0.05,0.8){0.35}{45}{180}
 \psarc[linewidth=0.9pt,linecolor=black,border=0pt]
   (-0.05,-0.25){0.35}{180}{315}
 \psline(-0.4,-0.25)(-0.4,0.8)
\psline{->}(-0.4,0.8)(-0.4,0.1)
  \rput[bl]{0}(1.1,-0.91){$b_\shortrightarrow$}
  \rput[bl]{0}(0.25,-0.85){$\bar{b}$}
  \endpspicture
\nonumber \\
&=& \sum\limits_{b}\Pr\nolimits_{B}\left( b\right)
\widetilde{\text{Tr}}_{\bar{b}}\left| \bar{b}%
,b_{\shortrightarrow };1\right\rangle \left\langle \bar{b},b_{\shortrightarrow
};1\right|
\nonumber \\
&=& \widetilde{\text{Tr}}_{\overline{B}}\sum\limits_{b}\Pr\nolimits_{B}\left(
b\right) \left|
\bar{b},b_{\shortrightarrow };1\right\rangle \left\langle
\bar{b},b_{\shortrightarrow
};1\right|
\end{eqnarray}%
where we used Eq.~(\ref{eq:tracebub}) in the first step. 

%% file: 03c-Nprobes.tex
\subsection{N Probes}
  \label{sec:NProbes}

The result for $N$ initially unentangled identical probe particles sent
through the interferometer may now be easily produced by iterating the
single probe calculation. The string of measurement outcomes $\left(
s_{1},\ldots ,s_{N}\right) $ occurs with probability%
\begin{equation}
\Pr \left( s_{1},\ldots ,s_{N}\right) =\sum\limits_{a,c,f,\mu }\rho _{\left(
a,c;f,\mu \right) \left( a,c;f,\mu \right) }^{A}p_{aa1,B}^{s_{1}}\ldots
p_{aa1,B}^{s_{N}}
\end{equation}%
and results in the measured target anyon reduced density matrix%
\begin{eqnarray}
&& \rho ^{A}\left( s_{1},\ldots ,s_{N}\right) =\sum\limits_{\substack{ %
a,a^{\prime },c,c^{\prime },f,\mu ,\mu ^{\prime }
\\ e,\alpha ,\beta ,f^{\prime },\nu ,\nu ^{\prime } }}
\frac{\rho _{\left(a,c;f,\mu \right) \left( a^{\prime },c^{\prime };f,\mu ^{\prime }\right)
}^{A}}{\left( d_{f}d_{f^{\prime }} \right) ^{1/2} }
\frac{p_{aa^{\prime }e,B}^{s_{1}}\ldots p_{aa^{\prime }e,B}^{s_{N}}}{%
\Pr \left( s_{1},\ldots ,s_{N}\right) }  \notag \\
&& \times \left[ \left( F_{a^{\prime }c^{\prime }}^{ac}\right) ^{-1}\right]
_{\left( f,\mu ,\mu ^{\prime }\right) \left( e,\alpha ,\beta \right) }\left[
F_{a^{\prime }c^{\prime }}^{ac}\right] _{\left( e,\alpha ,\beta \right)
\left( f^{\prime },\nu ,\nu ^{\prime }\right) }\left| a,c;f^{\prime },\nu
\right\rangle \left\langle a^{\prime },c^{\prime };f^{\prime },\nu ^{\prime
}\right|
.
\end{eqnarray}%
It is apparent that the specific order of the measurement outcomes is not
important in the result, but that only the total number of outcomes of each
type matters, hence leading to a binomial distribution. We denote the total
number of $s_{j}=\shortrightarrow $ in the string of measurement outcomes as
$n$, and cluster together all results with the same $n$. Defining (for arbitrary
$p$ and $q$)%
\begin{equation}
W_{N}\left( n;p,q\right) =\frac{N!}{n!\left( N-n\right) !}p^{n}q^{N-n}
\end{equation}%
the probability of measuring $n$ of the $N$ probes at the horizontal
detector is%
\begin{equation}
\Pr\nolimits_{N}\left( n\right) =\sum\limits_{a,c,f,\mu }\rho _{\left(
a,c;f,\mu \right) \left( a,c;f,\mu \right) }^{A}W_{N}\left(
n;p_{aa1,B}^{\shortrightarrow },p_{aa1,B}^{\shortuparrow }\right)
\end{equation}%
and these measurements produce the target anyon reduced density matrix%
\begin{eqnarray}
&&\rho _{N}^{A}\left( n\right)  =\sum\limits_{\substack{ a,a^{\prime
},c,c^{\prime },f,\mu ,\mu ^{\prime } \\ e,\alpha ,\beta,
f^{\prime },\nu ,\nu ^{\prime } }}\frac{\rho _{\left(
a,c;f,\mu \right) \left( a^{\prime },c^{\prime };f,\mu ^{\prime }\right)
}^{A}}{\left( d_{f}d_{f^{\prime }} \right) ^{1/2} }
\frac{W_{N}\left( n;p_{aa^{\prime }e,B}^{\shortrightarrow },p_{aa^{\prime
}e,B}^{\shortuparrow }\right) }{\Pr\nolimits_{N}\left( n\right) }  \nonumber \\
&& \times \left[ \left( F_{a^{\prime }c^{\prime }}^{ac}\right) ^{-1}\right]
_{\left( f,\mu ,\mu ^{\prime }\right) \left( e,\alpha ,\beta \right) }\left[
F_{a^{\prime }c^{\prime }}^{ac}\right] _{\left( e,\alpha ,\beta \right)
\left( f^{\prime },\nu ,\nu ^{\prime }\right) }\left| a,c;f^{\prime },\nu
\right\rangle \left\langle a^{\prime },c^{\prime };f^{\prime },\nu ^{\prime
}\right|
.
 \label{eq:rho_n}
\end{eqnarray}

In Ref.~\cite{Bonderson07a}, we obtained the reduced density matrix that ignores
the measurement outcomes and describes the decoherence (rather than the
precise details of collapse) due to the probe measurements. We find this
density matrix by averaging over $n$, giving us the result in Eq.~(15c) of
\cite{Bonderson07a}, though for more general target and probe systems%
\begin{eqnarray}
&&\rho _{N}^{A} =\sum\limits_{n=0}^{N}\Pr\nolimits_{N}\left( n\right) \rho
_{N}^{A}\left( n\right)
\nonumber \\
&&=\sum\limits_{\substack{ a,a^{\prime },c,c^{\prime },f,\mu ,\mu ^{\prime }
\\ e,\alpha ,\beta ,f^{\prime },\nu ,\nu ^{\prime
} }}\frac{\rho _{\left(a,c;f,\mu \right) \left( a^{\prime },c^{\prime };
f,\mu ^{\prime }\right)}^{A}}{\left( d_{f}d_{f^{\prime }} \right) ^{1/2} }
\left( \left| t_{1}\right| ^{2}M_{eB}+\left|
r_{1}\right| ^{2}\right) ^{N}  \nonumber \\
&&\times \left[ \left( F_{a^{\prime }c^{\prime }}^{ac}\right) ^{-1}\right]
_{\left( f,\mu ,\mu ^{\prime }\right) \left( e,\alpha ,\beta \right) }\left[
F_{a^{\prime }c^{\prime }}^{ac}\right] _{\left( e,\alpha ,\beta \right)
\left( f^{\prime },\nu ,\nu ^{\prime }\right) }\left| a,c;f^{\prime },\nu
\right\rangle \left\langle a^{\prime },c^{\prime };f^{\prime },\nu ^{\prime
}\right|
\label{eq:average_density_matrix}
\end{eqnarray}%
where we used%
\begin{eqnarray}
\qquad \quad
\sum\limits_{n=0}^{N}W_{N}\left( n;p_{aa^{\prime }e,B}^{\shortrightarrow
},p_{aa^{\prime }e,B}^{\shortuparrow }\right) &=& \left( p_{aa^{\prime
}e,B}^{\shortrightarrow }+p_{aa^{\prime }e,B}^{\shortuparrow }\right) ^{N} \nonumber \\
&=& \left(\left| t_{1}\right| ^{2}M_{eB}+\left| r_{1}\right| ^{2}\right) ^{N}.
\end{eqnarray}

The interferometry experiment distinguishes anyonic charges in the target by
their values of $p_{aa1,B}^{s}$, which determine the possible measurement
distributions. Different anyonic charges with the same probability
distributions of probe outcomes are indistinguishable by such probes, and so
should be grouped together into distinguishable subsets. We define $\mathcal{%
C}_{\kappa }$ for $\kappa =1,\ldots ,m\leq \left| \mathcal{C}\right| $ to be
the maximal disjoint subsets of $\mathcal{C}$ such that $%
p_{aa1,B}^{\shortrightarrow }=p_{\kappa }$ for all $a\in \mathcal{C}_{\kappa }$,
i.e.%
\begin{equation}
\mathcal{C}_{\kappa } \equiv \left\{ a\in \mathcal{C}:p_{aa1,B}^{%
\shortrightarrow }=p_{\kappa }\right\}
.
\end{equation}%
Note that $p_{aa1,B}^{\shortrightarrow }=p_{a^{\prime }a^{\prime
}1,B}^{\shortrightarrow }$ (for two different charges $a$ and $a^{\prime }$) iff%
\begin{equation}
\text{Re}\left\{ t_{1}r_{1}^{\ast }r_{2}^{\ast }t_{2}^{\ast }e^{i\left(
\theta_{\text{I}}-\theta_{\text{II}}\right) }M_{aB}\right\} =\text{Re}\left\{
t_{1}r_{1}^{\ast }r_{2}^{\ast }t_{2}^{\ast }e^{i\left( \theta _{\text{I}}-\theta
_{\text{II}}\right) }M_{a^{\prime }B}\right\}
\end{equation}%
which occurs either when:

\noindent
(i) at least one of $t_{1}$, $t_{2}$, $r_{1}$, or $r_{2}$ is zero, or

\noindent
(ii) $\left| M_{aB}\right| \cos \left( \theta +\varphi _{a}\right) =\left|
M_{a^{\prime }B}\right| \cos \left( \theta +\varphi _{a^{\prime }}\right) $,
where $\theta =\arg \left( t_{1}r_{1}^{\ast }r_{2}^{\ast }t_{2}^{\ast
}e^{i\left( \theta _{\text{I}}-\theta _{\text{II}}\right) }\right) $ and $\varphi
_{a}=\arg \left( M_{aB}\right) $.

If condition (i) is satisfied, then there is no interference and $\mathcal{C}%
_{1}=\mathcal{C}$ (all target anyonic charges give the same probe
measurement distribution). Condition (ii) is generically\footnote{%
The term ``generic'' is used in this paper only in reference to the
collection of interferometer parameters $t_{j}$, $r_{j}$, $\theta _{\text{I}}$, and
$\theta _{\text{II}}$.} only satisfied when $M_{aB}=M_{a^{\prime }B}$, but may be
non-generically satisfied by setting $\theta =-\arg \left\{
M_{aB}-M_{a^{\prime }B}\right\} \pm \frac{\pi}{2} $. With this notation,
we may write%
\begin{eqnarray}
\qquad \qquad \quad
\Pr\nolimits_{N}\left( n\right) &=&\sum\limits_{\kappa
}\Pr\nolimits_{A}\left( \kappa \right) W_{N}\left( n;p_{\kappa },1-p_{\kappa
}\right) \\
\Pr\nolimits_{A}\left( \kappa \right) &=&\sum\limits_{a\in \mathcal{C}%
_{\kappa },c,f,\mu }\rho _{\left( a,c;f,\mu \right) \left( a,c;f,\mu
\right) }^{A}
.
\end{eqnarray}%
We emphasize that if the parameters $t_{j},r_{j}$ and $\theta $ in the
experiment are known and adjustable, then the measurements may be used to
gather information regarding the quantities $M_{ab}$, which, through its
relation to the topological $S$-matrix, may be used to properly identify the
anyon model that describes an unknown system \cite{Bonderson06b}.

In Sections~\ref{sec:large-N} and \ref{sec:p_and_q}, we will show that, as
$N\rightarrow \infty $, the
fraction $r=n/N$ of measurement outcomes will be found to go to $r=p_{\kappa
}$ with probability $\Pr\nolimits_{A}\left( \kappa \right) $, and the target
anyon density matrix will generically collapse onto the corresponding
``fixed states'' given by%
\begin{eqnarray}
&& \rho _{\kappa }^{A} =\sum\limits_{\substack{ a,a^{\prime },c,c^{\prime
},f,\mu ,\mu ^{\prime } \\ e,\alpha ,\beta ,f^{\prime
},\nu ,\nu ^{\prime } }} \frac{\rho _{\left(
a,c;f,\mu \right) \left( a^{\prime },c^{\prime };f,\mu ^{\prime }\right)
}^{A}}{\left( d_{f}d_{f^{\prime }} \right) ^{1/2} }
\Delta _{aa^{\prime}e,B}\left( p_{\kappa }\right)  \notag \\
&&\times \left[ \left( F_{a^{\prime }c^{\prime }}^{ac}\right) ^{-1}\right]
_{\left( f,\mu ,\mu ^{\prime }\right) \left( e,\alpha ,\beta \right) }\left[
F_{a^{\prime }c^{\prime }}^{ac}\right] _{\left( e,\alpha ,\beta \right)
\left( f^{\prime },\nu ,\nu ^{\prime }\right) }\left| a,c;f^{\prime },\nu
\right\rangle \left\langle a^{\prime },c^{\prime };f^{\prime },\nu ^{\prime
}\right|
\end{eqnarray}%
where%
\begin{equation}
\Delta _{aa^{\prime }e,B}\left( p_{\kappa }\right) =\left\{
\begin{array}{cc}
\frac{1}{\Pr\nolimits_{A}\left( \kappa \right) } & \text{if }p_{aa^{\prime
}e,B}^{\shortrightarrow }=1-p_{aa^{\prime }e,B}^{\shortuparrow }=p_{\kappa }\text{ and
}a,a^{\prime }\in \mathcal{C}_{\kappa } \\
0 & \text{otherwise}%
\end{array}%
\right. .
\end{equation}%
Fixed state density matrices are left unchanged by probe measurements. We
also emphasize that the condition: $p_{aa^{\prime }e,B}^{\shortrightarrow
}=1-p_{aa^{\prime }e,B}^{\shortuparrow }=p_{\kappa }$ is equivalent to $M_{eB}=1$
(noting that $M_{eB}=1$ implies $M_{aB}=M_{a^{\prime}B}$ and
$a,a^{\prime }\in \mathcal{C}_{\kappa }$). This gives the interpretation that the
probes have the effect of collapsing superpositions of anyonic charges
$a$ and $a^{\prime}$ in the target that they can distinguish by monodromy
($M_{aB} \neq M_{a^{\prime}B}$), \textit{and} decohering all anyonic charge
entanglement between the target anyon $A$ and the anyons $C$ outside the central
interferometry region that the probes can ``see'' by monodromy, i.e. removing the
components of the density matrix corresponding to $e$-channels with $M_{eB}\neq 1$.
Non-generically, it is also possible to collapse onto
``rogue states,'' for which the diagonal density matrix elements are all
fixed and some of the
off-diagonal elements have fixed magnitude, but phases that change depending on
the measurement outcome (i.e. are ``quasi-fixed''). Because rogue states occur
only for specific, exactly precise experimental parameters, they will not
actually survive measurement in realistic experiments. We note that if $M_{eB}=1$ only for
$e=1$, then the probe distinguishes all charges, and the fixed states are given
by
\begin{equation}
\rho _{\kappa_{a} }^{A} = \sum\limits_{c} \frac{\Pr\nolimits _{A} \left(c | a
\right)}{ d_{a}d_{c} } \,\, \mathbb{I}_{ac}
=
\sum\limits_{c,f^{\prime},\nu} \frac{\Pr\nolimits _{A} \left(c | a
\right)}{ d_{a}d_{c} } \,\,\left| a,c;f^{\prime},\nu
\right\rangle \left\langle a,c;f^{\prime},\nu \right|
\end{equation}
where
\begin{equation}
\Pr\nolimits _{A} \left(c | a \right) = \frac{ \sum\limits_{f,\mu}\rho _{\left(
a,c;f,\mu \right) \left( a,c;f,\mu \right)}^{A} }{\sum\limits_{c,f,\mu}\rho _{\left(
a,c;f,\mu \right) \left( a,c;f,\mu \right)}^{A}}
,
\end{equation}%
for which the target anyon $A$ has definite charge and no entanglement with $C$.
We give examples of fixed state density matrices for several significant
anyon models in Section~\ref{sec:Examples}.

In principle, one may also consider the ``many-to-many'' experiment
described in Ref.~\cite{Overbosch01}, where the target anyon system is replaced
with a fresh one (described by the same initial density matrix) after each
probe measurement. For this type of experiment, the result for each probe is
described by the single probe outcome probability,
Eq.~(\ref{eq:single_probability}):%
\begin{equation}
\Pr \left( s\right) =\sum\limits_{a,c,f,\mu }\rho _{\left( a,c;f,\mu \right)
\left( a,c;f,\mu \right) }^{A}p_{aa1,B}^{s}.
\end{equation}%
Thus, for $N$ such probe measurements, the number $n$ of $s=\shortrightarrow$ measurement outcomes will have the binomial distribution: $%
W_{N}\left( n;\Pr \left( \shortrightarrow \right) ,\Pr \left( \shortuparrow \right)
\right) $. 

%% file: 03d-W.tex
\subsubsection{Large N}
  \label{sec:large-N}

We would like to analyze the large $N$ behavior of the measurements.
This is essentially determined by
$W_{N}\left( n;p_{\kappa },1-p_{\kappa }\right)$ and
$\frac{W_{N}\left( n;p_{aa^{\prime }e,B}^{\shortrightarrow },p_{aa^{\prime
}e,B}^{\shortuparrow }\right) }{\Pr\nolimits_{N}\left( n\right) }$, so we now
consider these in detail. Of course, $W_{N}\left( n;p_{\kappa },1-p_{\kappa
}\right) $ is just a familiar binomial distribution. Changing variables to
the fraction $r=n/N$ of total probe measurement outcomes in the horizontal
detector, the distribution in $r$ is given by%
\begin{equation}
w_{N}\left( r;p_{\kappa },1-p_{\kappa }\right) =W_{N}\left( rN;p_{\kappa
},1-p_{\kappa }\right) N
\end{equation}%
and has mean and standard deviation%
\begin{eqnarray}
\qquad \qquad \qquad \qquad
\overline{r} &=&p_{\kappa } \\
\Delta r &=&\sigma _{\kappa }\equiv \sqrt{p_{\kappa }\left( 1-p_{\kappa
}\right) /N}.
\end{eqnarray}%
Taking $N$ large and using Stirling's formula, this may be approximated by a
Gaussian distribution%
\begin{equation}
w_{N}\left( r;p_{\kappa },1-p_{\kappa }\right) \simeq \frac{1}{\sqrt{2\pi
\sigma _{\kappa }^{2}}}e^{-\frac{\left( r-p_{\kappa }\right) ^{2}}{2\sigma
_{\kappa }^{2}}}.
\end{equation}%
Taking the limit $N\rightarrow \infty $ gives%
\begin{equation}
\lim_{N\rightarrow \infty }w_{N}\left( r;p_{\kappa },1-p_{\kappa }\right)
=\delta \left( r-p_{\kappa }\right)
\end{equation}%
(defined such that $\int_{0}^{1}\delta \left( r-p\right) dr=1$, when $p=0$
or $1$), so the resulting probability distribution for the measurement
outcomes is%
\begin{equation}
\Pr \left( r\right) =\lim_{N\rightarrow \infty }\Pr\nolimits_{N}\left(
r\right) =\sum\limits_{\kappa }\Pr\nolimits_{A}\left( \kappa \right) \delta
\left( r-p_{\kappa }\right)
\end{equation}%
Thus, as $N\rightarrow \infty $, we will find the fraction of measurement
outcomes $r\rightarrow p_{\kappa }$ with probability $\Pr\nolimits_{A}\left(
\kappa \right) $.

Though the probability of obtaining the outcome $r$ which is away from the
closest $p_{\kappa }$ vanishes as%
\begin{equation}
w_{N}\left( r;p_{\kappa },1-p_{\kappa }\right) \sim \sqrt{N}\left( \frac{%
p_{\kappa }^{r}\left( 1-p_{\kappa }\right) ^{1-r}}{r^{r}\left( 1-r\right)
^{1-r}}\right) ^{N}
\end{equation}%
for large $N$, the resulting density matrix should still be well defined for
all $r$ (at least for large, but finite $N$). In particular, we will use the
positivity property of density matrices, in the
form of the Cauchy-Schwarz type inequality $\rho _{\mu \mu }\rho _{\nu \nu
}\geq \left| \rho _{\mu \nu }\right| ^{2}$, to evince their large $N$
behavior in terms of conditions on $p_{aa^{\prime }e,B}^{s}$. From the quantity
\begin{eqnarray}
\Delta _{N;aa^{\prime }e,B}\left( r\right) &\equiv &\frac{W_{N}\left(
rN;p_{aa^{\prime }e,B}^{\shortrightarrow },p_{aa^{\prime }e,B}^{\shortuparrow }\right)
}{\Pr\nolimits_{N}\left( rN\right) }
\nonumber \\
&=&\left\{ \sum\limits_{\kappa ^{\prime }}\Pr\nolimits_{A}\left( \kappa
^{\prime }\right) \left[ \left( \frac{p_{\kappa ^{\prime }}}{p_{aa^{\prime
}e,B}^{\shortrightarrow }}\right) ^{r}\left( \frac{1-p_{\kappa ^{\prime }}}{%
p_{aa^{\prime }e,B}^{\shortuparrow }}\right) ^{1-r}\right] ^{N}\right\} ^{-1}
\label{eq:Delta-r}
\end{eqnarray}%
we can see that as $N\rightarrow \infty $, the $e=1$ terms (those that
determine the ``diagonal'' elements) behave as:

\noindent
(i) $\Delta _{N;aa1,B}\left( r\right) \rightarrow \frac{1}{%
\Pr\nolimits_{A}\left( \kappa _{1}\right) }$ for $a\in \mathcal{C}_{\kappa
_{1}}$, if $\Pr\nolimits_{A}\left( \kappa _{1}\right) \neq 0$ and $p_{\kappa
_{1}}^{r}\left( 1-p_{\kappa _{1}}\right) ^{1-r}>p_{\kappa }^{r}\left(
1-p_{\kappa }\right) ^{1-r}$ for all $\kappa \neq \kappa _{1}$,

\noindent
(ii) $\Delta _{N;aa1,B}\left( r\right) \rightarrow \frac{1}{%
\Pr\nolimits_{A}\left( \kappa _{1}\right) +\Pr\nolimits_{A}\left( \kappa
_{2}\right) }$ for $a\in \mathcal{C}_{\kappa _{1}}\cup \mathcal{C}_{\kappa
_{2}}$, if $\Pr\nolimits_{A}\left( \kappa _{1}\right)
+\Pr\nolimits_{A}\left( \kappa _{2}\right) \neq 0$ and $p_{\kappa
_{1}}^{r}\left( 1-p_{\kappa _{1}}\right) ^{1-r}=p_{\kappa _{2}}^{r}\left(
1-p_{\kappa _{2}}\right) ^{1-r}>p_{\kappa }^{r}\left( 1-p_{\kappa }\right)
^{1-r}$ for all $\kappa \neq \kappa _{1},\kappa _{2}$, or

\noindent
(iii) $\Delta _{N;aa1,B}\left( r\right) \rightarrow 0$ for $a\in \mathcal{C}%
_{\kappa _{1}}$, if there is some $\kappa $ with $\Pr\nolimits_{A}\left(
\kappa \right) \neq 0$ and $p_{\kappa }^{r}\left( 1-p_{\kappa }\right)
^{1-r}>p_{\kappa _{1}}^{r}\left( 1-p_{\kappa _{1}}\right) ^{1-r}$.

If $a\in \mathcal{C}_{\kappa _{1}}$, where $p_{\kappa _{1}}^{r}\left(
1-p_{\kappa _{1}}\right) ^{1-r}>p_{\kappa }^{r}\left( 1-p_{\kappa }\right)
^{1-r}$ for all $\kappa \neq \kappa _{1}$, but $\Pr\nolimits_{A}\left(
\kappa _{1}\right) =0$, then $\Delta _{N;aa1,B}\left( r\right) \rightarrow
\infty $. However, $\Pr\nolimits_{A}\left( \kappa _{1}\right) =0$ also
implies that the density matrix coefficients involving $a$ are strictly
zero, so we need not worry about this case.

We note that for each $\kappa $%
, the variable $r$ has a closed interval $I_{\kappa }$, containing
$p_{\kappa }$ in its interior, such that $p_{\kappa }^{r}\left( 1-p_{\kappa
}\right) ^{1-r}\geq p_{\kappa ^{\prime }}^{r}\left( 1-p_{\kappa ^{\prime
}}\right) ^{1-r}$ for all $\kappa ^{\prime }\neq \kappa $ (i.e. $I_{\kappa }$
satisfies (i) in its interior and (ii) at its endpoints). We say that $r$ is
congruous with $\mathcal{C}_{\kappa }$ in this interval $I_{\kappa }$ (and
congruous to two different $\mathcal{C}_{\kappa }$ at the intersecting
endpoints of such intervals).

For arbitrary (in particular, the ``off-diagonal'') terms, the positivity
condition combined with Eq.~(\ref{eq:Delta-r}) as $N\rightarrow \infty $ tells
us that we
must have $\Delta _{N;aa^{\prime }e,B}\left( r\right) \rightarrow 0$, except
when $r$ is congruous with both $a$ and $a^{\prime }$, in which case $\left|
\Delta _{N;aa^{\prime }e,B}\left( r\right) \right| \leq \Delta
_{N;aa1,B}\left( r\right) $, and $\Delta _{N;aa^{\prime }e,B}\left( r\right)
\rightarrow \infty $ should not be allowed (except when the density matrix
elements involving $a$ or $a^{\prime }$ are strictly zero, making it
irrelevant). From this we find that either:

\noindent
(a) there is some $\kappa $ (possibly even with $a$ and/or $a^{\prime }$ in $%
\mathcal{C}_{\kappa }$) with $\Pr\nolimits_{A}\left( \kappa \right) \neq 0$
and $p_{\kappa }^{r}\left( 1-p_{\kappa }\right) ^{1-r}>\left| \
p_{aa^{\prime }e,B}^{\shortrightarrow }\right| ^{r}\left| p_{aa^{\prime
}e,B}^{\shortuparrow }\right| ^{1-r}$, in which case $\Delta _{N;aa^{\prime
}e,B}\left( r\right) \rightarrow 0$, or

\noindent
(b) $\left| \ p_{aa^{\prime }e,B}^{\shortrightarrow }\right| ^{r}\left|
p_{aa^{\prime }e,B}^{\shortuparrow }\right| ^{1-r}=\left( p_{aa1,B}^{\shortrightarrow
}\right) ^{r}\left( p_{aa1,B}^{\shortuparrow }\right) ^{1-r}=\left( p_{a^{\prime
}a^{\prime }1,B}^{\shortrightarrow }\right) ^{r}\left( p_{a^{\prime }a^{\prime
}1,B}^{\shortuparrow }\right) ^{1-r}$ with $r$ congruous with both $a$ and $%
a^{\prime }$, in which case $\left| \Delta _{N;aa^{\prime }e,B}\left(
r\right) \right| \rightarrow \Delta _{N;aa1,B}\left( r\right) $.

Case (b) deserves some further inspection. First, we note that we have%
\begin{equation}
\left| \ p_{aa^{\prime }e,B}^{\shortrightarrow }\right| ^{r}\left| p_{aa^{\prime
}e,B}^{\shortuparrow }\right| ^{1-r}\leq p_{\kappa }^{r}\left( 1-p_{\kappa
}\right) ^{1-r}
\end{equation}%
on the entire interval $I_{\kappa }$ congruous with $a\in \mathcal{C}%
_{\kappa }$. If there is some point $r_{\ast }$ in the interior of $%
I_{\kappa }$ for which%
\begin{equation}
\left| \ p_{aa^{\prime }e,B}^{\shortrightarrow }\right| ^{r_{\ast }}\left|
p_{aa^{\prime }e,B}^{\shortuparrow }\right| ^{1-r_{\ast }}=p_{\kappa }^{r_{\ast
}}\left( 1-p_{\kappa }\right) ^{1-r_{\ast }},
\end{equation}%
then in order not to violate the inequality when $r$ is increased or
decreased from $r_{\ast }$, we must have%
\begin{equation}
\frac{\left| \ p_{aa^{\prime }e,B}^{\shortrightarrow }\right| }{\left|
p_{aa^{\prime }e,B}^{\shortuparrow }\right| }=\frac{p_{\kappa }}{1-p_{\kappa }}.
\end{equation}%
It follows that%
\begin{equation}
\left| \ p_{aa^{\prime }e,B}^{\shortrightarrow }\right| ^{r}\left| p_{aa^{\prime
}e,B}^{\shortuparrow }\right| ^{1-r}=p_{\kappa }^{r}\left( 1-p_{\kappa }\right)
^{1-r}
\end{equation}%
on the entire interval $I_{\kappa }$, and, more significantly, that%
\begin{equation}
\left| \ p_{aa^{\prime }e,B}^{\shortrightarrow }\right| =1-\left| p_{aa^{\prime
}e,B}^{\shortuparrow }\right| =p_{\kappa }.
\end{equation}%
The same argument holds with respect to $a^{\prime }$ instead of $a$,
giving the additional condition $a,a^{\prime }\in \mathcal{C}_{\kappa }$.
Hence, even at exponentially suppressed $r$, superpositions of anyonic
charges from different $\mathcal{C}_{\kappa }$ do not survive measurement.

Pushing this a bit further, we note that for $r\in \left[ 0,1\right] $ and
fixed $p\in \left[ 0,1\right] $%
\begin{equation}
r^{r}\left( 1-r\right) ^{1-r}\geq p^{r}\left( 1-p\right) ^{1-r}
\end{equation}%
with equality at $r=p$. The positivity condition gave us (rewriting (a) and
(b))%
\begin{equation}
\max_{\kappa }\left\{ p_{\kappa }^{r}\left( 1-p_{\kappa }\right)
^{1-r}\right\} \geq \left| \ p_{aa^{\prime }e,B}^{\shortrightarrow }\right|
^{r}\left| p_{aa^{\prime }e,B}^{\shortuparrow }\right| ^{1-r}
\end{equation}%
with equality for $r\in I_{\kappa }$ if $a,a^{\prime }\in \mathcal{C}%
_{\kappa }$ and $\left| \ p_{aa^{\prime }e,B}^{\shortrightarrow }\right|
=1-\left| p_{aa^{\prime }e,B}^{\shortuparrow }\right| =p_{\kappa }$. Combining
these, we have%
\begin{equation}
r^{r}\left( 1-r\right) ^{1-r}\geq \left| \ p_{aa^{\prime }e,B}^{\shortrightarrow
}\right| ^{r}\left| p_{aa^{\prime }e,B}^{\shortuparrow }\right| ^{1-r}
\label{eq:pq-inequality}
\end{equation}%
for all $r$, with equality occurring at $r=\left| \ p_{aa^{\prime
}e,B}^{\shortrightarrow }\right| $ only when $p_{a}=p_{a^{\prime }}=\left| \
p_{aa^{\prime }e,B}^{\shortrightarrow }\right| =1-\left| p_{aa^{\prime
}e,B}^{\shortuparrow }\right| $. If $\left| \ p_{aa^{\prime }e,B}^{\shortrightarrow
}\right| +\left| p_{aa^{\prime }e,B}^{\shortuparrow }\right| >1$, then there is
some $r$ (e.g. $r=\left| \ p_{aa^{\prime }e,B}^{\shortrightarrow }\right| $) for
which $r^{r}\left( 1-r\right) ^{1-r}<\left| \ p_{aa^{\prime
}e,B}^{\shortrightarrow }\right| ^{r}\left| p_{aa^{\prime }e,B}^{\shortuparrow
}\right| ^{1-r}$, violating Eq.~(\ref{eq:pq-inequality}). If $\left| \
p_{aa^{\prime
}e,B}^{\shortrightarrow }\right| +\left| p_{aa^{\prime }e,B}^{\shortuparrow }\right|
=1 $, then $r^{r}\left( 1-r\right) ^{1-r}=\left| \ p_{aa^{\prime
}e,B}^{\shortrightarrow }\right| ^{r}\left| p_{aa^{\prime }e,B}^{\shortuparrow
}\right| ^{1-r}$ at $r=\left| \ p_{aa^{\prime }e,B}^{\shortrightarrow }\right| $.
Hence, we have%
\begin{equation}
\left| \ p_{aa^{\prime }e,B}^{\shortrightarrow }\right| +\left| p_{aa^{\prime
}e,B}^{\shortuparrow }\right| \leq 1
\end{equation}%
with equality only if $p_{a}=p_{a^{\prime }}=\left| \ p_{aa^{\prime
}e,B}^{\shortrightarrow }\right| =1-\left| p_{aa^{\prime }e,B}^{\shortuparrow }\right|
$. We believe one should be able to to show that this condition on $p_{aa^{\prime
}e,b}^{s}$\ follows directly from the properties of anyon models, in which case
these arguments could be made in the opposite direction, i.e. that
positivity of the density matrix being preserved by these probe measurements
follows from properties of anyon models; however, we have been unable to
succeed in doing so.

{}For $p_{aa^{\prime }e,B}^{\shortrightarrow }=p_{\kappa }e^{i\alpha _{aa^{\prime}e,B}}$
and $p_{aa^{\prime }e,B}^{\shortuparrow }=\left( 1-p_{k}\right) e^{i\beta
_{aa^{\prime }e,B}}$, we see that if $\alpha _{aa^{\prime
}e,B}=\beta _{aa^{\prime }e,B}$, then
\begin{equation}
\left| t_{1}\right| ^{2}M_{eB}+\left| r_{1}\right| ^{2}=
p_{aa^{\prime}e,B}^{\shortrightarrow }+p_{aa^{\prime }e,B}^{\shortuparrow }=
e^{i\alpha _{aa^{\prime}e,B}}
\end{equation}%
implies that either: (a) $r_{1}=0$ and $M_{eB}=e^{i\alpha
_{aa^{\prime }e,B}}$, or (b) $M_{eB}=1$ and $\alpha _{aa^{\prime }e,B}=\beta
_{aa^{\prime }e,B}=0$.

One might also find it instructive to consider a large $N$ expansion (using
Stirling's formula) around $p_{\kappa }$ to get%
\begin{eqnarray}
w_{N}\left( r;p_{aa^{\prime }e,B}^{\shortrightarrow },p_{aa^{\prime
}e,B}^{\shortuparrow }\right) &\simeq &w_{N}\left( r;p_{\kappa },1-p_{\kappa
}\right) e^{-G_{N}\left( r;p_{aa^{\prime }e,B}^{\shortrightarrow },p_{aa^{\prime
}e,B}^{\shortuparrow }\right) } \\
\Delta _{N;aa^{\prime }e,B}\left( r\right) &\simeq &\frac{1}{%
\Pr\nolimits_{A}\left( \kappa \right) }e^{-G_{N}\left( r;p_{aa^{\prime
}e,B}^{\shortrightarrow },p_{aa^{\prime }e,B}^{\shortuparrow }\right) } \\
G_{N}\left( r;p,q\right) &\approx &N\left[ p_{\kappa }\ln \left( \frac{%
p_{\kappa }}{p}\right) +\left( 1-p_{\kappa }\right) \ln \left( \frac{%
1-p_{\kappa }}{q}\right) \right]  \notag \\
&&+N\left( r-p_{\kappa }\right) \ln \left( \frac{p_{\kappa }}{p}\frac{q}{%
\left( 1-p_{\kappa }\right) }\right)
.
\end{eqnarray}%
Clearly, $e^{-G_{N}\left( r;p,q\right) }$ gives exponential suppression in $%
N $, unless $p=p_{\kappa }e^{i\alpha }$ and $q=\left( 1-p_{k}\right)
e^{i\beta }$, in which case
\begin{equation}
e^{-G_{N}\left( r;p,q\right) }=e^{i\left[ \alpha p_{\kappa }+\beta \left(
1-p_{\kappa }\right) \right] N}e^{i\left( \alpha -\beta \right) N\left(
r-p_{\kappa }\right) }
\end{equation}%
(which is equal to $1$, when $\alpha =\beta =0$). We also note that
integrating the quantity
$w_{N}\left( r;p_{aa^{\prime }e,B}^{\shortrightarrow },p_{aa^{\prime}e,B}^{\shortuparrow }\right) $
over $r$ vanishes exponentially in $N$, unless
$\alpha_{aa^{\prime }e,B}=\beta _{aa^{\prime }e,B}=0$, which is why
such quasi-fixed terms do not appear in Eq.~(\ref{eq:average_density_matrix}),
the density matrix obtained by ignoring measurement outcomes (except in the case
when $r_{1}=0$).

To summarize, we found that, for large $N$, the quantity $\Delta
_{N;aa^{\prime }e,B}\left( r\right) $ vanishes exponentially unless $r$ is
congruous with $a,a^{\prime }\in \mathcal{C}_{\kappa }$ and $\left| \
p_{aa^{\prime }e,B}^{\shortrightarrow }\right| =1-\left| p_{aa^{\prime
}e,B}^{\shortuparrow }\right| =p_{\kappa }$. This means a measurement outcome
fraction $r$ exponentially collapses the density matrix onto one that has
support only in $\mathcal{C}_{\kappa }$, and consequently will drive $r$
toward $p_{\kappa }$. The resulting target anyon reduced density matrix%
\begin{eqnarray}
&&
\rho ^{A}\left( r\right)  =\sum\limits_{\substack{ a,a^{\prime
},c,c^{\prime },f,\mu ,\mu ^{\prime } \\ e,\alpha ,\beta ,
f^{\prime },\nu ,\nu ^{\prime } }} \frac{\rho _{\left( a,c;f,\mu
\right) \left( a^{\prime },c^{\prime };f,\mu ^{\prime }\right) }^{A}}
{\left( d_{f}d_{f^{\prime }} \right) ^{1/2}}
\Delta_{aa^{\prime }e,B}\left( r\right)  \notag \\
&&
\times \left[ \left( F_{a^{\prime }c^{\prime }}^{ac}\right) ^{-1}\right]
_{\left( f,\mu ,\mu ^{\prime }\right) \left( e,\alpha ,\beta \right) }\left[
F_{a^{\prime }c^{\prime }}^{ac}\right] _{\left( e,\alpha ,\beta \right)
\left( f^{\prime },\nu ,\nu ^{\prime }\right) }\left| a,c;f^{\prime },\nu
\right\rangle \left\langle a^{\prime },c^{\prime };f^{\prime },\nu ^{\prime
}\right|,
\end{eqnarray}
where
\begin{equation}
\Delta _{aa^{\prime }e,B}\left( r\right)  =\lim_{N\rightarrow \infty
}\Delta _{N;aa^{\prime }e,B}\left( r\right),
\end{equation}%
is found with the probability distribution%
\begin{equation}
\Pr \left( r\right) =\sum\limits_{\kappa }\Pr\nolimits_{A}\left( \kappa
\right) \delta \left( r-p_{\kappa }\right) .
\end{equation}%
The resulting density matrices are of two forms:

\noindent
(1) \textit{fixed states}, for which all non-zero elements of the density matrix
correspond to $p_{aa^{\prime }e,B}^{\shortrightarrow }=1-p_{aa^{\prime
}e,B}^{\shortuparrow }=p_{\kappa }$, and

\noindent
(2) \textit{rogues states }(or \textit{quasi-fixed states}), for which all
elements of the density matrix correspond to $\left| \ p_{aa^{\prime
}e,B}^{\shortrightarrow }\right| =1-\left| p_{aa^{\prime }e,B}^{\shortuparrow }\right|
=p_{\kappa }$, but for some of the ``off-diagonal'' elements ($e\neq 1$)
with $p_{aa^{\prime }e,B}^{\shortrightarrow }=p_{\kappa }e^{i\alpha _{aa^{\prime
}e,B}}$ and $p_{aa^{\prime }e,B}^{\shortuparrow }=\left( 1-p_{k}\right) e^{i\beta
_{aa^{\prime }e,B}}$, where $\alpha _{aa^{\prime }e,B}$ and $\beta
_{aa^{\prime }e,B}$ are non-zero (unless $r_1 = 0$).

Fixed states have the property that probe measurements leave their density
matrix invariant. Rogue states have the property that probe measurements
leave their ``diagonal'' elements and possibly some of their ``off-diagonal''
elements invariant, while some of their ``off-diagonal'' elements are
unchanged in magnitude, but have a changing phase. We will see in
Section~\ref{sec:p_and_q} that satisfying the conditions for rogue states
requires non-generic experimental parameters. 

%% file: 03e-psandqs.tex
\subsubsection{Minding our p's}
  \label{sec:p_and_q}

In Section~\ref{sec:large-N}, we have shown that performing many probe
measurements collapses the target density matrix onto its elements which
correspond to $p_{aa^{\prime }e,B}^{s}$ satisfying%
\begin{equation}
\left| p_{aa^{\prime }e,B}^{\shortrightarrow }\right| =1-\left| p_{aa^{\prime
}e,B}^{_{\shortuparrow }}\right| =p_{\kappa }
\end{equation}%
for $a,a^{\prime }\in \mathcal{C}_{\kappa }$, so we would like to determine
when this condition is satisfied.

For completeness, we first list the results for the trivial cases where there is no
actual interferometry (for which $\mathcal{C}_{1}=\mathcal{C}$):

\noindent
(i) When $t_{1}=0$, we have $p_{aa^{\prime }e,B}^{\shortrightarrow }=\left|
t_{2}\right| ^{2}$ and $p_{aa^{\prime }e,B}^{_{\shortuparrow }}=\left|
r_{2}\right| ^{2}$, so all elements are fixed.

\noindent
(ii) When $r_{1}=0$, we have $p_{aa^{\prime }e,B}^{\shortrightarrow }=\left|
r_{2}\right| ^{2}M_{eB}$ and $p_{aa^{\prime }e,B}^{_{\shortuparrow }}=\left|
t_{2}\right| ^{2}M_{eB}$, so elements with $M_{eB}=e^{i\varphi _{eB}}$\ ($%
\varphi _{eB}\neq 0$) are quasi-fixed, and those with $M_{eB}=1$ are fixed.

\noindent
(iii) When $t_{2}=0$ (and $t_{1}\neq 0$), we have $p_{aa^{\prime
}e,B}^{\shortrightarrow }=\left| t_{1}\right| ^{2}M_{eB}$ and $p_{aa^{\prime
}e,B}^{_{\shortuparrow }}=\left| r_{1}\right| ^{2}$, so elements with $%
M_{eB}=e^{i\varphi _{eB}}$\ ($\varphi _{eB}\neq 0$) are quasi-fixed, and
those with $M_{eB}=1$ are fixed.

\noindent
(iv) When $r_{2}=0$ (and $t_{1}\neq 0$), we have $p_{aa^{\prime
}e,B}^{\shortrightarrow }=\left| r_{1}\right| ^{2}$ and $p_{aa^{\prime
}e,B}^{_{\shortuparrow }}=\left| t_{1}\right| ^{2}M_{eB}$, so elements with $%
M_{eB}=e^{i\varphi _{eB}}$\ ($\varphi _{eB}\neq 0$) are quasi-fixed, and
those with $M_{eB}=1$ are fixed.

{}From here on, we assume that $\left| t_{1}r_{1}t_{2}r_{2}\right| \neq 0$
(unless explicitly stated otherwise). We begin by considering the more
stringent condition necessary for fixed elements. Using
$p_{aa^{\prime }e,b}^{\shortrightarrow }+p_{aa^{\prime }e,b}^{_{\shortuparrow }}=\left|
t_{1}\right| ^{2}M_{eB}+\left| r_{1}\right| ^{2}$, and
Eq.~(\ref{eq:fused_monodromy}), we have:

\noindent
(v) (When $t_{1}\neq 0$) An element is fixed, with $p_{aa^{\prime
}e,b}^{\shortrightarrow }=1-p_{aa^{\prime }e,b}^{_{\shortuparrow }}=p_{\kappa }$, iff $%
M_{eB}=1$, and this implies $M_{aB}=M_{a^{\prime }B}$ and $a,a^{\prime }\in
\mathcal{C}_{\kappa }$.

Thus, even without initially requiring $a,a^{\prime }\in \mathcal{C}%
_{\kappa }$ (from positivity), we find that it is a necessary condition for
fixed elements.

Now, we examine the conditions that give quasi-fixed elements. Such terms
have $p_{aa^{\prime }e,B}^{\shortrightarrow }=p_{\kappa }e^{i\alpha _{aa^{\prime
}e,B}}$ and $p_{aa^{\prime }e,B}^{\shortuparrow }=\left( 1-p_{\kappa }\right)
e^{i\beta _{aa^{\prime }e,B}}$, with $\alpha _{aa^{\prime }e,B}\neq \beta
_{aa^{\prime }e,B}$ and $a,a^{\prime }\in \mathcal{C}_{\kappa }$. (Recall
that if $\alpha _{aa^{\prime }e,B}=\beta _{aa^{\prime }e,B}$ and $r_{1}\neq
0 $, then $\alpha _{aa^{\prime }e,B}=\beta _{aa^{\prime }e,B}=0$.) Examining
these conditions for $M_{aB}=M_{a^{\prime }B}$, we find%
\begin{eqnarray}
\qquad
0 &=&\left| t_{2}\right| ^{2}\left( \left| p_{aa^{\prime }e,B}^{\shortrightarrow
}\right| ^{2}-p_{\kappa }^{2}\right) +\left| r_{2}\right| ^{2}\left( \left|
p_{aa^{\prime }e,B}^{_{\shortuparrow }}\right| ^{2}-\left( 1-p_{\kappa }\right)
^{2}\right)  \notag \\
&=&\left| t_{1}\right| ^{4}\left| t_{2}\right| ^{2}\left| r_{2}\right|
^{2}\left( \left| M_{eB}\right| ^{2}-1\right) \notag \\
&& +2\left| t_{1}\right|
^{2}\left| r_{1}\right| ^{2}\left| t_{2}\right| ^{2}\left| r_{2}\right|
^{2}\left( \text{Re}\left\{ M_{eB}\right\} -1\right)
\end{eqnarray}%
which requires $M_{eB}=1$ and, hence, gives us:

\noindent
(vi) (When $\left| t_{1}r_{1}t_{2}r_{2}\right| \neq 0$) There are no
quasi-fixed elements for $p_{aa^{\prime }e,B}^{s}$ with $M_{aB}=M_{a^{\prime
}B}$, only fixed ones. (In particular, this applies to $a=a^{\prime }$.)

For $M_{aB}\neq M_{a^{\prime }B}$, we can only have $a,a^{\prime }\in
\mathcal{C}_{\kappa }$ (i.e. $p_{aa1,B}^{s}=p_{a^{\prime }a^{\prime
}1,B}^{s} $) when the experimental parameters are tuned to $\theta =-\arg
\left\{ M_{aB}-M_{a^{\prime }B}\right\} \pm \frac{\pi}{2} $, so quasi-fixed
elements only occur non-generically. From the
conditions on $p_{aa^{\prime }e,B}^{s}$, at these values of $\theta $, we
find%
\begin{eqnarray}
\qquad
0 &=&\left| p_{aa^{\prime }e,B}^{\shortrightarrow }\right| ^{2}-p_{\kappa
}^{2}-\left| p_{aa^{\prime }e,B}^{_{\shortuparrow }}\right| ^{2}+\left(
1-p_{\kappa }\right) ^{2}  \notag \\
&=&\left| t_{1}\right| ^{4}\left( \left| t_{2}\right| ^{2}-\left|
r_{2}\right| ^{2}\right) \left( 1-\left| M_{eB}\right| ^{2}\right) \notag \\
&& -2\left|
t_{1}\right| ^{2}\left( 1-\text{Re}\left\{ M_{eB}\right\} \right) 2\left|
t_{1}r_{1}t_{2}r_{2}\right| \text{Re}\left\{ e^{i\theta }M_{aB}\right\}
\notag \\
&&+2\left| t_{1}\right| ^{2}\text{Im}\left\{ M_{eB}\right\} \left|
t_{1}r_{1}t_{2}r_{2}\right| \text{Im}\left\{ e^{i\theta }M_{aB}+e^{-i\theta
}M_{a^{\prime }B}^{\ast }\right\}
\end{eqnarray}%
and%
\begin{eqnarray}
0 &=&\left| t_{2}\right| ^{2}\left( \left| p_{aa^{\prime }e,B}^{\shortrightarrow
}\right| ^{2}-p_{\kappa }^{2}\right) +\left| r_{2}\right| ^{2}\left( \left|
p_{aa^{\prime }e,B}^{_{\shortuparrow }}\right| ^{2}-\left( 1-p_{\kappa }\right)
^{2}\right)  \notag \\
&=&\left| t_{1}\right| ^{4}\left| t_{2}\right| ^{2}\left| r_{2}\right|
^{2}\left( \left| M_{eB}\right| ^{2}-1\right) +2\left| t_{1}\right|
^{2}\left| r_{1}\right| ^{2}\left| t_{2}\right| ^{2}\left| r_{2}\right|
^{2}\left( \text{Re}\left\{ M_{eB}\right\} -1\right)  \notag \\
&&+\left( \left| t_{1}r_{1}t_{2}r_{2}\right| \text{Im}\left\{ e^{i\theta
}M_{aB}+e^{-i\theta }M_{a^{\prime }B}^{\ast }\right\} \right) ^{2}
\end{eqnarray}%
which may be rewritten to give:

\noindent
(vii) Quasi-fixed elements with $p_{aa^{\prime }e,B}^{s}$ only occur
non-generically, and the conditions (when $\left|
t_{1}r_{1}t_{2}r_{2}\right| \neq 0$) that must be satisfied for them to
occur are:%
\begin{equation}
\theta =-\arg \left\{ M_{aB}-M_{a^{\prime }B}\right\} \pm \frac{\pi}{2}
\end{equation}%
\begin{equation}
\left[ \text{Im}\left\{ e^{i\theta }M_{aB}+e^{-i\theta }M_{a^{\prime }B}^{\ast
}\right\} \right]^{2} = \frac{\left| t_{1}\right| ^{2}}{\left| r_{1}\right|
^{2}}\left( 1-\left| M_{eB}\right| ^{2}\right) +2\left( 1-\text{Re}\left\{
M_{eB}\right\} \right)
\end{equation}%
\begin{eqnarray}
&&\text{Re}\left\{ e^{i\theta }M_{aB}\right\} = \left[ \frac{\left| t_{1}\right|
}{4\left| r_{1}\right| }\left( \frac{\left| t_{2}\right| }{\left|
r_{2}\right| }-\frac{\left| r_{2}\right| }{\left| t_{2}\right| }\right)
\left( 1-\left| M_{eB}\right| ^{2}\right) \right. \notag \\
&& + \left. \frac{1}{2} \text{Im}\left\{ M_{eB}\right\}
\text{Im}\left\{ e^{i\theta }M_{aB}+e^{-i\theta }M_{a^{\prime }B}^{\ast}
\right\} \right]
\left( 1-\text{Re}\left\{ M_{eB}\right\} \right)^{-1} .
\end{eqnarray}

To demonstrate that it is, in fact, sometimes possible to satisfy the conditions
for quasi-fixed elements given in (vii), we present the following example:

Consider an anyon model which has at least two different Abelian anyons $a$
and $a^{\prime }$, and some anyon $b$ for which $M_{ab}=e^{i\varphi _{ab}}$
and $M_{a^{\prime }b}=e^{i\varphi _{a^{\prime }b}}$ are not equal (for
example, almost any $\mathbb{Z}_{N}$ model, such as
$\mathbb{Z}_{2}^{\left( 1/2\right)} $ or $\mathbb{Z}_{3}^{\left( 1\right)} $,
is sufficient). The difference
charge $e$ is uniquely determined (since $a$ and $a^{\prime }$ are Abelian)
and has $M_{eb}=e^{i\varphi _{eb}}=e^{i\left( \varphi _{ab}-\varphi
_{a^{\prime }b}\right) }$. Setting $\theta =-\frac{1}{2}\left( \varphi
_{ab}+\varphi _{a^{\prime }b}\right) +n\pi $ gives%
\begin{eqnarray}
\quad
p_{aa^{\prime }e,b}^{\shortrightarrow } &=&\left( \left| t_{1}\right| \left|
r_{2}\right| e^{i\left( \frac{\varphi _{eb}}{2}+n\pi \right) }+\left|
r_{1}\right| \left| t_{2}\right| \right) ^{2} \\
p_{aa^{\prime }e,b}^{\shortuparrow } &=&\left( -\left| t_{1}\right| \left|
t_{2}\right| e^{i\left( \frac{\varphi _{eb}}{2}+n\pi \right) }+\left|
r_{1}\right| \left| r_{2}\right| \right) ^{2} \\
p_{aa1,b}^{\shortrightarrow } &=&p_{a^{\prime }a^{\prime }1,b}^{\shortrightarrow
}=\left| p_{aa^{\prime }e,b}^{\shortrightarrow }\right| =1-\left| p_{aa^{\prime
}e,b}^{\shortuparrow }\right|   \notag \\
&=&\left| t_{1}\right| ^{2}\left| r_{2}\right| ^{2}+2\left|
t_{1}r_{1}t_{2}r_{2}\right| \cos \left( \frac{\varphi _{eb}}{2}+n\pi \right)
+\left| r_{1}\right| ^{2}\left| t_{2}\right| ^{2}.
\end{eqnarray}%
In fact, it turns out this example is the only way to satisfy the
conditions for quasi-fixed elements with $\left| M_{eB}\right| =1$. Indeed,
this can even be shown without initially requiring
$a,a^{\prime }\in \mathcal{C}_{\kappa }$
from positivity. It seems rather difficult to satisfy the conditions for
quasi-fixed elements when $\left| M_{eB}\right| \neq 1$, and we suspect (but
are unable to prove) that it may, in general, actually be impossible. It is
certainly not possible to have quasi-fixed elements with
$\left| M_{eB}\right| \neq 1$ for arbitrary non-Abelian anyon models, as one can
check that they do not exist for either the Ising or Fib anyon models, for
example.

%% file: 03f-distinguishability.tex
\subsection{Distinguishability}

We would like to know how many probe anyons should be used to establish a
desired level of confidence in distinguishing between the various possible
outcomes. For a confidence level $1-\alpha $, the margin of error around $%
p_{\kappa }$ is specified as%
\begin{equation}
E_{\kappa }=z_{\alpha /2}^{\ast }\sigma _{\kappa },
\end{equation}%
i.e. the interval $\left[ p_{\kappa }-E_{\kappa },p_{\kappa }+E_{\kappa }%
\right] $ contains $1-\alpha$ of the probability distribution, where $z_{\alpha
/2}^{\ast }$ is defined by%
\begin{equation}
1-\alpha =\text{erf}\left( \frac{z_{\alpha /2}^{\ast }}{\sqrt{2}}\right) .
\end{equation}%
To achieve this level of confidence in distinguishing two values, $p_{1}$
and $p_{2}$, we pick $N$ so that these intervals have no overlap%
\begin{eqnarray}
\Delta p =\left| p_{1}-p_{2}\right| \gtrsim E_{1}+E_{2}&=&z_{\alpha
/2}^{\ast }\left( \sigma _{1}+\sigma _{2}\right)  \notag \\
&=&z_{\alpha /2}^{\ast }\left( \sqrt{\frac{p_{1}\left( 1-p_{1}\right) }{N}}+%
\sqrt{\frac{p_{2}\left( 1-p_{2}\right) }{N}}\right)
\end{eqnarray}%
which gives the estimated number of probes needed as%
\begin{equation}
N\gtrsim \left( \frac{z_{\alpha /2}^{\ast }\left( \sqrt{p_{1}\left(
1-p_{1}\right) }+\sqrt{p_{2}\left( 1-p_{2}\right) }\right) }{\Delta p}%
\right) ^{2}.
\end{equation}%
Since $p\left( 1-p\right) \leq \frac{1}{4}$, we could conservatively
estimate this for arbitrary $p_{j}$ as%
\begin{equation}
N\gtrsim \left( \frac{z_{\alpha /2}^{\ast }}{\Delta p}\right) ^{2}.
\end{equation}%
On the other hand, if $p_{1}$ and $p_{2}$ are of order $\left| t_{1}\right|
^{2}\sim \left| t_{2}\right| ^{2}\sim t^{2}\ll 1$, and $\Delta p$ is of
order $2t^{2}\Delta M$, where\ $\Delta M=\left| M_{a_{1}B}-M_{a_{2}B}\right|
$, (i.e. employing $\theta $ such that $\Delta p$ is as large as it can be)
then we can estimate%
\begin{equation}
  \label{eq:N_est}
N\gtrsim \left( \frac{z_{\alpha /2}^{\ast }}{t\Delta M}\right) ^{2}.
\end{equation}%
We note that for any two outcome probabilities, $p_{1}$ and $p_{2}$, there are
always two values of $\theta $ (i.e. non-generic conditions) that make
$p_{1}=p_{2}$, and hence indistinguishable. Here are the values of
$z_{\alpha /2}^{\ast }$ for some typical levels of confidence%
\begin{equation*}
\begin{array}{c||c|c|c|c|c}
1-\alpha & .6827 & .9545 & .99 & .999 & .9999 \\
\hline
z_{\alpha /2}^{\ast } & 1 & 2 & 2.576 & 3.2905 & 3.89059%
\end{array}%
\end{equation*}
For greater confidence, the number of probes needed roughly scales as $N \sim -\log \alpha$.

A special case of interest exists when $\left| t_{1}\right| =\left|
t_{2}\right| $ and $\left| M_{a_{1}B}\right| =1$ for one of two probabilities
that we wish to distinguish. In this case, using $\theta =\pi -\arg \left\{
M_{a_{1}B}\right\} $ gives $p_{1}=0$, so any measurement outcome $%
s=\rightarrow $ automatically tells us the target's anyonic charge is not in
$\mathcal{C}_{1}$. If the alternative outcome has $p_{2}\neq 0,1$, then $%
\mathcal{C}_{1}$ and $\mathcal{C}_{2}$ are said to be sometimes perfectly
distinguishable, since a $s=\rightarrow $ outcomes tells us the target's
anyonic charge is in $\mathcal{C}_{2}$. If $M_{a_{1}B}=-M_{a_{2}B}$ and we
also have $\left| t_{j}\right| ^{2}=1/2$, then $p_{2}=1$, and $\mathcal{C}%
_{1}$ and $\mathcal{C}_{2}$ are always perfectly distinguishable, since any
single probe measurement will indicate whether the target's anyonic charge is in
$\mathcal{C}_{1}$ or in $\mathcal{C}_{2}$. 

%% file: 03g-targetgeneralizations.tex
\subsection{Target System Configuration}
  \label{sec:targetgen}

In this section, we consider the effect of locating the anyons $C$ that are 
entangled with the target $A$ in different regions outside the central 
interferometry region. If $C$ is located above the central interferometer region, we would have%
\begin{equation}
V=\left[
\begin{array}{cc}
R_{BC} & 0 \\
0 & R_{BC}%
\end{array}%
\right] ,
\end{equation}%
for which similar diagrammatic evaluation gives%
\begin{eqnarray}
\qquad \qquad
p_{aa^{\prime }e,B}^{\shortrightarrow } &=&\left| t_{1}\right| ^{2}\left|
r_{2}\right| ^{2}+t_{1}r_{1}^{\ast }r_{2}^{\ast }t_{2}^{\ast
}e^{i\left( \theta _{\text{I}}-\theta _{\text{II}}\right) }M_{a^{\prime}B}  \notag \\
&&+t_{1}^{\ast }r_{1}t_{2}r_{2}e^{-i\left( \theta _{\text{I}}-\theta _{\text{II}}\right)
}M_{aB}^{\ast }+\left| r_{1}\right| ^{2}\left| t_{2}\right| ^{2} M_{eB}^{\ast} , \\
p_{aa^{\prime }e,B}^{\shortuparrow } &=&\left| t_{1}\right| ^{2}\left|
t_{2}\right| ^{2}-t_{1}r_{1}^{\ast }r_{2}^{\ast }t_{2}^{\ast
}e^{i\left( \theta _{\text{I}}-\theta _{\text{II}}\right) }M_{a^{\prime}B}  \notag \\
&&-t_{1}^{\ast }r_{1}t_{2}r_{2}e^{-i\left( \theta _{\text{I}}-\theta _{\text{II}}\right)
}M_{aB}^{\ast }+\left| r_{1}\right| ^{2}\left| r_{2}\right| ^{2} M_{eB}^{\ast}.
\end{eqnarray}
instead of Eqs.~(\ref{eq:p_right},\ref{eq:p_up},\ref{eq:p_B}). 
If $C$ is located between the output legs of the interferometer, we would instead have%
\begin{equation}
V=\left[
\begin{array}{cc}
R_{BC} & 0 \\
0 & R_{CB}^{-1}%
\end{array}%
\right] .
\end{equation}%
The resulting diagrammatic evaluation in this case gives%
\begin{eqnarray}
\qquad \qquad
p_{aa^{\prime }e,B}^{\shortrightarrow } &=&\left| t_{1}\right| ^{2}\left|
r_{2}\right| ^{2}+t_{1}r_{1}^{\ast }r_{2}^{\ast }t_{2}^{\ast
}e^{i\left( \theta _{\text{I}}-\theta _{\text{II}}\right) }M_{a^{\prime}B}  \notag \\
&&+t_{1}^{\ast }r_{1}t_{2}r_{2}e^{-i\left( \theta _{\text{I}}-\theta _{\text{II}}\right)
}M_{aB}^{\ast }+\left| r_{1}\right| ^{2}\left| t_{2}\right| ^{2} M_{eB}^{\ast} , \\
p_{aa^{\prime }e,B}^{\shortuparrow } &=&\left| t_{1}\right| ^{2}\left|
t_{2}\right| ^{2}M_{eB}-t_{1}r_{1}^{\ast }r_{2}^{\ast }t_{2}^{\ast
}e^{i\left( \theta _{\text{I}}-\theta _{\text{II}}\right) }M_{aB}  \notag \\
&&-t_{1}^{\ast }r_{1}t_{2}r_{2}e^{-i\left( \theta _{\text{I}}-\theta _{\text{II}}\right)
}M_{a^{\prime }B}^{\ast }+\left| r_{1}\right| ^{2}\left| r_{2}\right| ^{2}.
\end{eqnarray}%
For both of these cases, the arguments from before apply directly and limiting behavior 
is exactly the same. One can also envision more complicated situations, such as having 
the $C$ anyons distributed amongst all the regions outside the central one. The resulting 
calculations are straightforward, but too cumbersome to display here explicitly. 
However, the limiting behavior is essentially the same as before, as one would expect from 
the previous analysis: Interferometry measurement generically collapses the target system onto 
fixed states, which are characterized as having the target anyons $A$ (those which the probes 
interfere around) in a charge subset that the probe cannot distinguish by monodromy 
(i.e. $a,a^{\prime}$ such that $M_{aB}=M_{a^{\prime}b}$), and the anyons in distinct regions 
having no coherent anyonic charge entanglement crossing the probe anyons' beam paths that the probes can 
``see'' by monodromy (i.e. anyonic charge entanglement characterized by charge $e$ can 
entangle anyons in distinct regions only if $M_{eB}=1$).

%% file: 03g-probegeneralizations.tex
\subsection{Probe Generalizations}
  \label{sec:probegen}

In this section, we examine the effects of using probe systems that
are even more general than those employed so far. We will first consider
generalizing the input direction, so that probes may enter in arbitrary
superpositions of the two input directions. Then we will consider the use of
probes that are not identical, so that each probe system is described by a
different density matrix. For both of these, the probe systems and target system
are all still initially unentangled. One may also consider cases where there
is nontrivial initial entanglement between these systems, or post-interferometer
charge projections, but these typically lead to qualitatively different
behavior, and greatly increase the complexity of analysis, so we will not
consider them here.

\subsubsection{Generalized Input Directions}

For probes that are allowed to enter the interferometer through either
of the input legs, possibly even in superposition, the probe systems' density
matrices take the form%
\begin{equation}
\rho ^{B}=\sum\limits_{b,b^{\prime },d,d^{\prime },h,\lambda ,\lambda
^{\prime },r,r^{\prime }}\rho _{\left( d,b_{r};h,\lambda \right) \left(
d^{\prime },b_{r^{\prime }}^{\prime };h,\lambda ^{\prime }\right)
}^{B} \frac{1}{d_h} \left| d,b_{r};h,\lambda \right\rangle \left\langle d^{\prime
},b_{r^{\prime }}^{\prime };h,\lambda ^{\prime }\right|
.
\end{equation}%
Using this, we find the same result as before, except, instead of 
Eqs.~(\ref{eq:p_right},\ref{eq:p_up},\ref{eq:p_B}), the values of $p_{aa^{\prime}e,B}^{s}$ 
are given by%
\begin{equation}
p_{aa^{\prime }e,B}^{s}=\sum\limits_{d,h,\lambda ,b,r,r^{\prime }}\rho
_{\left( d,b_{r};h,\lambda \right) \left( d,b_{r^{\prime
}};h,\lambda \right) }^{B}p_{aa^{\prime }e,b,r,r^{\prime
}}^{s}
\end{equation}%
where%
\begin{eqnarray}
\qquad \qquad
p_{aa^{\prime }e,b,\shortrightarrow ,\shortrightarrow }^{\shortrightarrow } &=&\left|
t_{1}\right| ^{2}\left| r_{2}\right| ^{2}M_{eb}+t_{1}r_{1}^{\ast
}t_{2}^{\ast }r_{2}^{\ast }e^{i\left( \theta _{\text{I}}-\theta _{\text{II}}\right) }M_{ab}
\notag \\
&&+t_{1}^{\ast }r_{1}t_{2}r_{2}e^{-i\left( \theta _{\text{I}}-\theta _{\text{II}}\right)
}M_{a^{\prime }b}^{\ast }+\left| r_{1}\right| ^{2}\left| t_{2}\right| ^{2} \\
p_{aa^{\prime }e,b,\shortrightarrow ,\shortuparrow }^{\shortrightarrow }
&=&t_{1}r_{1}\left| r_{2}\right| ^{2}M_{eb}-t_{1}t_{1}t_{2}^{\ast
}r_{2}^{\ast }e^{i\left( \theta _{\text{I}}-\theta _{\text{II}}\right) }M_{ab}  \notag \\
&&+r_{1}r_{1}t_{2}r_{2}e^{-i\left( \theta _{\text{I}}-\theta _{\text{II}}\right)
}M_{a^{\prime }b}^{\ast }-t_{1}r_{1}\left| t_{2}\right| ^{2} \\
p_{aa^{\prime }e,b,\shortuparrow ,\shortrightarrow }^{\shortrightarrow } &=&t_{1}^{\ast
}r_{1}^{\ast }\left| r_{2}\right| ^{2}M_{eb}+r_{1}^{\ast }r_{1}^{\ast
}t_{2}^{\ast }r_{2}^{\ast }e^{i\left( \theta _{\text{I}}-\theta _{\text{II}}\right) }M_{ab}
\notag \\
&&-t_{1}^{\ast }t_{1}^{\ast }t_{2}r_{2}e^{-i\left( \theta _{\text{I}}-\theta
_{\text{II}}\right) }M_{a^{\prime }b}^{\ast }-t_{1}^{\ast }r_{1}^{\ast }\left|
t_{2}\right| ^{2} \\
p_{aa^{\prime }e,b,\shortuparrow ,\shortuparrow }^{\shortrightarrow } &=&\left|
r_{1}\right| ^{2}\left| r_{2}\right| ^{2}M_{eb}-t_{1}r_{1}^{\ast
}t_{2}^{\ast }r_{2}^{\ast }e^{i\left( \theta _{\text{I}}-\theta _{\text{II}}\right) }M_{ab}
\notag \\
&&-t_{1}^{\ast }r_{1}t_{2}r_{2}e^{-i\left( \theta _{\text{I}}-\theta _{\text{II}}\right)
}M_{a^{\prime }b}^{\ast }+\left| t_{1}\right| ^{2}\left| t_{2}\right| ^{2}
\end{eqnarray}%
and%
\begin{eqnarray}
\qquad \qquad
p_{aa^{\prime }e,b,\shortrightarrow ,\shortrightarrow }^{\shortuparrow } &=&\left|
t_{1}\right| ^{2}\left| t_{2}\right| ^{2}M_{eb}-t_{1}r_{1}^{\ast
}t_{2}^{\ast }r_{2}^{\ast }e^{i\left( \theta _{\text{I}}-\theta _{\text{II}}\right) }M_{ab}
\notag \\
&&-t_{1}^{\ast }r_{1}t_{2}r_{2}e^{-i\left( \theta _{\text{I}}-\theta _{\text{II}}\right)
}M_{a^{\prime }b}^{\ast }+\left| r_{1}\right| ^{2}\left| r_{2}\right| ^{2} \\
p_{aa^{\prime }e,b,\shortrightarrow ,\shortuparrow }^{\shortuparrow } &=&t_{1}r_{1}\left|
t_{2}\right| ^{2}M_{eb}+t_{1}t_{1}t_{2}^{\ast }r_{2}^{\ast }e^{i\left(
\theta _{\text{I}}-\theta _{\text{II}}\right) }M_{ab}  \notag \\
&&-r_{1}r_{1}t_{2}r_{2}e^{-i\left( \theta _{\text{I}}-\theta _{\text{II}}\right)
}M_{a^{\prime }b}^{\ast }-t_{1}r_{1}\left| r_{2}\right| ^{2} \\
p_{aa^{\prime }e,b,\shortuparrow ,\shortrightarrow }^{\shortuparrow } &=&t_{1}^{\ast
}r_{1}^{\ast }\left| t_{2}\right| ^{2}M_{eb}-r_{1}^{\ast }r_{1}^{\ast
}t_{2}^{\ast }r_{2}^{\ast }e^{i\left( \theta _{\text{I}}-\theta _{\text{II}}\right) }M_{ab}
\notag \\
&&+t_{1}^{\ast }t_{1}^{\ast }t_{2}r_{2}e^{-i\left( \theta _{\text{I}}-\theta
_{\text{II}}\right) }M_{a^{\prime }b}^{\ast }-t_{1}^{\ast }r_{1}^{\ast }\left|
r_{2}\right| ^{2} \\
p_{aa^{\prime }e,b,\shortuparrow ,\shortuparrow }^{\shortuparrow } &=&\left| r_{1}\right|
^{2}\left| t_{2}\right| ^{2}M_{eb}+t_{1}r_{1}^{\ast }t_{2}^{\ast
}r_{2}^{\ast }e^{i\left( \theta _{\text{I}}-\theta _{\text{II}}\right) }M_{ab}  \notag \\
&&+t_{1}^{\ast }r_{1}t_{2}r_{2}e^{-i\left( \theta _{\text{I}}-\theta _{\text{II}}\right)
}M_{a^{\prime }b}^{\ast }+\left| t_{1}\right| ^{2}\left| r_{2}\right| ^{2}
.
\end{eqnarray}%
It is straightforward to check that%
\begin{equation}
p_{aa1,B}^{\shortrightarrow }+p_{aa1,B}^{\shortuparrow
}=\sum\limits_{d,h,\lambda ,b,r}\rho _{\left( d,b_{r};h,\lambda \right)
\left( d,b_{r};h,\lambda \right) }^{B}=1
,
\end{equation}%
and one can see that, generically, the only terms in the target anyons' density
matrix that will survive many probe measurements are those in $e$-channels with%
\begin{equation}
M_{eB}= \sum\limits_{d,h,\lambda ,b,r}\rho _{\left( d,b_{r};h,\lambda \right)
\left( d,b_{r};h,\lambda \right) }^{B} M_{eb} =1
.
\end{equation}%

\subsubsection{Non-Identical Probes}

When the probes $B_{1},\ldots ,B_{N}$ are described by different density
matrices $\rho ^{B_{j}}$ (though are all still unentangled with each other
and with the target system), we must use%
\begin{eqnarray}
\qquad \qquad \qquad \qquad
p_{aa^{\prime }e,B_{j}}^{s} &=&\sum\limits_{b}\Pr\nolimits_{B_{j}}\left(
b\right) p_{aa^{\prime }e,b}^{s} \\
\Pr\nolimits_{B_{j}}\left( b\right)  &=&\sum\limits_{d,h,\lambda }\rho
_{\left( d,b_{\shortrightarrow};h,\lambda \right) \left( d,b_{%
\shortrightarrow};h,\lambda \right) }^{B_{j}}
\end{eqnarray}%
for each probe. This gives us the probability for the string of measurement
outcomes $\left( s_{1},\ldots ,s_{N}\right) $ to occur as%
\begin{equation}
\Pr \left( s_{1},\ldots ,s_{N}\right) =\sum\limits_{a,c,f,\mu }\rho _{\left(
a,c;f,\mu \right) \left( a,c;f,\mu \right) }^{A}p_{aa1,B_{1}}^{s_{1}}\ldots
p_{aa1,B_{N}}^{s_{N}}
,
\end{equation}%
with the resulting target anyon density matrix%
\begin{eqnarray}
&& \rho ^{A}\left( s_{1},\ldots ,s_{N}\right) =\sum\limits_{\substack{ %
a,a^{\prime },c,c^{\prime },f,\mu ,\mu ^{\prime } \\ e,\alpha ,\beta ,
f^{\prime },\nu ,\nu ^{\prime } }}\frac{\rho _{\left(
a,c;f,\mu \right) \left( a^{\prime },c^{\prime };f,\mu ^{\prime }\right)
}^{A}}{\left( d_{f}d_{f^{\prime }}\right) ^{1/2}}\frac{p_{aa^{\prime
}e,B_{1}}^{s_{1}}\ldots p_{aa^{\prime }e,B_{N}}^{s_{N}}}{\Pr \left(
s_{1},\ldots ,s_{N}\right) } \notag \\
&& \times \left[ \left( F_{a^{\prime }c^{\prime
}}^{ac}\right) ^{-1}\right] _{\left( f,\mu ,\mu ^{\prime }\right) \left(
e,\alpha ,\beta \right) } \left[ F_{a^{\prime }c^{\prime
}}^{ac}\right] _{\left( e,\alpha ,\beta \right) \left( f^{\prime },\nu ,\nu
^{\prime }\right) }\left| a,c;f^{\prime },\nu \right\rangle \left\langle
a^{\prime },c^{\prime };f^{\prime },\nu ^{\prime }\right|
.
\end{eqnarray}%
With this generalization, we find that the order of measurement outcomes
does, in fact, matter. This is obstructive to providing a quantitative
description of the large $N$ behavior; however, the qualitative behavior
should be transparent after the analysis in previous sections for the
identical probes. Each probe measurement will execute some amount of
projection, to some extent collapsing superpositions of anyonic charges that the
probe is able to distinguish by monodromy. 

%% file: 03h-Examples.tex
\subsection{Examples}
  \label{sec:Examples}

In this section, we apply the general results to some important examples,
specifically: the $\mathbb{Z}_{N}$, Fibonacci, and Ising anyon models. (The application to
some additional important examples, such as $\text{SU} \left( 2 \right)_{k}$ and
$\text{D}\left( \mathbb{Z}_{N} \right)$, may be found in~\cite{Bonderson07b}.)
All of these have $N_{ab}^{c}=0,1$, so we will drop the fusion/splitting
spaces' basis labels (greek indices), with the understanding that any symbol
involving a prohibited fusion vertex is set to zero. Anyon models are completely
specified by their $F$-symbols and $R$-symbols, so we will provide these, as
well as list some additional important quantities that can be derived from them,
for convenience. To relate these to interferometry experiments, we give the
corresponding fixed state probabilities $p_{\kappa }$ and density matrices
$\rho _{\kappa}^{A}$, as described in Section~\ref{sec:NProbes}.

\subsubsection{$\mathbb{Z}_{N}^{\left( w\right)}$}
  \label{sec:Z_N}

The Abelian $\mathbb{Z}_{N}$ anyon models~\cite{Moore89b} have anyonic
charges $\mathcal{C}=\left\{ 0,1,\ldots ,N-1\right\} $, where $0$ here
designates the vacuum charge. The fusion rules are just given by $\mathbb{Z}_{N}$ addition,
and, to denote this, we define $\left[ n\right]_{N}\in \mathcal{C}$ as the least residue of $\ n \text{ mod }N$. The $\mathbb{Z}_{N}^{\left( w\right)}$ anyon models, where $w=n$ for $N$ odd and $w=n$ and $n+\frac{1}{2}$ for $N$ even, with $n=0,1,\ldots,N-1$, are described by:%
\begin{equation*}
\begin{tabular}[b]{|l|l|l|}
\hline
\multicolumn{3}{|l|}{$\mathcal{C}=\left\{ 0,1,\ldots ,N-1\right\}, \quad a \times b =\left[a+b\right] _{N}$} \\ \hline
\multicolumn{3}{|l|}{for $w=n$: $\left[ F_{\left[a+b+c\right] _{N}}^{abc}\right] _{\left[a+b\right] _{N} \left[b+c\right] _{N}}
=\left[ F_{c\left[a+b-c\right] _{N}}^{ab}\right] _{\left[a-c\right] _{N}\left[a+b\right] _{N_{\phantom{j}}}}=1$} \\ \hline
\multicolumn{3}{|l|}{for $w=n+\frac{1}{2}$: $\left[ F_{\left[a+b+c\right] _{N}}^{abc}\right] _{\left[a+b\right] _{N} \left[b+c\right] _{N}}
=e^{i\frac{\pi }{N}a\left( b+c-\left[ b+c\right] _{N}\right) },$} \\
\multicolumn{3}{|l|}{ $\quad \quad \quad \quad \quad \quad \left[ F_{c\left[a+b-c\right] _{N}}^{ab}\right] _{\left[a-c\right] _{N}\left[a+b\right] _{N_{\phantom{j}}}}
=e^{i\frac{\pi }{N}c\left( \left[a-c\right] _{N}+b-\left[ a+b-c\right] _{N}\right) }$} \\ \hline
$R_{\left[a+b\right] _{N_{\phantom{j}}}}^{ab}=e^{i\frac{2\pi w}{N}ab}$ &
$S_{ab}=\frac{1}{\sqrt{N}}e^{i\frac{4\pi w}{N}ab}$ &
$M_{ab}=e^{i\frac{4\pi w}{N}ab}$ \\ \hline
\multicolumn{2}{|l|}{$d_{a}=1, \quad \mathcal{D}=\sqrt{N}$} & $\theta _{a}=e^{i2\pi \frac{w}{N}a^{2}}$ \\ \hline
\end{tabular}%
\end{equation*}
These anyon models describe some Chern-Simons/WZW theories, e.g. $\text{SU}(N)_1$, for which the corresponding anyon models
are $\mathbb{Z}_{N}^{{((N-1)/2)}}$ for $N$ odd and
$\mathbb{Z}_{N}^{{(N/2-1)}}$ for $N$ even; and $\text{U}(1)_{k}$, for which the
corresponding anyon models are $\mathbb{Z}_{2k}^{(1/2)}$ for $2k$ even and
$\mathbb{Z}_{4k}^{(1)}$ for $2k$ odd.

Of course, for Abelian anyon models such as these,
each physical quasiparticle excitation has a specific anyonic charge and all
fusion channels are uniquely determined, so superpositions of anyonic charge
are not actually possible, but such models might occur as a
subset of a non-Abelian anyon model, in which case superpositions of these
charges could potentially occur. In any case, one may still perform interferometry experiments
in these models to determine the charge of a target anyon. Using $b$ probes, we have:%
\begin{equation}
p_{a}=p_{aa0,b}^{\shortrightarrow }=\left| t_{1}\right| ^{2}\left| r_{2}\right|
^{2}+2\left| t_{1}r_{1}t_{2}r_{2}\right| \cos \left( \theta +\frac{4\pi w}{N}%
ab\right) +\left| r_{1}\right| ^{2}\left| t_{2}\right| ^{2}
\end{equation}
and
\begin{equation}
\Pr\nolimits_{A}\left( \kappa \right) = \sum\limits_{a\in C_{\kappa
},f}\rho _{\left( a,f-a;f\right) \left( a,f-a;f\right) }
\end{equation}
\begin{equation}
\rho _{\kappa }^{A} = \sum\limits_{a,a^{\prime }\in C_{\kappa },f}
\frac{\rho_{\left( a,f-a;f\right) \left( a^{\prime },f-a^{\prime };f\right) } }
{\Pr\nolimits_{A}\left( \kappa \right)}
\left| a,f-a;f\right\rangle \left\langle a^{\prime },f-a^{\prime };f\right|
\end{equation}%
For $\mathbb{Z}_{N}^{\left( w\right) }$ with $\gcd(2w,N)=1$ (i.e. the modular
$\mathbb{Z}_{N}$ models), the charge classes are singletons
$\mathcal{C}_{a}=\left\{ a\right\} $, so $a=a^{\prime }$ in the fixed state
density matrices.

\subsubsection{Fib}
  \label{sec:Fib}

The Fibonacci (Fib) anyon model (also known as SO$(3)_{3}$, since it may be
obtained from the SU$\left( 2\right) _{3}$ anyon model by restricting to
integer spins)\footnote{%
As a Chern-Simons or WZW theory, this is properly denoted as $\overline{\left( \text{G}%
_{2}\right)} _{1}$, since SO$\left( 3\right) _{k}$ is only allowed for
$k=0~{\rm mod}~4$.} is known to be universal for topological quantum
computation~\cite{Freedman02b}. It has two charges
$\mathcal{C}=\left\{1,\varepsilon \right\} $ and is described by (listing only
the non-trivial $F$-symbols and $R$-symbols, i.e. those not listed are equal to
one if their vertices are permitted by fusion, and equal to zero if they are not
permitted):%
\begin{equation*}
\begin{tabular}{|l|l|}
\hline
\multicolumn{2}{|l|}{$\mathcal{C}=\left\{1,\varepsilon \right\}, \quad 1\times 1=1,\quad 1\times \varepsilon=\varepsilon,\quad \varepsilon\times \varepsilon=1+\varepsilon$} \\
\hline
\multicolumn{2}{|l|}{$\left[ F_{\varepsilon}^{\varepsilon \varepsilon \varepsilon}\right] _{ef}=\left[ F_{\varepsilon \varepsilon}^{\varepsilon \varepsilon}\right] _{ef}
=\left[
\begin{array}{cc}
\phi ^{-1} & \phi ^{-1/2} \\
\phi ^{-1/2} & -\phi ^{-1}%
\end{array}%
\right] _{ef_{\phantom{j}}}^{\phantom{T}}$} \\ \hline
\multicolumn{2}{|l|}{$R_{1}^{\varepsilon \varepsilon}=e^{-i4\pi /5},\quad R_{\varepsilon}^{\varepsilon \varepsilon}=e^{i3\pi /5}$} \\ \hline
$S=\frac{1}{\sqrt{\phi +2}}\left[
\begin{array}{rr}
1 & \phi \\
\phi & -1%
\end{array}%
\right]^{\phantom{T}}_{\phantom{j}} $ & $M=\left[
\begin{array}{cc}
1 & 1 \\
1 & -\phi ^{-2}%
\end{array}%
\right] $ \\ \hline
$d_{1}=1,\quad d_{\varepsilon}=\phi,\quad \mathcal{D}=\sqrt{\phi+2}$ & $\theta _{1}=1,\quad \theta _{\varepsilon}=e^{i\frac{%
4\pi }{5}}$ \\ \hline
\end{tabular}%
\end{equation*}%
where $\phi =\frac{1+\sqrt{5}}{2}$ is the Golden ratio. We denote the anyon
model given by this with the complex conjugate values of the $R$-symbols and topological
spins as $\overline{\text{Fib}}$.

For $b=\varepsilon$ probes, we have $\mathcal{C}_{1}=\left\{ 1\right\} $,
$\mathcal{C}_{2}=\left\{ \varepsilon\right\} $ and%
\begin{eqnarray}
\qquad \quad
p_{1} &=&p_{111,\varepsilon}^{\shortrightarrow }=\left| t_{1}\right| ^{2}\left|
r_{2}\right| ^{2}+2\left| t_{1}r_{1}t_{2}r_{2}\right| \cos \theta +\left|
r_{1}\right| ^{2}\left| t_{2}\right| ^{2} \\
p_{2} &=&p_{\varepsilon \varepsilon 1,\varepsilon}^{\shortrightarrow }=\left| t_{1}\right| ^{2}\left|
r_{2}\right| ^{2}-2\phi ^{-2}\left| t_{1}r_{1}t_{2}r_{2}\right| \cos \theta +\left|
r_{1}\right| ^{2}\left| t_{2}\right| ^{2}
\end{eqnarray}%
\begin{equation}
\Pr\nolimits_{A}\left( 1\right) =\rho _{\left( 1,1;1\right)
\left( 1,1;1\right) } + \rho _{\left( 1,\varepsilon;\varepsilon\right)
\left( 1,\varepsilon;\varepsilon\right) }
\end{equation}%
\begin{eqnarray}
\qquad \qquad
\rho _{1}^{A} &=&\frac{1}{\Pr\nolimits_{A}\left( 1\right) }\left\{ \rho _{\left(
1,1;1\right) \left( 1,1;1\right) }\left| 1,1;1\right\rangle \left\langle
1,1;1\right| \right.  \notag \\
&&\left. \quad \quad \quad \quad +\phi ^{-1}\rho _{\left( 1,\varepsilon;\varepsilon\right)
\left( 1,\varepsilon;\varepsilon\right) }\left| 1,\varepsilon;\varepsilon\right\rangle \left\langle 1,\varepsilon;\varepsilon\right|
\right\}
\end{eqnarray}%
\begin{equation}
\Pr\nolimits_{A}\left( 2\right) = \rho _{\left(
\varepsilon,1;\varepsilon\right) \left(
\varepsilon,1;\varepsilon\right) }+\rho _{\left( \varepsilon,\varepsilon;1\right) \left( \varepsilon,\varepsilon;1\right) }+\rho
_{\left( \varepsilon,\varepsilon;\varepsilon\right) \left(
\varepsilon,\varepsilon;\varepsilon\right) }
\end{equation}%
\begin{eqnarray}
\qquad
\rho _{2}^{A} &=&\frac{1}{\Pr\nolimits_{A}\left( 2\right) }\left\{ \phi ^{-1}
\rho _{\left(\varepsilon,1;\varepsilon\right) \left( \varepsilon,1;\varepsilon\right) }\left| \varepsilon,1;\varepsilon\right\rangle \left\langle
\varepsilon,1;\varepsilon\right| \right.  \notag \\
&&\left. \quad \quad \quad \quad +\phi ^{-2}\left( \rho _{\left(
\varepsilon,\varepsilon;1\right) \left( \varepsilon,\varepsilon;1\right) }+\rho _{\left( \varepsilon,\varepsilon;\varepsilon\right) \left(
\varepsilon,\varepsilon;\varepsilon\right) }\right) \right. \nonumber \\
&&\left. \quad \quad \quad \quad \quad \phantom{\phi^{-1}} \times \left[ \left|
\varepsilon,\varepsilon;1\right\rangle \left\langle \varepsilon,\varepsilon;1\right| +\left| \varepsilon,\varepsilon;\varepsilon\right\rangle
\left\langle \varepsilon,\varepsilon;\varepsilon\right| \right] \right\}
\end{eqnarray}

We note that one can sometimes (approximately $69\%$ of the time, when the
target charge is not vacuum) perfectly distinguish the charges $1$ and
$\varepsilon$ with a single $b=\varepsilon$ probe measurement by setting the
experimental parameters to:
$\left| t_{1}\right| ^{2}=\left| t_{2}\right| ^{2}=1/2$ and
$\theta =\pi$, which give $p_{1}=0$ and $p_{2}=1-\frac{1}{2\phi }\simeq .69$.

\subsubsection{Ising}
  \label{sec:Ising}

The Ising anyon model is derived from the CFT that
describes the Ising model at criticality~\cite{Moore89b}. It has anyonic
charges $\mathcal{C}=\left\{1,\sigma,\psi \right\}$ (which respectively
correspond to vacuum, spin, and Majorana fermions in the CFT). The anyon model
is described by (listing only the non-trivial $F$-symbols and $R$-symbols):%
\begin{equation*}
\begin{tabular}{|l|l|}
\hline
\multicolumn{2}{|l|}{$\mathcal{C}=\left\{1,\sigma,\psi \right\}, \quad 1\times a=a,\quad \sigma \times \sigma%
=1+\psi,\quad \sigma \times \psi=\sigma,\quad \psi \times \psi=1$} \\ \hline
\multicolumn{2}{|l|}{$\left[ F_{\sigma}^{\sigma \sigma \sigma}\right] _{ef}=
\left[ F_{\sigma \sigma}^{\sigma \sigma}\right] _{ef}=
\left[
\begin{array}{rr}
\frac{1}{\sqrt{2}} & \frac{1}{\sqrt{2}} \\
\frac{1}{\sqrt{2}} & \frac{-1}{\sqrt{2}}%
\end{array}\right] _{ef}^{\phantom{T}}$} \\
\multicolumn{2}{|l|}{$\left[ F_{\psi}^{\sigma \psi \sigma}\right] _{\sigma \sigma}=%
\left[ F_{\sigma}^{\psi \sigma \psi}\right] _{\sigma \sigma_{\phantom{j}}}\!\!=
\left[ F_{\psi \sigma}^{\sigma \psi}\right] _{\sigma \sigma}=
\left[ F_{\sigma \psi}^{\psi \sigma}\right] _{\sigma \sigma}=-1 $} \\ \hline
\multicolumn{2}{|l|}{
$R_{1}^{\sigma \sigma}=e^{-i\frac{\pi }{8}},\quad R_{\psi}^{\sigma \sigma}=e^{i\frac{3\pi }{8}},
\quad R_{\sigma}^{\sigma \psi}=R_{\sigma}^{\psi \sigma}=e^{-i\frac{\pi }{2}},\quad R_{1}^{\psi \psi}=-1$} \\ \hline
$S=\frac{1}{2}\left[
\begin{array}{rrr}
1 & \sqrt{2} & 1 \\
\sqrt{2} & 0 & -\sqrt{2} \\
1 & -\sqrt{2} & 1%
\end{array}%
\right]^{\phantom{T}}_{\phantom{j}} $ & $M=\left[
\begin{array}{rrr}
1 & 1 & 1 \\
1 & 0 & -1 \\
1 & -1 & 1%
\end{array}%
\right] $ \\ \hline
$d_{1}=d_{\psi}=1,\quad d_{\sigma_{\phantom{j}}}\!\!=\sqrt{2}, \quad \mathcal{D}=2$ & $\theta _{1}=1,\quad \theta
_{\sigma}=e^{i\frac{\pi }{8}},\quad \theta _{\psi}=-1$ \\ \hline
\end{tabular}%
\end{equation*}%
where $e,f\in \left\{ 1,\psi\right\} $.

For $b=\psi$ probes, we have $\mathcal{C}_{1}=\left\{ 1,\psi\right\} $,
$\mathcal{C}_{2}=\left\{ \sigma \right\} $, and%
\begin{eqnarray}
p_{1} &=&p_{111,\psi}^{\shortrightarrow }=p_{\psi \psi 1,\psi}^{\shortrightarrow
}=p_{1 \psi \psi, \psi}^{\shortrightarrow }=p_{\psi 1 \psi, \psi}^{\shortrightarrow }  \notag \\
&=&\left| t_{1}\right| ^{2}\left| r_{2}\right| ^{2}+2\left|
t_{1}r_{1}r_{2}t_{2}\right| \cos \theta +\left| r_{1}\right| ^{2}\left|
t_{2}\right| ^{2} \\
p_{2} &=&p_{\sigma \sigma 1, \psi}^{\shortrightarrow }=p_{\sigma \sigma \psi, \psi}^{\shortrightarrow }=\left| t_{1}\right| ^{2}\left| r_{2}\right|
^{2}-2\left| t_{1}r_{1}r_{2}t_{2}\right| \cos \theta +\left| r_{1}\right|
^{2}\left| t_{2}\right| ^{2}
\end{eqnarray}%
\begin{eqnarray}
\qquad
\Pr\nolimits_{A}\left( 1\right) &=&
\rho _{\left(1,1;1\right) \left( 1,1;1\right) }
+ \rho _{\left(1,\sigma;\sigma\right) \left( 1,\sigma;\sigma\right) }
+\rho _{\left(1,\psi;\psi\right) \left( 1,\psi;\psi\right) } \notag \\
&& +\rho _{\left( \psi,1;\psi\right) \left(\psi,1;\psi\right) }
+\rho _{\left( \psi,\sigma;\sigma \right) \left(\psi,\sigma;\sigma\right) }
+\rho _{\left( \psi,\psi;1 \right) \left(\psi,\psi;1 \right) }
\end{eqnarray}%
\begin{eqnarray}
&&\rho _{1}^{A} =\frac{1}{\Pr\nolimits_{A}\left( 1\right)} \left\{
\rho _{\left( 1,1;1\right) \left( 1,1;1\right) }
\left|1,1;1\right\rangle \left\langle 1,1;1\right|
+\rho _{\left( 1,1;1\right) \left( \psi,\psi;1\right) }
\left|1,1;1\right\rangle \left\langle \psi,\psi;1\right| \right. \notag \\
&& \quad \quad \quad \quad +\rho _{\left( \psi,\psi;1\right) \left( 1,1;1\right) }
\left|\psi,\psi;1\right\rangle \left\langle 1,1;1\right|
+\rho _{\left( \psi,\psi;1\right) \left( \psi,\psi;1\right) }
\left|\psi,\psi;1\right\rangle \left\langle \psi,\psi;1\right| \notag \\
&&+\frac{1}{\sqrt{2}} \left( \rho _{\left( 1,\sigma;\sigma\right) \left( 1,\sigma;\sigma\right) }
\left|1,\sigma;\sigma\right\rangle \left\langle 1,\sigma;\sigma\right|
+\rho _{\left( 1,\sigma;\sigma\right) \left( \psi,\sigma;\sigma\right) }
\left|1,\sigma;\sigma\right\rangle \left\langle \psi,\sigma;\sigma\right| \right. \notag \\
&& \quad \quad \left. + \rho _{\left( \psi,\sigma;\sigma\right) \left( 1,\sigma;\sigma\right) }
\left|\psi,\sigma;\sigma\right\rangle \left\langle 1,\sigma;\sigma\right|
+\rho _{\left( \psi,\sigma;\sigma\right) \left( \psi,\sigma;\sigma\right) }
\left|\psi,\sigma;\sigma\right\rangle \left\langle \psi,\sigma;\sigma\right| \right) \notag\\
&&+\rho _{\left( 1,\psi;\psi\right) \left( 1,\psi;\psi\right) }
\left|1,\psi;\psi\right\rangle \left\langle 1,\psi;\psi\right|
+\rho _{\left( 1,\psi;\psi\right) \left( \psi,1;\psi\right) }
\left|1,\psi;\psi\right\rangle \left\langle \psi,1;\psi\right| \notag \\
&&+\rho _{\left( \psi,1;\psi\right) \left( 1,\psi;\psi\right) }
\left|\psi,1;\psi\right\rangle \left\langle 1,\psi;\psi\right|
+\left. \rho _{\left( \psi,1;\psi\right) \left( \psi,1;\psi\right) }
\left|\psi,1;\psi\right\rangle \left\langle \psi,1;\psi\right| \right\}
\end{eqnarray}%
\begin{equation}
\Pr\nolimits_{A}\left( 2\right)
=\rho _{\left( \sigma,1;\sigma\right)\left( \sigma,1;\sigma\right) }
+\rho _{\left( \sigma,\sigma;1\right)\left( \sigma,\sigma;1\right) }
+\rho _{\left( \sigma,\sigma;\psi\right)\left( \sigma,\sigma;\psi\right) }
+\rho _{\left( \sigma,\psi;\sigma\right)\left( \sigma,\psi;\sigma\right) }
\end{equation}
\begin{eqnarray}
&&\rho _{2}^{A} =\frac{1}{\Pr\nolimits_{A}\left( 2\right)} \left\{
\rho _{\left( \sigma,\sigma;1\right) \left( \sigma,\sigma;1\right) }
\left| \sigma,\sigma;1\right\rangle \left\langle \sigma ,\sigma;1\right| \right. \notag \\
&&+ \frac{1}{\sqrt{2}} \left( \rho _{\left( \sigma,1;\sigma\right) \left( \sigma,1;\sigma\right) }
\left| \sigma,1;\sigma\right\rangle \left\langle \sigma ,1;\sigma\right|
+\rho _{\left( \sigma,1;\sigma\right) \left( \sigma,\psi;\sigma\right) }
\left| \sigma,1;\sigma\right\rangle \left\langle \sigma ,\psi;\sigma\right| \right. \notag \\
&&\left. +\rho _{\left( \sigma,\psi;\sigma\right) \left( \sigma,1;\sigma\right) }
\left| \sigma,\psi;\sigma\right\rangle \left\langle \sigma ,1;\sigma\right|
+\rho _{\left( \sigma,\psi;\sigma\right) \left( \sigma,\psi; \sigma \right) }
\left| \sigma,\psi;\sigma\right\rangle \left\langle \sigma ,1;\sigma\right| \right) \notag \\
&& \left. \quad \quad \quad \quad \quad \quad \quad \quad \quad \quad \quad \quad \quad \quad
+\rho _{\left( \sigma,\sigma;\psi\right) \left( \sigma,\sigma;\psi\right) }
\left| \sigma,\sigma;\psi\right\rangle \left\langle \sigma ,\sigma;\psi\right| \right\}
\end{eqnarray}

For $b=\sigma$ probes, we have $\mathcal{C}_{1}=\left\{ 1\right\} $,
$\mathcal{C}_{2}=\left\{ \sigma\right\} $, $\mathcal{C}_{3}=\left\{ \psi\right\} $, and%
\begin{eqnarray}
\qquad
p_{1} &=&p_{111,\sigma}^{\shortrightarrow }=\left| t_{1}\right| ^{2}\left|
r_{2}\right| ^{2}+2\left| t_{1}r_{1}r_{2}t_{2}\right| \cos \theta +\left|
r_{1}\right| ^{2}\left| t_{2}\right| ^{2} \\
p_{2} &=&p_{\sigma \sigma 1 , \sigma}^{\shortrightarrow }=\left|
t_{1}\right| ^{2}\left| r_{2}\right| ^{2}+\left| r_{1}\right| ^{2}\left|
t_{2}\right| ^{2} \\
p_{3} &=&p_{\psi \psi 1, \sigma}^{\shortrightarrow }=\left| t_{1}\right| ^{2}\left|
r_{2}\right| ^{2}-2\left| t_{1}r_{1}r_{2}t_{2}\right| \cos \theta +\left|
r_{1}\right| ^{2}\left| t_{2}\right| ^{2}
\end{eqnarray}%
\begin{equation}
\Pr\nolimits_{A}\left( 1\right) = \rho _{\left( 1,1;1\right)\left(
1,1;1\right) }
+\rho _{\left( 1,\sigma;\sigma\right)\left( 1,\sigma;\sigma\right) }
+\rho _{\left( 1,\psi;\psi\right)\left( 1,\psi;\psi\right) }
\end{equation}
\begin{eqnarray}
\qquad
\rho _{1}^{A} &=&\frac{1}{\Pr\nolimits_{A}\left( 1\right)} \left\{
\rho _{\left( 1,1;1\right) \left(1,1;1\right) }
\left|1,1;1\right\rangle \left\langle 1,1;1\right| \right. \notag \\
&& \quad \quad \quad \quad + \frac{1}{\sqrt{2}}\rho _{\left( 1,\sigma;\sigma\right) \left(1,\sigma;\sigma\right) }
\left|1,\sigma;\sigma\right\rangle \left\langle 1,\sigma;\sigma\right| \notag \\
&& \left. \quad \quad \quad \quad \quad \quad
+\rho _{\left( 1,\psi;\psi\right) \left(1,\psi;\psi\right) }
\left|1,\psi;\psi\right\rangle \left\langle 1,\psi;\psi\right| \right\}
\end{eqnarray}%
\begin{eqnarray}
\qquad
\Pr\nolimits_{A}\left( 2\right) &=&
\rho _{\left( \sigma,1;\sigma\right)\left( \sigma,1;\sigma\right) }
+\rho _{\left( \sigma,\sigma;1\right)\left( \sigma,\sigma;1\right) } \notag \\
&&+\rho _{\left( \sigma,\sigma;\psi\right)\left( \sigma,\sigma;\psi\right) }
+\rho _{\left( \sigma,\psi;\sigma\right)\left( \sigma,\psi;\sigma\right) }
\end{eqnarray}%
\begin{eqnarray}
\rho _{2}^{A} &=& \frac{1}{\Pr\nolimits_{A}\left( 2\right) }\left\{ \frac{1}{\sqrt{2}}
\rho_{\left( \sigma,1;\sigma\right) \left( \sigma,1,\sigma \right) }
\left| \sigma,1;\sigma\right\rangle \left\langle \sigma,1;\sigma\right| \right. \notag \\
&& \quad \quad \quad \quad +\frac{1}{\sqrt{2}}\rho _{\left( \sigma,\psi;\sigma\right) \left(
\sigma,\psi;\sigma\right) }
\left| \sigma,\psi;\sigma\right\rangle \left\langle \sigma,\psi;\sigma\right|  \notag \\
&& \quad \quad \quad \quad +\frac{1}{2}\left( \rho _{\left( \sigma,\sigma;1\right) \left(
\sigma,\sigma;1\right) }
+\rho _{\left( \sigma,\sigma;\psi\right) \left( \sigma,\sigma;\psi\right) }\right)  \notag \\
&& \quad \quad \quad \quad \quad \left. \phantom{\frac{1}{\sqrt{2}}} \times \left[ \left| \sigma,\sigma;1\right\rangle \left\langle \sigma,%
\sigma;1\right| +\left| \sigma,\sigma;\psi\right\rangle \left\langle \sigma,\sigma%
;\psi\right| \right] \right\}
\end{eqnarray}%
\begin{equation}
\Pr\nolimits_{A}\left( 3\right) = \rho _{\left(\psi,1;\psi\right) \left( \psi,1;\psi\right) }
+\rho _{\left(\psi,\sigma;\sigma\right) \left( \psi,\sigma;\sigma\right) }
+\rho _{\left(\psi,\psi;1\right) \left( \psi,\psi;1\right) }
\end{equation}
\begin{eqnarray}
\qquad
\rho _{3}^{A} &=&\frac{1}{\Pr\nolimits_{A}\left( 3\right)} \left\{
\rho _{\left( \psi,1;\psi\right) \left(\psi,1;\psi\right) }
\left|\psi,1;\psi\right\rangle \left\langle \psi,1;\psi\right| \right. \notag \\
&& \quad \quad \quad \quad + \frac{1}{\sqrt{2}}\rho _{\left( \psi,\sigma;\sigma\right) \left(\psi,\sigma;\sigma\right) }
\left|\psi,\sigma;\sigma\right\rangle \left\langle \psi,\sigma;\sigma\right| \notag \\
&& \quad \quad \quad \quad \quad \quad
\left. +\rho _{\left( \psi,\psi;1\right) \left(\psi,\psi;1\right) }
\left|\psi,\psi;1\right\rangle \left\langle \psi,\psi;1\right| \right\}
\end{eqnarray}

We note that one can always perfectly distinguish the charges $1$ and $\psi$
with a single $b=\sigma$ probe measurement by setting the experimental
parameters such that $\left| t_{1}\right| ^{2}=\left| t_{2}\right| ^{2}=1/2$
and $\theta =\pi $, which give $p_{1}=0$ and $p_{3}=1$. 

%% file: 04-FQH_2PC.tex
\section{Fractional Quantum Hall Double Point-Contact Interferometer}
  \label{sec:FQH_2PC}

After the detailed analysis of Section~\ref{sec:MZI}, one hopes
that it has application in physical systems, and not just to the abstract
idealizations that exist in our minds. In pursuing this hope, we turn our
attention to fractional quantum Hall systems, since they represent the most
likely candidates for possessing anyons and realizing braiding statistics
of either Abelian or non-Abelian nature.

Indeed, a setup that is rather similar to the Mach-Zehnder interferometer
described in Section~\ref{sec:MZI} has been experimentally realized in a
quantum Hall system~\cite{Ji03}. This interferometer has, so far, only
achieved functionality in the integer quantum Hall regime (though, even there,
the physical observations are
not completely understood~\cite{Neder05,Neder06}), but it should be able, in
principle, to detect the presence of braiding
statistics~\cite{Kane03,Jonckheere05,Law06}, and even discern whether a system
possesses non-Abelian statistics~\cite{Feldman06}.
Unfortunately however, there is a crucial and debilitating difference between
the FQH Mach-Zehnder interferometer of~\cite{Ji03} and the Mach-Zehnder
interferometer described in Section~\ref{sec:MZI}: because of
the chiral nature of FQH edge currents, one of the detectors and its drain are
unavoidably situated \emph{inside} the central interferometry region. As a
result, probe anyons accumulate in this region, effectively altering the target
anyon's charge. This effect renders the interferometer incapable of measuring a
target charge, and hence, useless for qubit readout in topological quantum
computation.

\begin{figure}[t!]
\begin{center}
\includegraphics[scale=0.85]{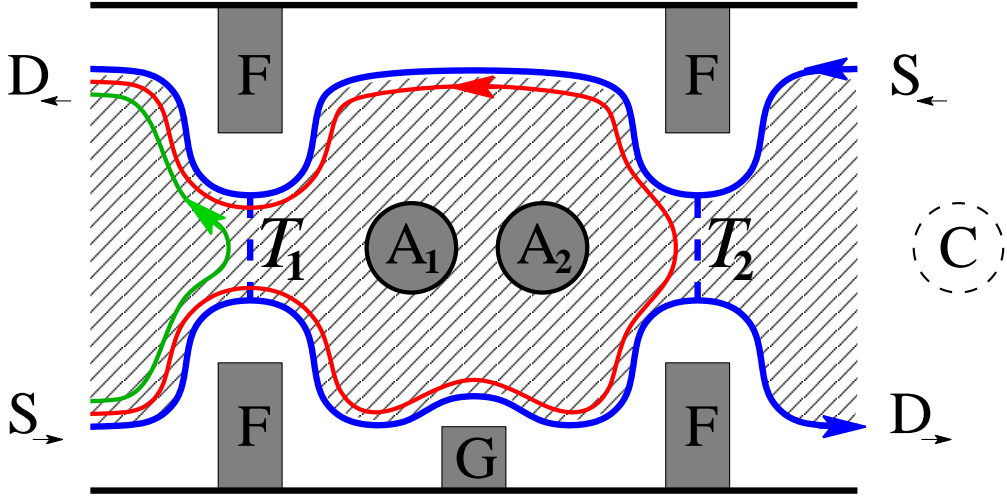}
\caption{A double point-contact interferometer for measuring braiding statistics in
fractional quantum Hall systems. The hatched region contains an incompressible
FQH liquid. $S_s$ and $D_s$ indicate the ``sources'' and ``detectors'' of edge
currents. The front gates (F) are used to bring the opposite edge currents
(indicated by arrows) close to each other to form two tunneling junctions.
Applying voltage to the central gate creates an antidot in the middle and
controls the number $n$ of quasiholes contained there. An additional side gate
(G) can be used to change the shape and the length of one of the paths in the
interferometer.}
\label{fig:two-point-interferometer}
\end{center}
\end{figure}
Fortunately, there is another type of interferometer that can be constructed
in quantum Hall systems which \emph{is} capable of measuring a target
charge: the double point-contact interferometer. Moreover, such interferometers,
which are of the Fabry--P\'{e}rot type~\cite{Fabry1897}, involving
higher orders of interference, have already achieved experimental
functionality in the fractional quantum Hall regime \cite{Goldman05a}.
The double point-contact interferometer was first proposed for use in FQH systems
in Ref.~\cite{Chamon97}, where it was analyzed for the Abelian states. It was
analyzed for the Moore--Read state~\cite{Moore91},
the most likely physical realization of non-Abelian statistics, expected to
occur at $\nu = 5/2$ and $7/2$ filling fractions, in
Refs.~\cite{Fradkin98,DasSarma05,Stern06a,Bonderson06a}. See
also~\cite{Grosfeld06b,Hou06,Fendley06a,Fendley07a,Ardonne07a,Overbosch07a,Rosenow07b} for related matters. It was
further analyzed for arbitrary anyon models, and specifically for all the
Read--Rezayi states~\cite{Read99}, in particular, the one expected to occur at $\nu = 12/5$ filling
fraction, in Ref.~\cite{Bonderson06b} (and subsequently analyzed for the
$\nu = 12/5$ Read--Rezayi state with homoplastic techniques in
Refs.~\cite{Chung06,Fidkowski07c}). In all of these previous analyses for
non-Abelian states, the results were given to lowest order in the tunneling
amplitude, and only for target anyons that were assumed to be in a state of
definite anyonic charge (i.e. already collapsed). In what follows, we provide
expressions including all orders of tunneling, both to explicitly display
the unitarity of the quantum evolution and to account for potentially
measureable corrections. Furthermore, we allow the target to be in a
superposition of different anyonic charges, and relate the results to the
analysis of Section \ref{sec:MZI}, so that we now have a proper
description of the measurement collapse behavior for these interferometers.
Experimental efforts in realization of the double point-contact interferometer
have been carried out for Abelian FQH states
\cite{Goldman05a,Camino05a,Camino05b,Camino06a,Camino07a,Camino07b}.
Whether or not these experiments have conclusively demonstrated fractional
statistics of excitations in the Abelian FQHE regime remains a topic of some
debate~\cite{Kim06e,Rosenow07a}\footnote{One of the reasons for the
uncertainty in interpreting the results of the experiments
testing the Abelian statistics in the FQH regime is the fact that the
statistical angle and the conventional Aharonov--Bohm phase
acquired by a charged quasiparticle in a magnetic field are not easy to tell
apart (this point is discussed in Refs.~\cite{Gefen91,Chamon97}).
{}From this perspective, a non-Abelian FQH state might have an advantage,
being that its effect from braiding statistics dramatically differs from the
charge-background field contribution.}.

The double point-contact interferometer consists of a quantum Hall bar with two
constrictions (point-contacts) and (at least) two antidots, $A_1$ and $A_2$,
in between them, as depicted in Fig.~\ref{fig:two-point-interferometer}.
The constrictions are created by applying voltage to the front gates (F) on
top of the Hall bar; by adjusting this voltage, one may control
the tunneling amplitudes $t_{1}$ and $t_{2}$. In the absence of inter-edge
tunneling, the gapped bulk of the FQH liquid gives rise to a quantized Hall
conductance: $G_{xy} = I/\left( V_{D_{\shortleftarrow}} -
V_{S_{\shortrightarrow}} \right) = \nu e^2/h$, where the current through the
Hall bar is $I = \left(I_{D_{\shortleftarrow}}-I_{S_{\shortrightarrow}}\right)$.
At the same time, the diagonal resistance vanishes:
$R_{xx} = \left(V_{D_{\shortleftarrow}}-V_{S_{\shortleftarrow}}\right)/I=0$.
Tunneling current between the opposite edges leads to a
deviation of $G_{xy}$ from its quantized value, or equivalently,
to the appearance of $G_{xx} \propto R_{xx}\neq 0$.  By measuring the diagonal
conductance $G_{xx}$, one effectively measures the interference between the two
tunneling paths around the antidot.  The tunneling amplitudes
$t_{1}$ and $t_{2}$ must be kept small,
to ensure that the tunneling current is completely due to quasiholes rather
than composite excitations. Treating tunneling as a perturbation, one can use
renormalization group (RG) methods to compare various contributions to
the overall current. Such analysis shows that in the weak tunneling regime,
the tunneling current at a single point-contact has the dependence $I\propto V^{4s-1}$ where $s$ is the
scaling dimension/spin of the corresponding fields/anyons
\cite{Wen92b,Fendley06a,Fendley07a}. It follows that the dominant contribution
in this regime is from the field with lowest scaling dimension, which, in FQH
systems, is the fundamental quasihole. It should be noted that the quasihole tunneling is
actually relevant in the RG sense, which, in more
physical terms, translates into the tendency of these point contacts to
become effectively pinched off in the limit of zero temperature and zero bias.
On a more mundane level, the quantum Hall liquid can be broken into separate
puddles by the introduction of a constriction due to purely electrostatic
effects (such as edges not being sufficiently sharp). In this regard, the
recent experimental evidence \cite{Miller07a}, indicating that it is possible to
construct a point-contact for which the $\nu=5/2$ state persists in the
tunneling region, is reassuring.

The two antidots are used to store two clusters of non-Abelian
quasiparticles, $A_1$ and $A_2$ respectively, whose combined anyonic charge
is being probed. The reason for two antidots, rather than just one (as
has been previously suggested 
in~\cite{Chamon97,Fradkin98,Stern06a,Bonderson06a,Bonderson06b}), is to allow
for the combined target to maintain a coherent superposition of
anyonic charges without decoherence from energetics that become important
at short range. In particular, the energy splitting between the
states of different anyonic charge on an antidot is expected to scale as
$L^{-1}$ (where $L$ is the linear size of the dot) due to both kinetic
(different angular momentum) and potential (different Coulomb energy) effects
\cite{Bonderson06a}. On the other hand, for two separated antidots, this
energy difference should vanish exponentially with the distance between
them, with suppression determined by the gap \cite{DasSarma05}.

In order to appropriately examine the resulting interference patterns, we
envision several experimentally variable parameters: (i) the central gate
voltages allowing one to control the number of quasiholes on the antidots,
(ii) the perpendicular magnetic field, (iii) the back gate voltage
controlling the uniform electron density, and (iv) a side gate (G) that can be
used to modify the shape of the edge (and, hence, total area and background flux
within) the central interferometry region. The reason for proposing all
these different controls is to be able to
separately vary the Abelian Aharonov-Bohm phase and the number of quasiholes
on the antidots. In fact, having all these different controls may turn out
to be redundant, but they may prove beneficial for experimental success.

The target anyon $A$, is the combination of the anyons
$A_{1}$, $A_{2}$, and all others (including strays) situated inside the central
interferometry region. In general, any edge excitation qualifies as a probe
anyon, but since tunneling is dominated by the fundamental quasiholes, we can
effectively allow the probes to have definite anyonic charge $b$ equal to that
of the fundamental quasihole. Letting $\left( 1,0\right) $ and
$\left( 0,1\right) $\ correspond to the top and bottom edge, respectively (also
denoted as $s=\shortleftarrow ,\shortrightarrow $, respectively), the unitary evolution
operator for a probe anyon $B$ entering the system along the edge is given by%
\begin{eqnarray}
U =\left[
\begin{array}{cc}
r_{1}^{\ast }r_{2}^{\ast }e^{i\theta _{\text{I}}}R_{AB}W_{AB}R_{CB} & \frac{1}{%
t_{1}^{\ast }}\left( 1-\left| r_{1}\right| ^{2}W_{BA}\right)  \\
R_{BC}\frac{1}{t_{2}}\left( -1+\left| r_{2}\right| ^{2}W_{AB}\right) R_{CB}
& r_{1}r_{2}e^{i\theta _{\text{II}}}R_{BC}R_{BA}W_{BA}%
\end{array}
\right]
,
\label{eq:unitary-evolution}
\end{eqnarray}%
when the $C$ anyons (those outside the central interferometry region that are
entangled with $A$) are in the region to the right of central, where we have
defined%
\begin{eqnarray}
\qquad \qquad \qquad
W_{AB} &=&\sum\limits_{n=0}^{\infty }\left( -t_{1}^{\ast }t_{2}e^{i\left(
\theta _{\text{I}}+\theta _{\text{II}}\right) }R_{BA}R_{AB}\right) ^{n}  \notag \\
&=&\left[ 1+t_{1}^{\ast }t_{2}e^{i\left( \theta _{\text{I}}+\theta _{\text{II}}\right)
}R_{BA}R_{AB}\right] ^{-1}
.
\end{eqnarray}%
The phases $\theta _{\text{I}}$ and $\theta _{\text{II}}$ are respectively
picked up from traveling
counter-clockwise along the top and bottom edge around the central
interferometry region, and include the contribution from the enclosed background
magnetic field. We note that when higher order terms are significant, it might
be the case that tunneling contributions from excitations other than the
fundamental quasiholes (which have different tunneling amplitudes) are also
important, but nevertheless proceed with considering all orders of tunneling in
this manner. The tunneling matrices are%
\begin{equation}
T_{j}=\left[
\begin{array}{cc}
r_{j}^{\ast } & t_{j} \\
-t_{j}^{\ast } & r_{j}%
\end{array}%
\right]
\end{equation}%
with $j=1,2$ for the left and right point contacts, respectively. We can
perform a similar density matrix calculation as for the Mach-Zehnder
interferometer, except with more complicated diagrams in this case. Sending
a single probe particle in from the bottom edge ($s=\shortrightarrow $) (which is
effectively done by applying a bias voltage across the edges), and detecting it
coming out at the bottom or top edge gives the same form for the resulting
density matrix as in Eq.~(\ref{eq:rho_n}), except with more complicated
$p_{aa^{\prime }e,b}^{s}$ that are determined by using $U$ of
Eq.~(\ref{eq:unitary-evolution}) for $VU$ in
Eqs.~(\ref{eq:rho_VU}--\ref{eq:measurment_projection}). To order
$\left|t\right| ^{2}$ (for $\left| t_{1}\right| \sim \left| t_{2}\right| $
small), we find%
\begin{eqnarray}
\quad
p_{aa^{\prime }e,b}^{\shortrightarrow } &\simeq &\left| r_{1}\right| ^{2}\left|
r_{2}\right| ^{2}\left( 1-t_{1}^{\ast }t_{2}e^{i\left( \theta _{\text{I}}+\theta
_{\text{II}}\right) }M_{ab}-t_{1}t_{2}^{\ast }e^{-i\left( \theta _{\text{I}}+\theta
_{\text{II}}\right) }M_{a^{\prime }b}^{\ast }\right)  \nonumber \\
&\simeq &1-\left| t_{1}\right| ^{2}-\left| t_{2}\right| ^{2}-\left|
t_{1}t_{2}\right| \left( e^{i\beta }M_{ab}+e^{-i\beta }M_{a^{\prime }b}^{\ast }\right)
\end{eqnarray}%
and%
\begin{eqnarray}
\qquad
p_{aa^{\prime }e,b}^{\shortleftarrow } &\simeq &\left| t_{1}\right| ^{2}+\left|
r_{1}\right| ^{2}t_{1}^{\ast }t_{2}e^{i\left( \theta _{\text{I}}+\theta
_{\text{II}}\right) }M_{ab}  \notag \\
&&+\left| r_{1}\right| ^{2}t_{1}t_{2}^{\ast }e^{-i\left( \theta _{\text{I}}+\theta
_{\text{II}}\right) }M_{a^{\prime }b}^{\ast }+\left| r_{1}\right| ^{4}\left|
t_{2}\right| ^{2}M_{eb} \nonumber \\
&\simeq &\left| t_{1}\right| ^{2}+\left| t_{1}t_{2}\right| \left( e^{i\beta }
M_{ab}+e^{-i\beta }M_{a^{\prime }b}^{\ast }\right)+\left|
t_{2}\right| ^{2}M_{eb}
\end{eqnarray}%
where we have defined $\beta =\arg \left\{ t_{1}^{\ast }t_{2}e^{i\left(
\theta _{\text{I}}+\theta _{\text{II}}\right) }\right\} $. We see that%
\begin{equation}
p_{aa^{\prime }e,b}^{\shortrightarrow } + p_{aa^{\prime }e,b}^{\shortleftarrow } \simeq
\left| t_{2}\right| ^{2} M_{eb} + \left|r_{2}\right| ^{2}
.
\end{equation}%
Here we have $\left| t_{2}\right| ^{2}$ as the probability of the probe
$B$ passing between anyons $A$ and $C$, rather than $\left| t_{1}\right| ^{2}$, as in the
case analyzed for the Mach-Zehnder interferometer, because of the location of $C$.
The values for the two outcome probabilities (i.e. the $e=1$ terms) to all
orders are%
\begin{eqnarray}
p_{aa1,b}^{\shortrightarrow } &=&\sum\limits_{c}N_{ab}^{c}\frac{d_{c}}{d_{a}d_{b}}%
\frac{\left| r_{1}\right| ^{2}\left| r_{2}\right| ^{2}}{\left| 1+t_{1}^{\ast
}t_{2}e^{i\left( \theta _{\text{I}}+\theta _{\text{II}}\right) }e^{i2\pi \left(
s_{c}-s_{a}-s_{b}\right) }\right| ^{2}} \nonumber \\
&=&\sum\limits_{c}N_{ab}^{c}\frac{d_{c}}{d_{a}d_{b}}\frac{\left|
r_{1}\right| ^{2}\left| r_{2}\right| ^{2}}{1+\left| t_{1}\right| ^{2}\left|
t_{2}\right| ^{2}+2\left| t_{1}t_{2}\right| \cos \left[ \beta +2\pi \left(
s_{c}-s_{a}-s_{b}\right) \right] } \\
&\simeq &1-\left| t_{1}\right| ^{2}-\left| t_{2}\right| ^{2}-2\left|
t_{1}t_{2}\right| \text{Re}\left\{ e^{i\beta }M_{ab}\right\}  \\
p_{aa1,b}^{\shortleftarrow } &=&1-p_{aa1,b}^{\shortrightarrow }.
\end{eqnarray}%
These are also the values of $p_{aa^{\prime }e,b}^{s}$ to all orders when $%
M_{eb}=1$, but in general $p_{aa^{\prime }e,b}^{s}$ does not have such a
nice form. One may also perform edge current tunneling calculations employing the conformal field
theoretic description of the edge modes in order to determine the effects of the source-drain voltage, the separation length between the two point-contacts, and the temperature~\cite{Chamon97} (see also~\cite{Wen92b,Fendley06a,Fendley07a,Ardonne07a,Fidkowski07c}).
The result of such considerations is essentially an interference suppressing $Q$-factor [recall the discussion following Eq.~(\ref{eq:visibility})] that decreases (with modulation) as any of these three
quantities increase. There may be additional sources of interference suppression, such as
switching noise~\cite{Grosfeld06b} or edge-bulk tunneling~\cite{Overbosch07a,Rosenow07b}. Though the suppression
factor is only close to $Q=1$ in certain regimes, we will
again ignore it, but keep its existence in the back of our minds.

As before, the target system collapses onto states with common
values of $p_{aa1,b}^{\shortrightarrow }$, generically producing a density matrix
with non-zero elements that correspond to difference charges $e$ with $%
M_{eb}=1$ (and $M_{ab}=M_{a^{\prime }b}$). To first order, the behavior is
essentially identical to that of the Mach-Zehnder interferometer which we
previously obtained, but the higher order terms may require more stringent
conditions for superpositions to survive measurement collapse than just
indistinguishability of monodromy scalar components (since this only guarantees
proper matching to first order). Specifically, for superpositions of $a$ and
$a^{\prime }$ to survive, they must have%
\begin{equation}
\sum\limits_{c}N_{ab}^{c}\frac{d_{c}}{d_{a}}\left( \frac{\theta _{c}}{\theta
_{a}}\right) ^{n}=\sum\limits_{c}N_{a^{\prime }b}^{c}\frac{d_{c}}{%
d_{a^{\prime }}}\left( \frac{\theta _{c}}{\theta _{a^{\prime }}}\right) ^{n}
\end{equation}%
for all $n$, and some much more cumbersome condition for the survival of
coherent superpositions corresponding to difference charge $e$. However,
it seems that this condition is often equivalent to
indistinguishability of monodromy scalar components for models of interest.
In order to have $p_{aa1,b}^{\shortleftarrow }=0$, i.e. producing
sometimes perfect distinguishability\footnote{One can never have \emph{always}
perfect distinguishability for this interferometer, since it must be in the
weak tunneling limit, which prevents ever having
$\left| t_{j}\right|^{2}=1/2$.}, we require equal tunneling probabilities
$\left| t_{1}\right| =\left| t_{2}\right| $, and $\cos \left[
\beta +2\pi \left( s_{c}-s_{a}-s_{b}\right) \right] =-1$ for all $%
N_{ab}^{c}\neq 0$. In Eq.~(\ref{eq:N_est}), we obtained an estimate for the total number of probes,
$N\gtrsim \left( \frac{z_{\alpha /2}^{\ast }}{t\Delta M}\right) ^{2}$, needed to collapse
and distinguish a superposition of two anyonic charges in the target, with some level of
confidence $1-\alpha$. Translating this into the amount of time $\tau$ necessary for such
a measurement, we get the estimate $\tau \gtrsim \frac{e}{ \left| I_{\text{tot}} \right| } \left( \frac{z_{\alpha /2}^{\ast }}{t\Delta M}\right) ^{2}$, where $I_{\text{tot}}$ is the total edge current.

{}From these results, we find that when the target is in a state of
definite charge $a$ (or, more generally, in a fixed state $\rho_{\kappa}$ with $a \in \mathcal{C}_{\kappa}$), the longitudinal conductance will be proportional to the
probability of the probe injected along the bottom edge to be ``detected''
exiting along the top edge:%
\begin{equation}
G _{xx}\varpropto p_{aa1,b}^{\shortleftarrow }\simeq \left| t_{1}\right|
^{2}+\left| t_{2}\right| ^{2}+2\left| t_{1}t_{2}\right| \text{Re}\left\{
e^{i\beta }M_{ab}\right\}
\end{equation}%
which is exactly Eq.~(7) in Ref.~\cite{Bonderson06b}. This is an experimentally
measurable quantity, found by measuring the voltage between
$S_{\shortrightarrow}$ and $D_{\shortleftarrow}$. Using the side gate (G), one
can vary $\beta$ and, from the resulting modulation in the conductance,
determine the amplitude of $M_{ab}$. Indeed, the measurement of this quantity may be
used to help properly identify the topological order of an unknown physical state.

\subsection{Predictions for FQH States}

The results of this section are applicable to any FQH state. Because of their relative significance, we
will only give the explicit details here for the Abelian hierarchy states, the Moore--Read state, and
the (particle-hole conjugate of the) $k=3,M=1$ Read--Rezayi state. The anyon models of these FQH states may be easily described in terms of those given in Section~\ref{sec:Examples}. The application of this section's results to the entire Read--Rezayi series and the NASS states of~\cite{Ardonne99} may be found in~\cite{Bonderson07b} (and partially in~\cite{Bonderson06b}). In all these FQH examples, it is important to remember that the electric charge and anyonic charge are coupled through Abelian terms for FQH states. Consequently, superselection of electric charge only permits superposition of anyonic charges that correspond to the same electric charge (i.e. are composed of the same number of fundamental quasiholes).

\subsubsection{The Abelian Hierarchy States $\left(\nu = n/m\right)$}

The Abelian fractional quantum Hall states can all be constructed from
$\mathbb{Z}_{N}$ models. The general formulation in terms of $K$
matrices may be found in \cite{Wen92a}, but we will describe the Laughlin
and hierarchy states \cite{Laughlin83,Haldane83,Halperin84,Jain89} that occur at
filling fractions $\nu = n/m$ (with $m$ odd and $n<m$). As shown in~\cite{Moore91}, the statistical factor of the fundamental quasihole in these states is
$\theta = \frac{ \pi p}{m}$ where $p$ is odd and $np \equiv 1$~mod~$m$ (which
uniquely defines $p$ modulo $2m$). It follows that these states are described by
$\mathbb{Z}_{2m}^{\left( p\right) }$ of Section~\ref{sec:Z_N}, in
which a fundamental quasihole has anyonic charge $\left[ 1\right] _{2m}$ and
electric charge $e/m$, while an electron (which has electric charge $-e$) has
anyonic charge $\left[ m\right] _{2m}$. Using $b=\left[ 1\right] _{2m}$ probes, we have
\begin{eqnarray}
\qquad \qquad \quad
p_{aa0,b}^{\shortleftarrow } &=&1-\frac{\left| r_{1}\right| ^{2}\left|
r_{2}\right| ^{2}}{\left| 1+\left| t_{1}t_{2}\right| e^{i\left( \beta +n%
\frac{2 \pi p }{m}\right) }\right| ^{2}} \\
&\simeq &\left| t_{1}\right| ^{2}+\left| t_{2}\right| ^{2}+2\left|
t_{1}t_{2}\right| \cos \left( \beta +n\frac{2 \pi p }{m}\right)
,
\end{eqnarray}%
and all permissible states are automatically fixed states with the target system's
anyons $A$ and $C$ having definite charge%
\begin{equation}
\rho _{a}^{A} = \left|a,c; \left[ a+c\right] _{2m}\right\rangle \left\langle a,c; \left[ a+c\right] _{2m}\right|.
\end{equation}

\subsubsection{The Moore--Read State $\left( \nu=5/2,7/2 \right)$}

The anyon model corresponding to the Moore--Read state, expected to describe the
$\nu=5/2,7/2$ plateaus, is given by~\cite{Bonderson07b}:%
\begin{eqnarray}
\qquad \qquad \quad
\text{MR} &=&\left. \text{Ising}\times \mathbb{Z}_{8}^{\left(
1/2\right) }\right| _{\mathcal{C}} \notag \\
\mathcal{C} &=&\left\{ \left( 1,\left[ 2m\right]
_{8}\right) ,\left( \sigma ,\left[ 2m+1\right] _{8}\right) ,\left( \psi ,%
\left[ 2m\right] _{8}\right) \right\}
\label{eq:MR}
\end{eqnarray}%
(for $m \in \mathbb{Z}$), the restriction of the direct product of the Ising and
$\mathbb{Z}_{8}^{\left(1/2\right)}$ anyon models to the charge spectrum
$\mathcal{C} \subset \mathcal{C}_{\text{Ising}} \times \mathcal{C}_{\mathbb{Z}_{8}^{\left(
1/2\right) }}$ in which the $1$ and $\psi$ Ising charges are paired with the even sector of $\mathbb{Z}_{8}^{\left(1/2\right)}$ and the $\sigma$ Ising charge is paired with the odd sector
of $\mathbb{Z}_{8}^{\left(1/2\right)}$. Writing
$a = \left( a_{\text{I}},a_{\mathbb{Z}} \right) \in \mathcal{C}$,
where $a_{\text{I}} \in \mathcal{C}_{\text{Ising}}$ and
$a_{\mathbb{Z}} \in \mathcal{C}_{\mathbb{Z}_{8}^{\left(1/2\right) }}$, this is more explicitly given by%
\begin{equation*}
\begin{tabular}{|l|l|l|}
\hline
\multicolumn{3}{|l|}{$\mathcal{C} =\left\{ \left( 1,\left[ 2m\right]
_{8}\right) ,\left( \sigma ,\left[ 2m+1\right] _{8}\right) ,\left( \psi ,%
\left[ 2m\right] _{8}\right) \right\}, \quad N_{ab}^{c} = N_{a_{\text{I}} b_{\text{I}} }^{c_{\text{I}} } N_{a_{\mathbb{Z}} b_{\mathbb{Z}} }^{c_{\mathbb{Z}} }$} \\ \hline
\multicolumn{3}{|l|}{$\left[ F_{d}^{abc}\right] _{ef}=
\left[ F_{d_{\text{I}}}^{a_{\text{I}}b_{\text{I}}c_{\text{I}}}\right] _{e_{\text{I}}f_{\text{I}}}
\left[ F_{d_{\mathbb{Z}}}^{a_{\mathbb{Z}}b_{\mathbb{Z}}c_{\mathbb{Z}}}\right] _{e_{\mathbb{Z}}f_{\mathbb{Z}}} , \quad
\left[ F_{cd}^{ab}\right] _{ef}=
\left[ F_{c_{\text{I}}d_{\text{I}}}^{a_{\text{I}}b_{\text{I}}}\right] _{e_{\text{I}}f_{\text{I}}}
\left[ F_{c_{\mathbb{Z}}d_{\mathbb{Z}}}^{a_{\mathbb{Z}}b_{\mathbb{Z}}}\right] _{e_{\mathbb{Z}}f_{\mathbb{Z} \phantom{j}}} $} \\ \hline
$R_{c}^{ab}=R_{c_{\text{I}}}^{a_{\text{I}}b_{\text{I}}}R_{c_{\mathbb{Z}}}^{a_{\mathbb{Z}}b_{\mathbb{Z}}}$ &
$S_{ab}=\sqrt{2}S_{a_{\text{I}}b_{\text{I}}} S_{a_{\mathbb{Z}}b_{\mathbb{Z}}}$ &
$M_{ab}=M_{a_{\text{I}}b_{\text{I}}} M_{a_{\mathbb{Z}}b_{\mathbb{Z}}}$ \\ \hline
\multicolumn{2}{|l|} { $d_{a} = d_{ a_{\text{I}}}, \quad \mathcal{D}=4$ } &
$\theta _{a} = \theta_{a_{\text{I}}} \theta_{a_{\mathbb{Z}}} $ \\ \hline
\end{tabular}%
\end{equation*}%
where all the symbols ($N,F,R,S,M,d,\theta$) labeled by subscript $\text{I}$ charges are those of
the Ising model given in the table of Section~\ref{sec:Ising}, and those with subscript
$\mathbb{Z}$ charges are those of the $\mathbb{Z}_{8}^{\left(1/2\right)}$ anyon model given in
the table of Section~\ref{sec:Z_N}. The factor of $\sqrt{2}$ arises in front of the product of $S$-matrices because of the restriction of the
charge spectrum. The fundamental quasihole has anyonic charge
$\left( \sigma,\left[ 1\right] _{8}\right) $ and electric charge $e/4$. The
electron has anyonic charge $\left( \psi ,\left[ 4\right] _{8}\right) $.

The target system's anyons $A$ and $C$ may by thought of as composed of $n$ and $m$
fundamental quasiholes, respectively (with electric charges $ne/4$ and $me/4$), where
$n,m$ are integers (possibly negative).
If $n$ is even, the target's total anyonic charge may be in some superposition of
$\left( 1, \left[ n \right]_{8} \right) $ and
$\left( \psi, \left[ n \right]_{8} \right) $. If $n$ is odd, the target anyon
has total anyonic charge $\left( \sigma, \left[ n \right]_{8} \right) $. The same holds for
the target's entangled partner $C$. We will employ the shorthand
$ x_{n} = \left( x, \left[ n \right]_{8} \right) $, with $x=1,\sigma,\psi$. For probe anyons that are fundamental quasiholes,
$b=\sigma_{1}$, this gives%
\begin{eqnarray}
\qquad \qquad
p_{1_{n}1_{n}1_{0},\sigma_{1}}^{\shortleftarrow } &=&1-\frac{\left| r_{1}\right| ^{2}\left|
r_{2}\right| ^{2}}{\left| 1+ \left|
t_{1}t_{2}\right| e^{i\left( \beta +n\frac{\pi }{4}\right) }\right| ^{2}} \\
&\simeq &\left| t_{1}\right| ^{2}+\left| t_{2}\right| ^{2}+
2\left| t_{1}t_{2}\right| \cos \left( \beta +n\frac{\pi }{4} \right) \\
p_{\psi_{n}\psi_{n}1_{0},\sigma_{1}}^{\shortleftarrow } &=&1-\frac{\left| r_{1}\right| ^{2}\left|
r_{2}\right| ^{2}}{\left| 1- \left|
t_{1}t_{2}\right| e^{i\left( \beta +n\frac{\pi }{4}\right) }\right| ^{2}} \\
&\simeq &\left| t_{1}\right| ^{2}+\left| t_{2}\right| ^{2}-
2\left| t_{1}t_{2}\right| \cos \left( \beta +n\frac{\pi }{4} \right) \\
p_{\sigma_{n}\sigma_{n}1_{0},\sigma_{1}}^{\shortleftarrow } &=&1-\frac{\left| r_{1}\right| ^{2}\left|
r_{2}\right| ^{2}\left( 1+\left| t_{1}t_{2}\right| ^{2}\right) }{\left|
1-\left( -1\right) ^{\frac{n-1}{2}}\left| t_{1}t_{2}\right| ^{2}e^{i2\beta
}\right| ^{2}} \\
&\simeq &\left| t_{1}\right| ^{2}+\left| t_{2}\right| ^{2}-2\left|
t_{1}t_{2}\right| ^{2}\left[ 1+\left( -1\right) ^{\frac{n-1}{2}}\cos \left(
2\beta \right) \right] .
\end{eqnarray}%
With these probes (and electric charge superselection), all charges are distinguishable,
so the charge classes $\mathcal{C}_{\kappa}$
are all singletons and interferometry will collapse any superposition of charge in the
target onto a definite charge state. Of specific note is that for $n$ odd, the interference is suppressed.
The leading order modulation occurs at fourth order in $t$, and has the twice the modulation frequency.
In fact, higher
order harmonics enter as modulations in $2j\beta $ that are $4j^{th}$ order in
$t$. When $m$ is even, the probabilities and fixed states are%
\begin{equation}
\Pr\nolimits_{A}\left( \kappa_{1_{n}}\right) =
\rho _{\left( 1_{n},1_{m}; 1_{n+m}\right) \left( 1_{n},1_{m}; 1_{n+m}\right) }
+\rho _{\left( 1_{n},\psi_{m}; \psi_{n+m}\right) \left( 1_{n},\psi_{m}; \psi_{n+m}\right) }
\end{equation}%
\begin{eqnarray}
\rho _{\kappa_{1_{n}}}^{A} &=&\frac{1}{\Pr\nolimits_{A}\left( \kappa_{1_{n}}\right)} \left\{
\rho _{\left( 1_{n},1_{m}; 1_{n+m}\right) \left( 1_{n},1_{m}; 1_{n+m}\right) }
\left|1_{n},1_{m}; 1_{n+m}\right\rangle \left\langle 1_{n},1_{m}; 1_{n+m}\right| \right. \notag \\
&& \left. \quad
+\rho _{
\left( 1_{n},\psi_{m}; \psi_{n+m}\right) \left( 1_{n},\psi_{m}; \psi_{n+m}\right) }
\left|1_{n},\psi_{m}; \psi_{n+m}\right\rangle \left\langle 1_{n},\psi_{m}; \psi_{n+m}\right| \right\}
\end{eqnarray}%
\begin{equation}
\Pr\nolimits_{A}\left( \kappa_{\sigma_{n}}\right) =
\rho _{\left( \sigma_{n},1_{m}; \sigma_{n+m}\right) \left( \sigma_{n},1_{m}; \sigma_{n+m}\right) }
+\rho _{\left( \sigma_{n},\psi_{m}; \sigma_{n+m}\right) \left( \sigma_{n},\psi_{m}; \sigma_{n+m}\right) }
\end{equation}%
\begin{eqnarray}
\rho _{\kappa_{\sigma_{n}}}^{A} &=& \frac{1}{\Pr\nolimits_{A}\left( \kappa_{\sigma_{n}}\right) }
\frac{1}{\sqrt{2}} \notag \\
&& \times \left\{
\rho _{\left( \sigma_{n},1_{m}; \sigma_{n+m}\right) \left( \sigma_{n},1_{m}; \sigma_{n+m}\right) }
\left| \sigma_{n},1_{m}; \sigma_{n+m}\right\rangle \left\langle \sigma_{n},1_{m}; \sigma_{n+m}\right| \right. \notag \\
&& \quad \left. +
\rho _{\left( \sigma_{n},\psi_{m}; \sigma_{n+m}\right) \left( \sigma_{n},\psi_{m}; \sigma_{n+m}\right) }
\left| \sigma_{n},\psi_{m}; \sigma_{n+m}\right\rangle \left\langle \sigma_{n},\psi_{m}; \sigma_{n+m}\right|  \right\}
\end{eqnarray}%
\begin{equation}
\Pr\nolimits_{A}\left( \kappa_{\psi_{n}}\right) =
\rho _{\left( \psi_{n},1_{m}; \psi_{n+m}\right) \left( \psi_{n},1_{m}; \psi_{n+m}\right) }
+\rho _{\left( \psi_{n},\psi_{m}; 1_{n+m}\right) \left( \psi_{n},\psi_{m}; 1_{n+m}\right) }
\end{equation}%
\begin{eqnarray}
\rho _{\kappa_{\psi_{n}}}^{A} &=&\frac{1}{\Pr\nolimits_{A}\left( \kappa_{\psi_{n}}\right)} \left\{
\rho _{\left( \psi_{n},1_{m}; \psi_{n+m}\right) \left( \psi_{n},1_{m}; \psi_{n+m}\right) }
\left|\psi_{n},1_{m}; \psi_{n+m}\right\rangle \left\langle \psi_{n},1_{m}; \psi_{n+m}\right| \right. \notag \\
&& \left. \quad
+\rho _{
\left( \psi_{n},\psi_{m}; 1_{n+m}\right) \left( \psi_{n},\psi_{m}; 1_{n+m}\right) }
\left|\psi_{n},\psi_{m}; 1_{n+m}\right\rangle \left\langle \psi_{n},\psi_{m}; 1_{n+m}\right| \right\}
\end{eqnarray}%
and when $m$ is odd, they are%
\begin{equation}
\Pr\nolimits_{A}\left( \kappa_{1_{n}}\right) =
\rho _{\left( 1_{n},\sigma_{m}; \sigma_{n+m}\right) \left( 1_{n},\sigma_{m}; \sigma_{n+m}\right) }
\end{equation}%
\begin{equation}
\rho _{\kappa_{1_{n}}}^{A} = \frac{1}{\sqrt{2}}
\left|1_{n},\sigma_{m}; \sigma_{n+m}\right\rangle \left\langle 1_{n},\sigma_{m}; \sigma_{n+m}\right|
\end{equation}%
\begin{equation}
\Pr\nolimits_{A}\left( \kappa_{\sigma_{n}}\right) =
\rho _{\left( \sigma_{n},\sigma_{m}; 1_{n+m}\right) \left( \sigma_{n},\sigma_{m}; 1_{n+m}\right) }
+\rho _{\left( \sigma_{n},\sigma_{m}; \psi_{n+m}\right) \left( \sigma_{n},\sigma_{m}; \psi_{n+m}\right) }
\end{equation}%
\begin{equation}
\rho _{\kappa_{\sigma_{n}}}^{A} = \frac{1}{2} \left[
\left|\sigma_{n},\sigma_{m}; 1_{n+m}\right\rangle \left\langle \sigma_{n},\sigma_{m}; 1_{n+m}\right|
+\left|\sigma_{n},\sigma_{m}; \psi_{n+m} \right\rangle \left\langle \sigma_{n},\sigma_{m}; \psi_{n+m}\right|
\right]
\end{equation}%
\begin{equation}
\Pr\nolimits_{A}\left( \kappa_{\psi_{n}}\right) =
\rho _{\left( \psi_{n},\sigma_{m}; \sigma_{n+m}\right) \left( \psi_{n},\sigma_{m}; \sigma_{n+m}\right) }
\end{equation}%
\begin{equation}
\rho _{\kappa_{\psi_{n}}}^{A} = \frac{1}{\sqrt{2}}
\left|\psi_{n},\sigma_{m}; \sigma_{n+m}\right\rangle \left\langle \psi_{n},\sigma_{m}; \sigma_{n+m}\right|
\end{equation}%

If one had a way to effectively suppress the tunneling of fundamental quasiholes,
then the next most dominant contribution to tunneling comes from excitations with anyonic charge
$b=\left( 1,\left[ 2 \right]_{8} \right) $, which are Abelian, and
give (with different values of $T_{j}$)%
\begin{eqnarray}
\qquad \qquad \quad
p_{x_{n}x_{n}1_{0},1_{2}}^{\shortleftarrow } &=&1-\frac{\left| r_{1}\right| ^{2}\left|
r_{2}\right| ^{2}}{\left| 1+\left| t_{1}t_{2}\right| e^{i\left( \beta +n%
\frac{\pi }{2}\right) }\right| ^{2}} \\
&\simeq &\left| t_{1}\right| ^{2}+\left| t_{2}\right| ^{2}+2\left|
t_{1}t_{2}\right| \cos \left( \beta +n\frac{\pi }{2}\right)
.
\end{eqnarray}%
The Ising charges are obviously indistinguishable when the probe has Ising charge $1$,
so superpositions of $1$ and $\psi$ will not be affected by these probes (i.e. all
allowed target density matrices are fixed states).
If we had sufficiently good precision and control over the experimental
variables to set them exactly to $\left| t_{1}\right| =\left| t_{2}\right| =t $ and
$\cos\left( \beta +n\frac{\pi }{4}\right) =-1$ for $a = \left(1, \left[ n \right]_{8}\right)$
and $b=\left( \sigma,\left[ 1 \right]_{8} \right) $, then we would find
$p_{aa1,b}^{\shortleftarrow }=0$\ to all orders, providing a method of suppressing
tunneling of fundamental quasiholes. (These settings would give
$p_{aa1,b}^{\shortleftarrow }=\frac{4 t^{2}}{\left( 1+ t^{2}\right) ^{2}}$\ 
for $a = \left(\psi, \left[ n \right]_{8}\right)$.)
Using this value of $\beta$ for $b=\left( 1,\left[ 2 \right]_{8} \right) $ probes gives%
\begin{eqnarray}
\qquad \qquad \quad
p_{1_{n}1_{n}1_{0},1_{2}}^{\shortleftarrow } &=&1-\frac{\left( 1- t^{2} \right) ^{2}}
{\left| 1- t^{2} e^{i n \frac{\pi }{4} }\right| ^{2}} \\
&\simeq & 2 t^{2} \left[ 1 - \cos \left( n \pi / 4 \right) \right]
.
\end{eqnarray}
Again, tunneling of these probes will be suppressed when $n \equiv 0$~mod~$8$, and the next
most dominant tunneling contribution will be from $b=\left( \psi,\left[ 2 \right]_{8} \right) $, giving 
$p_{1_{n}1_{n}1_{0},\psi_{2}}^{\shortleftarrow } = 4t^{2}$.

\subsubsection{The Read--Rezayi State $\left(\nu=12/5\right)$}

The anyon model corresponding to the state expected to describe the $\nu =12/5$ plateau is~\cite{Bonderson07b}:%
\begin{equation}
\overline{\text{RR}}_{3,1} = \overline{\text{Fib}} \times \mathbb{Z}_{10}^{\left( 3\right) },
\end{equation}%
the particle-hole conjugate of the $k=3,M=1$ Read--Rezayi state\footnote{In general, the anyon models
describing the $k,M$ odd Read--Rezayi states may be written neatly as the direct product of anyon models
$\text{RR}_{k,M} = \text{SO}\left(3\right)_{k} \times
\mathbb{Z}_{2 \left( kM+2 \right)}^{\left( \left( k \left( kM+2 \right) - M \right)/2 \right)}$~\cite{Bonderson07b}.}. Writing
$a = \left( a_{\text{F}},a_{\mathbb{Z}} \right) \in \mathcal{C}$,
where $a_{\text{F}} \in \mathcal{C}_{\overline{\text{Fib}}}$ and
$a_{\mathbb{Z}} \in \mathcal{C}_{\mathbb{Z}_{10}^{\left(3\right) }}$, this is more explicitly given by%
\begin{equation*}
\begin{tabular}{|l|l|l|}
\hline
\multicolumn{3}{|l|}{$\mathcal{C} =\left\{ \left( 1,\left[ n\right]
_{10}\right) ,\left( \varepsilon,\left[ n\right] _{10}\right) \right\}, \quad N_{ab}^{c} = N_{a_{\text{F}} b_{\text{F}} }^{c_{\text{F}} } N_{a_{\mathbb{Z}} b_{\mathbb{Z}} }^{c_{\mathbb{Z}} }$} \\ \hline
\multicolumn{3}{|l|}{$\left[ F_{d}^{abc}\right] _{ef}=
\left[ F_{d_{\text{F}}}^{a_{\text{F}}b_{\text{F}}c_{\text{F}}}\right] _{e_{\text{F}}f_{\text{F}}}
\left[ F_{d_{\mathbb{Z}}}^{a_{\mathbb{Z}}b_{\mathbb{Z}}c_{\mathbb{Z}}}\right] _{e_{\mathbb{Z}}f_{\mathbb{Z}}} , \quad
\left[ F_{cd}^{ab}\right] _{ef}=
\left[ F_{c_{\text{F}}d_{\text{F}}}^{a_{\text{F}}b_{\text{F}}}\right] _{e_{\text{F}}f_{\text{F}}}
\left[ F_{c_{\mathbb{Z}}d_{\mathbb{Z}}}^{a_{\mathbb{Z}}b_{\mathbb{Z}}}\right] _{e_{\mathbb{Z}}f_{\mathbb{Z} \phantom{j}}} $} \\ \hline
$R_{c}^{ab}=R_{c_{\text{F}}}^{a_{\text{F}}b_{\text{F}}}R_{c_{\mathbb{Z}}}^{a_{\mathbb{Z}}b_{\mathbb{Z}}}$ &
$S_{ab}=S_{a_{\text{F}}b_{\text{F}}} S_{a_{\mathbb{Z}}b_{\mathbb{Z}}}$ &
$M_{ab}=M_{a_{\text{F}}b_{\text{F}}} M_{a_{\mathbb{Z}}b_{\mathbb{Z}}}$ \\ \hline
\multicolumn{2}{|l|} { $d_{a} = d_{ a_{\text{F}}}, \quad \mathcal{D}=\sqrt{10\left(\phi + 2\right)}$ } &
$\theta _{a} = \theta_{a_{\text{F}}} \theta_{a_{\mathbb{Z}}} $ \\ \hline
\end{tabular}%
\end{equation*}%
where $n \in \mathbb{Z}$, and all the symbols ($N,F,R,S,M,d,\theta$) labeled by subscript $\text{F}$ charges are those of
the $\overline{\text{Fib}}$ given as the complex conjugate of those in the table of Section~\ref{sec:Fib}, and those with subscript $\mathbb{Z}$ charges are those of the $\mathbb{Z}_{10}^{\left(3\right)}$ anyon model given in
the table of Section~\ref{sec:Z_N}. The fundamental quasihole has anyonic charge
$\left( \varepsilon ,\left[ 1\right]_{10}\right) $ and electric charge $e/5$.
The electron has anyonic charge $\left( 1,\left[ 5\right] _{10}\right) $. Being
a direct product of $\overline{\text{Fib}}$ and an Abelian theory, universal topological quantum
computation could be achieved through braiding quasiholes of this system.

The target system's anyons $A$ and $C$ carrying total electric charges $ne/5$ and $me/5$, respectively,
have $a_{\mathbb{Z}}=\left[ n\right]_{10}$ and $c_{\mathbb{Z}}=\left[ m\right]_{10}$, and may
be in superpositions of $a_{\text{F}},c_{\text{F}} = 1,\varepsilon $. We employ the shorthand
(similar to before) $ x_{n} = \left( x, \left[ n \right]_{10} \right) $, with $x=1,\varepsilon$. For probe anyons that are fundamental quasiholes,
$b=\varepsilon_{1}$, this gives%
\begin{eqnarray}
\qquad \quad
p_{1_{n}1_{n}1_{0},\varepsilon_{1}}^{\shortleftarrow } &\simeq& \left| t_{1}\right| ^{2}
+\left| t_{2}\right|^{2}+2\left| t_{1}t_{2}\right| \cos \left( \beta -n\frac{4\pi }{5}\right) \\
p_{\varepsilon_{n}\varepsilon_{n}1_{0},\varepsilon_{1}}^{\shortleftarrow } &\simeq&
\left| t_{1}\right| ^{2}+\left| t_{2}\right|^{2}- 2 \phi ^{-2} \left| t_{1}t_{2}\right|
\cos \left( \beta -n\frac{4\pi }{5}\right)
.
\end{eqnarray}%
With these probes (and electric charge superselection), all charges are distinguishable,
so the charge classes $\mathcal{C}_{\kappa}$ are all singletons and interferometry will collapse any superposition of charge in the target onto a definite charge state. We note that when the target has $\overline{\text{Fib}}$ charge $\varepsilon$, the interference is suppressed, though still second order in $t$. The probabilities and fixed states are%
\begin{equation}
\Pr\nolimits_{A}\left( \kappa_{1_{n}}\right) =
\rho _{\left( 1_{n},1_{m};1_{n+m}\right)\left( 1_{n},1_{m};1_{n+m}\right) }
+ \rho _{\left( 1_{n},\varepsilon_{m};\varepsilon_{n+m}\right)\left( 1_{n},\varepsilon_{m};\varepsilon_{n+m}\right) }
\end{equation}%
\begin{eqnarray}
\rho _{1}^{A} &=&\frac{1}{\Pr\nolimits_{A}\left( \kappa_{1_{n}}\right) }\left\{
\rho _{\left(1_{n},1_{m};1_{n+m}\right) \left( 1_{n},1_{m};1_{n+m}\right) }
\left| 1_{n},1_{m};1_{n+m}\right\rangle \left\langle 1_{n},1_{m};1_{n+m}\right| \right.  \notag \\
&&\left. \quad \quad +\phi ^{-1}\rho _{\left( 1_{n},\varepsilon_{m};\varepsilon_{n+m}\right)
\left( 1_{n},\varepsilon_{m};\varepsilon_{n+m}\right) }
\left| 1_{n},\varepsilon_{m};\varepsilon_{n+m}\right\rangle \left\langle 1_{n},\varepsilon_{m};\varepsilon_{n+m}\right|
\right\}
\end{eqnarray}%
\begin{eqnarray}
\Pr\nolimits_{A}\left( \kappa_{\varepsilon_{n}}\right) &=&
\rho _{\left(\varepsilon_{n},1_{m};\varepsilon_{n+m}\right) \left(
\varepsilon_{n},1_{m};\varepsilon_{n+m}\right) } \notag \\
&& \quad +\rho _{\left( \varepsilon_{n},\varepsilon_{m};1_{n+m}\right) \left( \varepsilon_{n},\varepsilon_{m};1_{n+m}\right) }
+\rho_{\left( \varepsilon_{n},\varepsilon_{m};\varepsilon_{n+m}\right) \left(
\varepsilon_{n},\varepsilon_{m};\varepsilon_{n+m}\right) }
\end{eqnarray}%
\begin{eqnarray}
&&\rho _{\kappa_{\varepsilon_{n}}}^{A} =\frac{1}{\Pr\nolimits_{A}\left( \kappa_{\varepsilon_{n}}\right) }\left\{ \phi ^{-1}
\rho _{\left(\varepsilon_{n},1_{m};\varepsilon_{n+m}\right) \left( \varepsilon_{n},1_{m};\varepsilon_{n+m}\right) }
\left| \varepsilon_{n},1_{m};\varepsilon_{n+m}\right\rangle \left\langle
\varepsilon_{n},1_{m};\varepsilon_{n+m}\right| \right.  \notag \\
&& \qquad \left. +\phi ^{-2}\left( \rho _{\left(
\varepsilon_{n},\varepsilon_{m};1_{n+m}\right) \left( \varepsilon_{n},\varepsilon_{m};1_{n+m}\right) }+\rho _{\left( \varepsilon_{n},\varepsilon_{m};\varepsilon_{n+m}\right) \left(
\varepsilon_{n},\varepsilon_{m};\varepsilon_{n+m}\right) }\right) \right. \nonumber \\
&&\left. \! \! \phantom{\phi^{-1}} \times \left[ \left|
\varepsilon_{n},\varepsilon_{m};1_{n+m}\right\rangle \left\langle \varepsilon_{n},\varepsilon_{m};1_{n+m}\right|
+\left| \varepsilon_{n},\varepsilon_{m};\varepsilon_{n+m}\right\rangle
\left\langle \varepsilon_{n},\varepsilon_{m};\varepsilon_{n+m}\right| \right] \right\}
\end{eqnarray}

By varying $\beta $, one can distinguish whether $a_{\text{F}}$, the
$\overline{\text{Fib}}$ charge of a target anyon, is $1$ or $\varepsilon $, without
needing to know the precise value of the phase involved, because the interference
fringe amplitude is suppressed by a factor of $\phi ^{-2}\approx .38$ for
$\varepsilon $. We emphasize that this provides the $\overline{\text{RR}}_{3,1}$
state with a distinct advantage over the Moore-Read state with respect to being
able to distinguish the non-Abelian anyonic charges that would be used in these
systems as the computational basis states for topological qubits (i.e. $1$ and
$\psi $ for MR vs. $1$ and $\varepsilon $ for $\overline{\text{RR}}_{3,1}$).